\RequirePackage{fix-cm}
\documentclass[smallextended]{svjour3}       
\smartqed  
\usepackage{graphicx}
\usepackage{dcolumn}
\usepackage{bm}
\usepackage{url}
\usepackage{epstopdf}
\usepackage{amsfonts} 
\usepackage{relsize}
\usepackage{multirow}
\usepackage{physics}
\usepackage{makecell}
\usepackage{hyperref}
\hypersetup{
    colorlinks=true,
    linkcolor=blue,
    filecolor=magenta,      
    urlcolor=cyan,
    citecolor=blue,
}
\newtheorem{myprob}{Problem}
\usepackage{cite}
\usepackage{scalerel}

\newcommand{\diff}[1]{\ensuremath{\mathrm{d}#1}}

\begin{document}

\title{Quantum Path Computing:  Computing Architecture with  Propagation Paths in Multiple Plane Diffraction of Classical Sources of Fermion and Boson Particles}

\titlerunning{Quantum Path Computing}         

\author{Burhan Gulbahar        
}

\institute{B. Gulbahar \at
              Department of Electrical and Electronics Engineering \\
              Tel.: +90-236-2330131\
              Fax: +90-216-5649999\\
              \email{burhan.gulbahar@ozyegin.edu.tr}           
}

\date{Received: date / Accepted: date}

\maketitle

\begin{abstract}
Quantum computing (QC) architectures utilizing classical or coherent resources with Gaussian transformations are classically simulable as an indicator of the lack of QC power. Simple optical setups utilizing wave-particle duality and interferometers achieve QC speed-up with the cost of exponential complexity of resources in time, space or energy. However, linear optical networks composed of single photon inputs and photon number measurements such as boson sampling achieve solving problems which are not efficiently solvable by classical computers while emphasizing the power of linear optics. In this article, quantum path computing (QPC)  setup is introduced as the simplest optical QC satisfying five fundamental properties all-in-one: exploiting only the coherent sources being either fermion or boson, i.e., Gaussian wave packet of standard laser, simple setup of multiple plane diffraction (MPD) with multiple slits by creating distinct propagation paths, standard intensity measurement on the detector, energy efficient design and practical problem solving capability. MPD is unique with non-Gaussian transformations by realizing an exponentially increasing number of highly interfering propagation paths while making classical simulation significantly hard. It does not require single photon resources or number resolving detection mechanisms making the experimental implementation of QC significantly low complexity. QPC setup is utilized for the solutions of   specific instances of  two practical and hard number theoretical problems: partial sum of Riemann theta function and period finding to solve Diophantine approximation. Quantumness of MPD with negative volume of Wigner function is numerically analyzed and open issues  for the best utilization of QPC are discussed.
\end{abstract}

\section{Introduction}
\label{Section1} 

The Young's double slit experiment is at the heart of quantum mechanics (QM) with wave-particle duality as emphasized by Feynman \cite{feynman}.  Previous quantum computing (QC) systems utilizing classical optics, wave-particle duality or interferometer structures targeting a low complexity hardware design achieve QC speed-up for factoring problems, Gauss sum, generalized truncated Fourier sums and similar problems  \cite{puentes2004optical, vedral2010elusive, vcerny1993quantum, haist2007optical, rangelov2009factorizing}. However, they have apparently the cost of exponential complexity of resources in time, space or energy making their utilization impractical for problem solving.  On the other hand, more powerful architectures based on linear optics such as boson sampling achieve solving problems which are not efficiently solvable by classical computers \cite{aaronson2011computational}. They utilize challenging single photon  sources \cite{wang2018toward, flamini2018photonic} and the targeted problems are not directly practical such as matrix permanents  \cite{aaronson2011computational}. The design for a  significantly low hardware complexity optical QC is a promising dream which combines all-in-one targets: a) the practical problem solving capability, b) promising QC advantages with energy efficient processing of sources and measurement not requiring exponentially increasing resources with respect to the problem size, c) using only coherent or classical particle sources including both bosons and fermions, e.g., standard laser sources with Gaussian wave packets, d) transforming the emitted source through the simple classical optics and e) intensity measurement with simple and traditional particle detectors such as photodetectors or electron detectors for photon and electron, respectively. 

In this article, multi-plane diffraction (MPD) based design is proposed  as a simple extension of single plane  double-slit interference setup  while satisfying all the required properties of the ultimate target. It utilizes a novel resource for computation, i.e., exponentially increasing number of particle propagation paths bringing an exponentially large Hilbert space in time or history in an analogical manner to multi-particle entanglement in space. MPD allows energy efficient scaling of the problem size with increased number of slits and conventional sampling of the intensities of interfering paths \cite{gulbahar2018quantum}.
 
\begin{figure}[!t]
\centering
\includegraphics[width=4in]{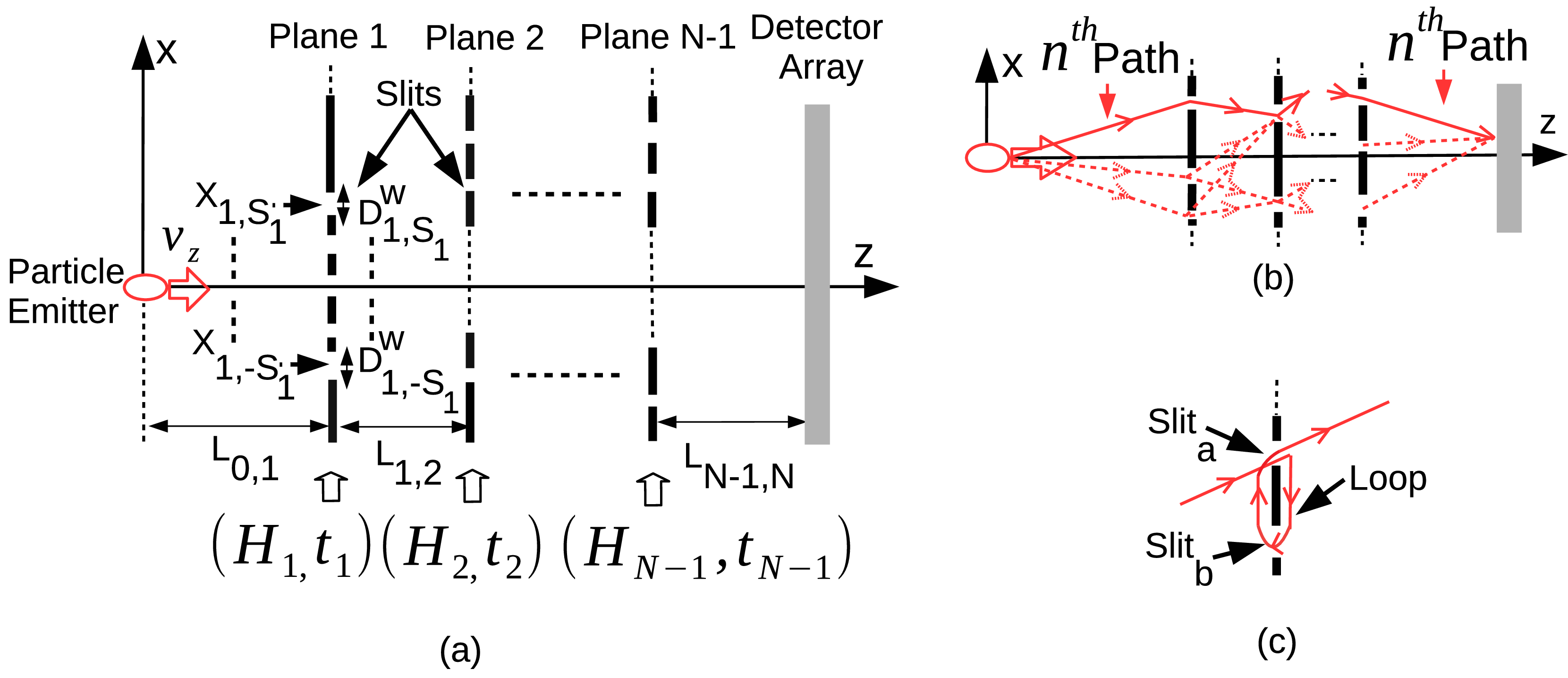}\\
\caption{(a) QPC architecture with  consecutive and parallel planes of slits,  Hilbert subspaces ($\mathcal{H}_j$ for $j \, \in [0, N-1]$) for the trajectories including diffraction at the event times $t_j$, (b) $n$th path interfering on the sensor plane, and (c) non-classical (exotic) path making  a loop between two neighbor slits  \cite{exotic, exotic2}.}
\label{Figure1}
\end{figure} 

 Feynman's history state idea  \cite{feynman2} entangles the time steps in the computation and clock register where the simulation wave function is represented as a complex superposition of the time steps of the  computation \cite{kitaev, bausch, tempelclock}.  The entire history of QC becomes the ground state of the local Hamiltonian as a superposition entangled with time. It is utilized to show equivalence of adiabatic and gate based QC in \cite{ahoronov2008adiabatic}. The history state is formulated as follows  \cite{kitaev}:
\begin{equation}
\ket{\Psi_{h,0}}  \equiv \frac{1}{N}\sum_{i = 0}^{N-1} \ket{\Psi_{t_i, 0}}  \equiv    \frac{1}{N}\sum_{i = 0}^{N-1}  U_{t_i} \,  U_{t_{i-1}}  \hdots U_{t_1} \ket{\Psi_{0}} \otimes \ket{t_i}
\end{equation}
where $\otimes$ is the tensor product, $\ket{\Psi_{0}}$ is the initial data register state, $\ket{t_i}$ is the clock register state and the operations $U_{t_i}$ for $i \in [1, N-1]$ denote  $N-1$ gates operating on the initial state.
Furthermore, the clock  is realized by a hopping Hamiltonian with a mapping to quantum walk (QW) as another important application of the Feynman-Kitaev construction \cite{caha2018clocks}. QPC includes the states of the wave function at different times  corresponding to the diffraction operations on each plane in an analogical manner to the entangled time steps.  The propagation of the initial coherent or classical wave function is tracked as a superposition through diffractions at specific time steps allowing to model the final wave function (instead of the final Hamiltonian as in \cite{ahoronov2008adiabatic}) in terms of the system parameters. It requires further analysis to construct a computational relation between history state based formulation with clock registers and QPC similar to the relation realized for adiabatic computing and QWs. QPC is much  more clearly modeled with consistent histories \cite{griffiths1984consistent, griffiths1993consistent, griffiths2003consistent} or entangled history formulation \cite{cotler2016, cotler2015} of QM as thoroughly formulated in \cite{gulbahar2018quantum} where  history state for the particle diffracting through planes is defined as follows:
\begin{equation}
\label{eq_0_9b3d_4dff} 
\vert \Psi_{N-1} ) = \sum_{n}  \pi_n \,  \mathbf{P}_{N-1,n_{N-1}}  \otimes \mathbf{P}_{N-2,n_{N-2}}  \otimes ...  \otimes \mathbf{P}_{1,n_1}
\end{equation}
where $\vert. )$ is the notation  introduced in \cite{cotler2016, cotler2015} for some history state between times $t_1$ and $t_{N-1}$,   the projector $\mathbf{P}_{j,i}$ denotes the diffraction where $i \, \in [-S_j, S_j]$  for the slit indices on $j$th plane and $\pi_n \, = \,1$ is the equal superposition of histories (or particle trajectories) indexed by $n$ as shown in Fig. \ref{Figure1}. Trajectory Hilbert space is defined as follows:
\begin{equation}
\label{hilbert1}
\mathcal{H} \equiv \mathcal{H}_{N-1} \otimes ... \otimes \mathcal{H}_{1}
\end{equation}
where $\mathcal{H}_j$ denotes the family of the projectors ($\mathbf{P}_{j,i}$) through $j$th plane with slits having central positions and widths of $X_{j, i}$ and $D^w_{j,i}$, respectively, where $j \, \in [1, N-1]$ and $i \, \in [-S_j, S_j]$. This provides the QC power with an entanglement relation in time domain \cite{gulbahar2018quantum} in analogy to the spatial entanglement while it is best exploited with Feynman's path integral (FPI)  formulation modeling the propagation histories  with a sum-over-paths approach.

Semi-classical approximation of the propagator kernel in FPI formalism  is utilized to analyze classical simulability of Clifford gates on stabilizer states and Gottesman-Knill theorem in \cite{kocia2017semiclassical}.  Propagator is defined as follows \cite{feynman, kocia2017semiclassical}: 
\begin{equation}
\bra{x} e^{-i \, H \, t \, / \, \hbar} \ket{x'} = \int e^{(i \, / \, \hbar) S({\color{black}x, \,x', \,t})} \mathcal{D}[x(t)]
\end{equation}
where $\mathcal{D}[x(t)]$ is the integral over all paths and $S({\color{black}x, \,x', \,t})$ is the action for the path between $x$ and $x'$ evolving under Hamiltonian $H$ at the time $t$. The propagator is approximated semi-classically with quadratic functional variations around the classical paths as the following \cite{tannor2007introduction, kocia2017semiclassical}:
\begin{equation} 
\int e^{(i \, / \, \hbar) S({\color{black}x, \,x', \,t})} \mathcal{D}[x(t)] \approx \sum_{\mbox{classical paths}} \mathcal{D}\, e^{(i \, / \, \hbar) (S \, + \, \delta^2 S \, / \, 2) }
\end{equation}
where $\delta S = 0$ for classical paths. In \cite{kocia2017semiclassical}, it is observed for continuous systems  that the propagation between Gaussian states with harmonic Hamiltonian is simulable classically by requiring only a single classical path without the importance of the relative phases of different classical contributions. The same idea is extended to the discrete case showing single path contribution for Clifford gates while requiring exponentially large number of sum-over-paths with respect to the number of qubits in case of including T gates. In addition, in \cite{koh2017computing}, it is stressed out that a sum-over-paths approach for general quantum circuits includes an exponential number of terms without any efficient classical algorithm to compute this sum. The importance of interferometry architectures as a non-classicality measure is emphasized in \cite{yuan2018unification} where double-slit experiment is regarded as a process to probe the phase difference between different paths. Non-classical or quantum behavior, including quantum entanglement, discord and coherence, is measured with the interferometry capability by checking whether the measurement outcome is independent of the phase information. In this article, the classical paths are hardwired to the system setup as distinct paths of propagation through slit based trajectories forcing us to calculate the phases for each path with significant interference among the paths.  Intensity calculation after diffraction through consecutive planes requires the computation of exponentially increasing number of path amplitudes and phases while making the classical simulation significantly hard. 

In this article, FPI based modeling of QM is preferred to characterize the effect of history for each trajectory in an easy way, i.e., the consecutive effects of the physical parameters of the diffraction slits and the travel time among the planes. FPI includes the history based formulation as an inherent element with propagation kernels  more suitable to the main resource utilized for QC purposes, i.e., Hilbert space composed of the histories of the diffractive projections at specific time instants. It allows to obtain the superposition wave function easily and better formulates the exponential number of paths to compute  \cite{feynman}. On the other hand, it is an interesting open issue to formulate MPD with universal quantum circuit gates to understand the computational capability of MPD, e.g., testing for universal QC or modeling the group of the  gates which can be implemented with QPC.

\subsection{The Comparison with Linear Optics based Implementations}

Classical simulability of linear optics implementations is achieved if  the evolution of the state can be modeled in terms of a unitary matrix rather than  an exponential complexity \cite{knill2001scheme, aaronson2011computational}. Coherent state or classical inputs, e.g., Gaussian wave packet or the output of a standard laser, and adaptive Gaussian measurements are simulated in classical polynomial time since the computing system maps the original Gaussian source into Gaussian output states, e.g., with operations in Clifford semi-group \cite{bartlett2003requirement, bartlett2002efficient, sasaki2006multimode}.   Such states are tracked by only exploiting the mean and covariance representation of states while non-Gaussian transformations make the  clever representation of Gaussian states in terms of means and variances not adequate for efficient classical simulation.  Knill, Laflamme and Milburn (KLM) scheme achieves optical QC by including adaptive measurements and photon counting in addition to the simple linear optical setup \cite{knill2001scheme}. Furthermore,  in Boson sampling \cite{aaronson2011computational},   single-photon inputs, which are highly challenging with the exponentially increasing difficulty of scaling for high photon numbers \cite{wang2018toward, flamini2018photonic}, and photon number measurements are utilized which are different from the coherent Gaussian sources discussed in \cite{bartlett2003requirement}.  Modified versions of Boson sampling with Gaussian input sources and number-resolved photodetection are  discussed  in \cite{lund2014boson,hamilton2017gaussian,  rhode2013sampling, kruse2018detailed} introducing Gaussian version with strong evidence of classically hard simulation. They show the importance of the operations and the measurements rather than only the source for computational capabilities in a computing system.  

In this article, non-Gaussian states are generated with non-Gaussian transformations, i.e., MPD, converting coherent wave packets into a superposition compared with Gaussian transformations in Clifford semi-group preserving the Gaussian nature of the wave.  On the other hand, each path of the particle through slits can be simulated  classically since Gaussian property is preserved for each path due to the diffraction through Gaussian slits. However, MPD requires tracking exponentially large number of classical operations since the number of propagation paths is exponentially increasing with the number of diffraction planes and the output is a superposition of these paths. Classical simulation of MPD requires exponentially increasing classical resources to track each Gaussian state in the superposition output. MPD has validity for both bosons and fermions with coherent Gaussian sources providing a significant experimental advantage compared with the difficulty to generate single photons. Besides that, it allows solutions for practical number theoretical problems including the partial sum of Riemann theta function and period finding for solutions of specific instances of Diophantine approximation problem as discussed  in Sections \ref{Section7} and \ref{Section8}. Quantumness of MPD with classical coherent sources is shown theoretically and with practical simulation parameters in \cite{gulbahar2018quantum} by violating Leggett-Garg inequality (LGI) which is the time domain analog of Bell's inequality as another supporting observation for non-classical character of QPC setup based on MPD. LGIs and Bell inequalities are utilized to test for quantumness of the systems including QC architectures.

\subsection{Quantumness and Negative Volume of Wigner Function}
 Positivity of Wigner function is another indicator proposed for classicality and classical simulation of system states.  Wigner function of a Gaussian state  is Gaussian which is positive leading to a  quasi-classical description of Gaussian inputs and Gaussian transformations. The negativity  is proposed as a measure of quantum correlations including entanglement in  \cite{arkhipov2018negativity, siyouri2016negativity,  dahl2006entanglement} and as a resource for QC in \cite{veitch2012negative, raussendorf2017contextuality, albarelli2018resource}. Continuous variable Gottesman-Knill theorem is extended to a large class of non-Gaussian mixed states with positive Wigner function such that  even non-Gaussian input states are not enough for QC advantages  \cite{veitch2013efficient}. The negative volume of Wigner function is defined as follows \cite{kenfack2004negativity}:
\begin{equation}
V_N \equiv  \frac{1}{2} \, \bigg( \int \int \vert W(x,p) \vert \, dx \, dp \, -  \, 1 \bigg) 
\end{equation}
where the Wigner function for the density function $\rho$ in the one-dimensional position basis is calculated as follows:
\begin{equation}
  W(x,p)  \equiv \frac{1}{\pi} \, \int_{-\infty}^{\infty}  \bra{x \, + \, y} \rho \ket{x \, - \, y} e^{-i \, 2 \, \pi \, \frac{p}{\hbar} y} \, dy
\end{equation}

In addition, negativity of  Wigner function is exploited in numerous fields. An entropic parameter with quantum nature is proposed in \cite{kowalewska2008wigner} as an indicator of quantum chaos based on the negative volume of Wigner function as a non-classicality parameter. It is shown that the defined entropic parameter shows fast and large changes in the regions corresponding to classical chaos. In \cite{siyouri2019markovian}, negativity of Wigner function is utilized 
detect and quantify quantum correlations in open multipartite quantum systems under the influence of both Markovian
and non-Markovian environments. It is shown that the negativity is sensitive to quantum discord in these systems. In \cite{quijandria2018steady}, it is shown that nonlinearity of a continuously driven two-level system (TLS) is enough to generate Wigner non-classical states of light by calculating Wigner function of one-dimensional  and steady-state
resonance fluorescence.  Furthermore, the capability of the setup for generating the class of states necessary for universal quantum computing is emphasized. A rapid and coherent mechanical squeezer is introduced in \cite{bennett2018rapid} by utilizing four optomechanical pulses while squeezing of arbitrary mechanical inputs, including non-Gaussian states, is discussed by preserving negativity even in the presence of decoherence.  They emphasize applications in quantum information technologies to enhance the storage of phononic Schr{\"o}dinger cat states.  

In this article, superposition of Gaussian wave packets for each trajectory of the particle has significant interference with large negative volume of Wigner function as shown in simulation studies in Section \ref{Section10} with simultaneously increasing volume of the negativity and the number of propagation paths.
  
\subsection{The Comparison with Quantum Walks}
  QW  based architectures present alternative systems to the standard circuit model for QC with the speed-up of search algorithms and universal QC capability \cite{childs2009universal}. QW is considered as an extension of the classical counterpart where a walker is jumping on the sites of a lattice  with a given probability \cite{sansoni2012two}. The  discrete and continuous QW types have the fundamental features of interference and superposition with non-classical dynamic evolution, i.e.,  Schr{\"o}dinger dynamics of the jumper particle. An analogy between QW and multiple-slit interference architecture is proposed in \cite{qwalk1}. Similarly, the role played by the interference effects in the dynamics of a quantum walker and  simulations based on interferometric devices are discussed in \cite{qwalk4}. There are also optical implementations of QWs resembling the structure of MPD with increasing numbers of trajectories such as multi-dimensional QWs implemented with classical optics \cite{goyal2015implementation, tang2018experimental, schreiber20122d}. Implementations of a QW on a line can be  described by classical physics \cite{knight2003quantum, jeong2004simulation}.  Single photon QWs are simulated by classical coherent waves with the measurement of light intensity since single and multiple photon problems can be described with the same probability distributions \cite{qi2014, perets2008realization}. 

MPD is analogical to QW models in terms of exploiting the classical and coherent wave sources, exponentially increasing number of trajectories, interference and superposition while with the following fundamental differences:
\begin{itemize}
\item MPD particle covers all lattice locations (diffraction slits on the $j$th plane at the time $t_j$) at a single time-step at once rather than adjacency based evolution. As an example, assume that a particle in a QW setup jumps to neighbor locations for a single line model. Then, there are $\prod_{i=1}^{N-1} (2 \, i) = 2^{N-1} (N-1)!$ paths at $(N-1)$th time step. Assume that  the number of slits on each plane for MPD is chosen as  $2 \,(N-1)+1$ corresponding to the maximum number of lattice sites at  $(N-1)$th QW step. Then, the number of paths in MPD grows as $ (2\,N-1)^i$ at $t_i$ compared with $2^i \, i!$ in QW such that MPD has exponentially larger number of paths with an exponentially larger Hilbert space compared with the fundamental QW model.  
\item MPD includes the effects of exotic paths, i.e., visiting the slits on the same plane as discussed in Section \ref{Section9} and as shown in Fig. \ref{Figure1}(c),  as another factor increasing the Hilbert space of MPD in an exponentially large manner.
\item The model of the problems for QWs and MPD are different, e.g., ballistic expansion of  the particle, exponentially faster hitting times, quantum search or graph isomorphism based problems in QW \cite{lovett2010universal} compared with numerical problems related to Riemann theta functions or hidden subgroup problems in QPC by introducing a novel set of practical problems.  
\item There is not any coin operation in MPD  to determine the next step movement. The physical properties of individual slits, i.e., the diameter and position in the proposed Gaussian slit model, combined with the history of the particle until the time of diffraction determines the probability of the particle to be diffracted through the slits on the next plane.
\end{itemize} 
As a final remark, QPC proposes a setup based on coherent particle source and linear optics requiring new approaches to understand the exact nature of resources for QC advantages in a QC system. MPD has analogies with boson sampling and QWs, and promising properties in terms of sum-over-paths complexity, negativity of the Wigner function and violation of Leggett-Garg inequality in \cite{gulbahar2018quantum}. It is an open question as clearly emphasized in \cite{venegas2012quantum, ferrie2011quasi}  to characterize the exclusive QM properties and operations which are enhancing computing capabilities. The resources for   QC are observed to be specific to the setup without allowing to simplify to a single resource or reason \cite{vedral2010elusive}.
 
\subsection{Contributions and Main Results}

The contributions in this article are summarized as follows:
\begin{enumerate}
\item  QPC as the simplest optical QC design achieves simultaneous targets all-in-one: a) practical problem solving capability with applications for partial Riemann theta sum and period finding for solutions of specific instances of Diophantine approximation problem, b) energy efficient processing of sources and measurement, c) exploiting coherent or classical particle sources including both bosons and fermions, d)  simple classical optics of MPD, and e)  intensity measurement with traditional detectors.
\item QPC, for the first time, utilizes particle propagation trajectory based Hilbert space for QC purposes as a solid example of the practical utilization of history based entanglement resources. 
\item Introduction and numerical simulation of a novel performance metric for the trade off between the problem complexity modeled as the number of the interfering paths and the total energy to realize interference pattern.
\item Theoretical modeling and numerical analysis of utilization of QPC for  specific instances of two important and hard number theoretical problems: partial sum of Riemann theta sum and period finding for simultaneous Diophantine approximation (SDA) problem.
\item Extending single plane  exotic path modeling and numerical analysis in   \cite{exotic, exotic2} to propagation through multiple planes with multiple slits.
\end{enumerate} 

QPC generates a black-box (BB) function $f_{BB}[k]$ with a promising special form as thoroughly discussed in Sections \ref{Section5} and \ref{Section7} to utilize in solutions of important and classically hard number theoretical problems as follows: 
 \begin{align}
 \label{blackboxfunc} 
\begin{split} 
  f_{BB}[k] \equiv \Bigg \vert \sum_{x_1 \in X_1} \hdots \sum_{x_{N-1} \in X_{N-1}}  \, e^{(A_{x} \,  + \, \imath \, B_x) (k \, T_s)^2} \,   \Upsilon_x  \, e^{  \overrightarrow{x}^T \, \mathbf{H_{x}}  \, \overrightarrow{x}  }   
 \,   e^{(\overrightarrow{h}_{x}^T  \,  \overrightarrow{x})  \, k \, T_s}      \Bigg \vert^2     & 
\end{split}
\end{align} 
where $k \in \mathbb{Z}$, $T_s \in \mathbb{R^+}$ is a sampling interval, $A_x \in \mathbb{R^-}$,  $B_x \in \mathbb{R^+}$,  $\Upsilon_x \, \in \mathbb{C}$, $ \overrightarrow{x} = [x_1 \, \, \, x_2 \hdots x_{N-1}]^T$ is a column vector composed of the slit positions  $x_j$ on each $j$th plane chosen from the corresponding set $X_j$ with a countable number of elements. The complex valued matrix   $\mathbf{H}_x \equiv \mathbf{H}_{R,x} \, + \, \imath\, \mathbf{H}_{I,x}$ and the vector $\overrightarrow{h}_x \equiv \overrightarrow{c}_x + \, \imath\,\overrightarrow{d}_x$ of the system setup have the values depending on the slit widths on each $j$th  plane  for $j \in [1, N-1]$  corresponding to the specific selection of slits in the path  $\overrightarrow{x}$, inter-plane durations for the particle propagation, particle mass $m$, beam width  $\sigma_0$ of the Gaussian source wave packet and Planck's constant $\hbar$. Each selection of the slits in $\overrightarrow{x}$ corresponds to a unique path for the particle to diffract. Therefore, the positions of the slits identify the index of a particular path or trajectory. In this article, the computational hardness of calculating (\ref{blackboxfunc}) in  an  efficient manner is discussed and two different methods utilizing (\ref{blackboxfunc}) for practical problems are introduced. 
  
The first method exploiting QPC calculates partial sum of Riemann theta function or multi-dimensional theta function as modeled in detail in Section \ref{Section7} with important applications in number theory and geometry \cite{riemann1857theorie, deconinck2004computing, mumford1983tata, osborne2002nonlinear, wahls2015fast, frauendiener2017efficient}. If the the slit  widths on each plane are constrained as being uniform specific to each plane, then the parameters  $A$, B, $\Upsilon$, $\mathbf{H}$ and $\overrightarrow{h}$ in (\ref{blackboxfunc}) become independent of the specific path $\overrightarrow{x}$. Then, BB function is converted to a form of partial sum of Riemann theta function.  The first utilization of QPC is to prepare a setup to solve specific groups of Riemann theta functions. Riemann theta function has important computational difficulties requiring complicated methods for the large number of contributions in the summation growing exponentially with $N$.  Therefore, the more complicated form  in (\ref{blackboxfunc}) with matrix and vector parameters depending on the path $\overrightarrow{x}$ has a much harder computational complexity. There is no apparent way of computing (\ref{blackboxfunc}) in a classically efficient manner for the specific sets of the matrices and vectors corresponding to a general experimental MPD setup with the user determined system parameters.

The second solution method based on QPC utilizes the phase in $ e^{(\overrightarrow{h}_{x}^T  \,  \overrightarrow{x})  \, k \, T_s}$  $=$  $e^{(\overrightarrow{c}_{x}^T  \,  \overrightarrow{x})  \, k \, T_s}  e^{(\imath \, \overrightarrow{d}_{x}^T  \,  \overrightarrow{x})  \, k \, T_s}$ for period finding and the solution of specific instances of SDA problems. QPC period finding algorithm is introduced in Section \ref{Section8} in analogy to QC period finding based on quantum gates \cite{nc}. Exponentially growing number of different $b[n] \in \mathbb{R}$ values are obtained with the multiplication $b[n] \propto \overrightarrow{d}_{x}^T  \,  \overrightarrow{x}$ varying for each path $\overrightarrow{x}$ indexed with $n$ as a classically hard SDA problem. Simple and classically solvable versions obtained with $b[n] \propto \overrightarrow{d}^T  \,  \overrightarrow{x}$ with path independent $\overrightarrow{d}$ {\color{black}are} numerically analyzed to understand the main idea in QPC based period finding.
 
In addition, a novel performance metric is introduced by emphasizing the trade off between the required number of particles or the amount of energy sources to accurately compute $f_{BB}[k]$ for solving a specific problem and the number of interfering paths. The non-classical properties of  MPD  is further analyzed and simulated by calculating the negative volume of Wigner function in comparison with the logarithmic number of the propagation paths.

Some open issues are discussed.  It is an open issue to find the sets of SDA problems which can be solved with an energy efficient QPC setup. Furthermore, designing the optimum algorithm to perform period finding in analogy to QC period finding algorithms utilizing continued fractions and inverse fast Fourier transform (IFFT) is an open issue \cite{nc}. Moreover, determining whether the problems whose solutions can be efficiently provided with QPC can also be efficiently solved with classical computers is another important open issue.  Formal complexity analysis of the QPC power obtained with  (\ref{blackboxfunc}) is an open issue. Besides that, it is an open issue to design a novel multiple time diffraction setup with different geometries rather than simple planar diffractions in a manner tuned to a specific target problem. On the other hand, the modeling of BB  function for the setups with arbitrary slits is an open issue compared with the Gaussian slit assumption in the article. The extension to arbitrary slits results in the solutions of different computational problems.

\subsection{Methodology}
Exponentially increasing number of interfering trajectories or paths {\color{black}are} utilized to define a novel resource for QC, i.e., Hilbert space of the particle propagation trajectories. A novel computing solution denoted by  QPC is defined by exploiting two special novel features:
\begin{enumerate}
 \item Consecutive and parallel diffraction planes with  multiple slits creating exponentially large number of particle trajectories until being detected on the final plane, i.e., sensor plane,  creating tensor product Hilbert subspaces of diffraction through each plane. Calculation of the exact intensity distribution on each plane requires exponentially increasing number of path integrals or summations making the classical simulation significantly difficult. It is valid for both bosons and fermions including electrons, photons, neutrons and even molecules. The particle source is assumed to be a Gaussian wave packet as the coherent or classical output of a standard laser.
 \item Computation capability of the special BB function in  (\ref{blackboxfunc}) or (\ref{qpcpowergeneral}) as the main computing power of the system design.  Energy-complexity trade off is analyzed based on the number of required summations of the paths on the sensor plane compared with the total probability of the measurement. Increasing number of slits with closely spaced spatial intervals results in an increase in both the complexity and the probability of the measurement as a unique power and advantage of MPD design. 
\end{enumerate}    

There is not any measurement regarding a specific trajectory but only interference pattern on the final plane without violating standard QM. Interference experiments are recently getting more attention to analyze non-classical (exotic) paths, e.g., passing through the slits on the same plane consecutively and even multiple times as shown in Fig. \ref{Figure1}(c), and Gouy phase effect in the measurement of Sorkin parameter \cite{exotic, exotic2}.  QPC extends, for the first time, previous formulation to MPD setups while simulating the effects of multiple exotic paths on multiple planes compared with previous studies utilizing single plane based diffraction and single exotic path \cite{exotic, exotic2}.

\subsection{Organization}
In Section \ref{Section2},   physical setup is presented. In Sections \ref{Section3} and \ref{Section4},  trajectory Hilbert space  and MPD modeling with FPIs are presented, respectively. QPC  BB function  and the computational hardness are discussed in Section \ref{Section5}. Energy flow versus complexity trade off is modeled in Section  \ref{Section6}. In Sections \ref{Section7} and \ref{Section8}, the application of QPC for partial sum of Riemann theta function and period finding are presented, respectively. In Section \ref{Section9}, effects of non-classical paths are modeled while in Section \ref{Section10}, numerical simulations are performed. Finally, in Sections \ref{Section11} and \ref{Section12}, open issues and conclusions are presented, respectively.

\section{Multi-plane Diffraction System Design}
\label{Section2}

There are $N-1$ planes of slits in front of a particle source and the interference pattern is observed by the sensor plane with the index $N$ as shown in Fig. \ref{Figure1}(a). Particles  are assumed to perform free space propagation  between the planes. The plane  with the index $j$ has   in total  $S_{j,T}  \equiv 2 \, S_j\, + \, 1$ slits   with $(.)_{j,T}$ representing the total and $S_j$ is utilized to index the slits with the numbers between $-S_j$ and $S_j$.  The central positions and  widths of slits are  given by $X_{j, i}$ and $D^w_{j,i}$, respectively, where $j \, \in [1, N-1]$ and $i \, \in [-S_j, S_j]$. The  set of ordered slit positions  on $j$th plane is denoted by the column vector $\overrightarrow{X}_j$  or with the set denoted by $X_j$.   Row vectors are represented with the transpose operation, i.e., $(.)^T$. The whole set of  slit positions on $N-1$ parallel planes are denoted by $\mathbf{X}_{N-1}$. Distance between  $i$th and $j$th planes is given by $L_{i,j}$ where the distances from particle emission source to the first plane and from $(N-1)$th plane to the detection plane are given by $L_{0,1}$ and $L_{N-1,N}$, respectively. Behavior of  the particle is assumed to be classical in $z$-axis with the velocity given by $v_z$ while  quantum superposition interference is assumed to be observed in $x$-axis as a one dimensional model to be easily extended to two dimensional (2D) systems.  
 
Time duration for the particle to travel between $(j-1)$th and $j$th planes is assumed to be $t_{j-1,j} \, = \, L_{j-1,j} \, / \, v_z  $ for  $j \, \in [1, N]$. Position in $x$-axis on  $j$th plane is denoted by $x_j$ while the wave functions of $n$th path and superposition of all paths on  $j$th plane are denoted by  $\Psi_{n,j}(x_j)$ and $\Psi_j(x_j)$, respectively.  Inter-plane distance and  duration vectors are represented by $\overrightarrow{L}^T =[L_{0,1} \, \hdots \,L_{N-1, N}]$ and $\overrightarrow{t}^T = [t_{0,1}\, \hdots \, t_{N-1,N}]$, respectively. Trajectories are indexed by $n$ for $n \in [0, N_p-1]$ as shown in Fig. \ref{Figure1}(b) where $N_p = \prod_{j=1}^{N-1} S_{j,T}$  is the total number of paths until to the sensor plane ($N$th plane) measurement and $Path_n \equiv \lbrace s_{n,1}, \, s_{n,2}, \, \hdots \, s_{n,N-1}; \, s_{n,j} \in [-S_j, S_j]  \rbrace$ is the indices of the slits for $n$th path.  Therefore,  slit position for $n$th path on $j$th plane is given by $X_{j, s_{n,j}}$.  Similarly, $N_{p,j} \equiv \prod_{i=1}^{j-1} S_{i,T}$ denotes the number of paths for the particle diffracting through the $j-1$th plane.

Calculation of  inter-plane durations by $t_{j-1,j} = L_{j-1, j}\, / \, v_z $ is accurate due to $L_{j-1,j} \gg D^w_{j-1,i}, X_{j-1,i}$ for $j  \, \in [2, {\color{black}N}]$ and $i \in [-S_{j{\color{black}-1}}, S_{j{\color{black}-1}}]$ such that  quantum effects are emphasized in  $x$-axis. Non-relativistic modeling of  particle behavior is assumed. Source is a single Gaussian wave function while  Gaussian slits are utilized with FPI approach \cite{feynman}. Next, trajectory Hilbert space as the resource for QPC is described.  

\section{Trajectory Hilbert Space}
\label{Section3}
 
QPC realizes subspaces analogical to  spatial qudits such that diffractive projection family through the set of the slits on $j$th plane results in a Hilbert subspace at $t_j$ as shown  in Fig. \ref{Figure1}(a). There is not any measurement to determine the diffracted slit positions in any trajectory. It does not violate   the standard interpretation of QM while utilizing superposition of trajectories with a tensor product space of diffraction events \cite{griffiths2003consistent}. FPI methodology results in the intensity measurement  $I_N(x)  =  \,  \big \vert \sum_{n=0}^{N_p-1} \Psi_{n,N}(x) \big \vert^2$ on $N$th plane as follows: 
 \begin{align} 
 \label{sQM}
\begin{split}
I_N(x)  = \, \bigg \vert \sum_{n=0}^{N_p-1} \int_{\overrightarrow{x}} K_n(x, \overrightarrow{x};  t_N, t_{N-1}, \hdots, t_0) \, \Psi_{0}(x_0) \, \diff \overrightarrow{x} \bigg \vert^2    
\end{split}
\end{align}
where  $t_0$ and $t_N$ are initial and the measurement times, respectively, $t_j$ for $j \in [1, N-1]$ is the diffraction time, $x_0$ and $x$ are the initial and the sensor plane position variables, respectively,  $\Psi_{0}(x)$ is the coherent source wave function, $\Psi_{n,N}(x)$  is the wave function of $n$th trajectory on the sensor plane, $\int_{\overrightarrow{x}} \diff \overrightarrow{x}$ denotes the integration with respect to $x_j$ for $j \, \in  [0, N-1]$ and $K_n(x, \overrightarrow{x};  t_N, t_{N-1}, \hdots, t_0)= K_n(x, x_{N-1}, \hdots, x_0;  t_N, t_{N-1}, \hdots, t_0)$ is the overall propagation kernel with the detailed models defined in Section \ref{Section4}.

The trajectory of the particle is defined as a sequence of projection operators corresponding to the diffraction through slits. Consecutive set of slits for $n$th trajectory is defined  as  $S \rightarrow X_{1, s_{n,1}} \rightarrow X_{2, s_{n,2}} \rightarrow  ...\rightarrow X_{N-1, s_{n,N-1}}$ where  $S$ is the initial state  at the source at $t_0$. Trajectory Hilbert space is defined  in (\ref{hilbert1}) in Section \ref{Section1}. Therefore, as the particle passes through multiple planes, each possible trajectory results in an interfering functional contribution on the final wave function on the sensor plane.  Projection operators denoting the particle to be in the Gaussian slit (for a one dimensional model for simplicity) are defined in a  coarse grained  sense as discussed in \cite{dowker1992quantum} as follows:
\begin{equation} 
P_{\beta_{j,i}}(X_{j, i}) \equiv  \int_{ -\infty}^{\infty}  \diff x \, \mbox{exp} \bigg ( - \frac{(x - X_{j, i})^2}{2 \,\beta_{j,i}^2 }\bigg)\ket{x} \bra{x}
\end{equation}
where the effective slit width is $D^w_{j,i} \equiv 2 \, \beta_{j,i}$, $j \, \in [1, N-1]$ and $i \, \in [-S_j, S_j]$. If the slit widths are uniform for each $j$th plane with $\beta_j$, then $\overrightarrow{\beta}  \equiv [\beta_1, \hdots, \beta_{N-1}]^T$ represents the vector of the slit widths. If the slit widths are different, then $S_{\beta}$ denotes the set with the elements $\beta_{j,i}$. The set of Gaussian slit projectors satisfies mutual exclusivity in an approximate sense since the integrals include  intersections of  slit intervals defined by the widths $\beta_{j,i}$. In simulations, slit distances are chosen large enough to satisfy $\mbox{exp} \big ( - (X_{j, m} - X_{j, l})^2 \, / \, (2 \,\beta_{j,m}^2) \big) \ll 1$ for $m \neq l$.  Next,  FPI  modeling of MPD is presented. 

\section{Multi-plane  Diffraction Modeling}
\label{Section4}
 
\begin{figure*}[!t]
\centering
\includegraphics[width=12.2cm, height = 1.2cm]{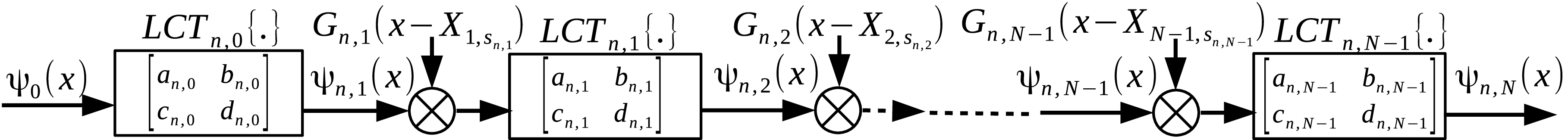}\\
\caption{Evolution of $\Psi_0 (x)$ in $n$th path as consecutive operations of $LCT_{n,0}\lbrace . \rbrace$ followed by the operations of $LCT_{n, j}\lbrace . \rbrace$ and multiplication by the effective slit functions $G_{n, j}(x_j)$ for $j \in [1, N-1]$ resulting in the final wave function of $\Psi_{n, N} (x)$.} 
\label{Figure2}
\end{figure*}
$\Psi_{n,N}(x)$ is calculated with free particle kernels \cite{feynman}. $K(x_1, t_1; x_0, t_0)$ denotes free particle kernel for the paths between  time-position values $(t_0, x_0)$ and $(t_1, x_1)$ defined as follows:
\begin{equation}
K(x_1, t_1; x_0, t_0) = \sqrt{m  /   (2 \, \pi \, \imath \, \hbar \, \Delta t)} \,  \mbox{exp}(\imath \, m \,  \Delta x^2 \, / \, (2 \, \hbar \, \Delta t))
\end{equation}
where $\Delta t = t_1 - t_0$ and $\Delta x = x_1 - x_0$ and $m$ is the free particle mass. 
If $\int_{\overrightarrow{x}} \diff \overrightarrow{x}$ denotes the integration with respect to  $x_j$ for $j \, \in  [0, N-1]$ between  $-\infty$ and $\infty$,  then $\Psi_{n,N}(x)$ is given as follows by describing $K_n(.)$  in (\ref{sQM}):
\newcommand*{\Scale}[2][4]{\scalebox{#1}{$#2$}}%
\begin{align}
\begin{split}
\label{eq1}
\Scale[1.45]{\int}_{\overrightarrow{x}} \, \Scale[0.95]{\diff \overrightarrow{x}   \, K(x, t_{N}; x_{N-1}, t_{N-1})  \, G_{n, N-1 }(x_{N-1} - X_{N-1, s_{n,N-1}})    } &\\
\Scale[0.93]{\big( \prod_{j = 1}^{N-2} K(x_{j+1}, t_{j+1}; x_j, t_j)  \,  G_{n,j}(x_{j}  - X_{j, s_{n,j}})  \big)  \,  K(x_1, t_1; x_0, t_0) \, \Psi_{0}(x_0)}  &
\end{split}
\end{align}
where $t_j = t_0 + \sum_{k=1}^{j} t_{k-1,k}$ for $j \, \in [0,N]$ and $G_{n,j}(x_j)$ denotes the effective function of the slit with  the index $s_{n,j}$ on $j$th  plane for $n$th path.  The result is described in terms of linear canonical transforms (LCTs). LCT of a function $f(x)$, i.e.,  $LCT_{a,b,c,d} \lbrace f(x) \rbrace$, is defined as   $ exp \lbrace-\imath \frac{\pi}{4} \rbrace$  $\sqrt{\eta}$   $\int_{-\infty}^{\infty} exp \lbrace \imath \pi \, (\alpha \, x^2 - 2 \, \eta \, x\, u  \, + \gamma \, u^2 ) \rbrace f(u) \, \diff u$ where  LCT matrix is  $\lbrace a, b, c, d \rbrace \equiv \lbrace  \gamma\, / \, \eta,       1 \ / \, \eta, (\alpha \, \gamma - \eta^2) \, / \, \eta, \alpha \, / \,\eta \rbrace$ and $a\, d - b \,c = 1$ for a given set of parameters $(\alpha, \gamma, \eta)$ \cite{haldun2}. Then, evolution of $\Psi_0 (x_0)$ is represented as shown in Fig. \ref{Figure2} where $LCT_{n,j}\lbrace . \rbrace$ denotes the LCT with the matrix $\lbrace a_{n,j}, b_{n,j}, c_{n,j}, d_{n,j} \rbrace \equiv \lbrace 1,  2 \, \pi \, \hbar \, t_{j, j+1} \, / \, m, 0,  1     \rbrace$ with the transformation parameters $\alpha = \gamma = \eta = m \, / \, (2 \, \pi \, \hbar \, t_{j, j+1})$ for $j \, \in [0,N-1]$ not depending on the path index $n$ due to the classical approximation in $z$-axis.  Next, QPC BB function and its computational hardness are discussed.

\section{Quantum Path Computing Black Box Function and Computational Hardness}
\label{Section5}
 
Interference on sensor plane  is transformed into a form to exploit  quantum superposition and computation of BB function for performing QC tasks. For simplicity of calculation, we firstly assume that the slit  widths are the same on a single plane, i.e., $G_{n,j}(x) = \mbox{exp}(- \, x^2 \, / \, (2 \, \beta_j^2) )$  with $D^w_{j,i} \equiv 2 \, \beta_j$ for $j$th plane. Then, it is extended to a general MPD setup with different slit widths, i.e., $D^w_{j,i} \equiv 2 \, \beta_{j,i}$.  The  source wave function is a Gaussian wave packet of the form $\Psi_0(x) \,  =  \, \mbox{exp}\big(- \, x^2 \, / \, (2 \, \sigma_0^2) \big) \, / \, \sqrt{\sigma_0 \,\sqrt{\pi }} $ \cite{feynman, exotic}.  Then, after taking the consecutive path integrals of $\Psi_0(x)$ through each path as shown in Appendix  \ref{AppendixA}, the following superposition  intensity $I_{N}(x)$   is obtained:
 \begin{align}
 \label{qpcpower}  
\begin{split}
 I_{N}(x) =  \, e^{2 \, A_{N-1} \, x^2} \, 
 \bigg \vert \sum_{n=0}^{N_p-1}  \Upsilon_N\, e^{ \overrightarrow{x}_{n}^T \, \mathbf{H}_{\color{black}R}  \, \overrightarrow{x}_{n} } \,e^{ \imath\, \overrightarrow{x}_{n}^T \, \mathbf{H}_{\color{black}I}  \, \overrightarrow{x}_{n} } \, e^{\overrightarrow{c}^T   \overrightarrow{x}_{n} \, x}  \,e^{\imath \, \overrightarrow{d}^T    \overrightarrow{x}_{n} \, x} \bigg \vert^2     & 
\end{split}
\end{align} 
where $\mathbf{H} = \mathbf{H}_{\color{black}R} \, + \, \imath \, \mathbf{H}_{\color{black}I}$ is $N-1 \times N-1$ matrix which is composed of correlated  real  $\mathbf{H}_{\color{black}R}$ and imaginary parts $\mathbf{H}_{\color{black}I}$,   $A_{N-1}$ is a real negative variable, $\Upsilon_N$ is a complex variable,   $\overrightarrow{c}$ and $\overrightarrow{d}$ are $N-1$ dimensional column vectors, and the slit position vector for $n$th trajectory is $\overrightarrow{x}_{n} \equiv \left[ X_{1, s_{n,1}} \, \hdots \,  X_{N-1, s_{n,N-1}}  \right]^T$. The parameters $\Upsilon_N$, $\mathbf{H}$, $\overrightarrow{c}$, $\overrightarrow{d}$  and  $A_{N-1}$ depend on  $\overrightarrow{\beta}$, inter plane duration vector $\overrightarrow{t}$, the source parameter $\sigma_0$, Planck's constant $\hbar$ and particle mass $m$. They do not depend on trajectory index $n$  or  $\overrightarrow{x}_{n}$  due to the assumption of uniform slit widths. The relaxation of uniform slit widths results in trajectory dependent parameter sets as shown next.   The combined design of $\overrightarrow{x}_{n}$, $\Upsilon_N$, $\mathbf{H}$, $\overrightarrow{c}$, $\overrightarrow{d}$ and $A_{N-1}$ while choosing adapted set of $x$-axis samples promises a solution to optimization problems with significantly large $N_p$. 

If the constraint of the uniform slit width on each plane is relaxed, then  $\Upsilon_N$, $\mathbf{H}$,  $\overrightarrow{c}$, $\overrightarrow{d}$ and $A_{N-1}$ will all depend on the path index $n$ since there is a consecutive set of different $\beta_{j,i}$ values along each path effecting the final output value. If the new parameters depending on $n$ are denoted with $A_{N-1,n}$, $B_{N-1,n}$, $\Upsilon_{N,n}$, $\mathbf{H}_{N-1,n} \equiv \mathbf{H}_{R,N-1,n} \, + \, \imath \, \mathbf{H}_{{\color{black}I,N-1,n}}  $,  $\overrightarrow{c}_{N-1,n}$ and  $\overrightarrow{d}_{N-1,n}$, then the general form of the output intensity denoted by $I_{N}^G(x)$ is given as follows:
 \begin{align}
 \label{qpcpowergeneral}
\begin{split}
  I_{N}^G(x) =  \,  
 \Bigg \vert \sum_{n=0}^{N_p-1} \bigg( e^{(A_{N-1,n} \, + \,  \imath \, B_{N-1,n} ) \, x^2} \, \Upsilon_{N,n} \,   e^{ \overrightarrow{x}_{n}^T \, \mathbf{H}_{{\color{black}R,N-1,n}}  \, \overrightarrow{x}_{n} }\hspace{1.3cm} & \\
 \, \times  \,e^{ \imath\, \overrightarrow{x}_{n}^T \, \mathbf{H}_{{\color{black}I,N-1,n}}  \, \overrightarrow{x}_{n} }  \, e^{\overrightarrow{c}_{N-1,n}^T   \overrightarrow{x}_{n} \, x}  \,e^{\imath \, \overrightarrow{d}_{N-1,n}^T    \overrightarrow{x}_{n} \, x}  \bigg) \Bigg \vert^2      & 
\end{split} 
\end{align}
$I_{N}^G(x)$ has a much more complicated form without any apparent and efficient classical way to calculate compared to the approximation methods of Riemann theta function in (\ref{qpcpower}) which is already extremely hard to calculate classically. The matrices $\mathbf{H}_{N-1,n}$ changing for each trajectory make the problem significantly difficult. It is an open issue whether there exists a polynomial complexity solution to calculate $I_{N}^G(x)$ by exploiting the correlation among $\Upsilon_{N,n}$, $\mathbf{H}_{N-1,n}$,  $\overrightarrow{c}_{N-1,n}$, $\overrightarrow{d}_{N-1,n}$,  $A_{N-1,n}$ and $B_{N-1,n}$ for the hardest case of non-uniform slit widths and non-uniform slit position space including the path vectors  $\overrightarrow{x}_{n}$ obtained from  $X_{j, i}$ where $j \, \in [1, N-1]$ and $i \, \in [-S_j, S_j]$. In Section \ref{Section9}, the effects of exotic paths are modeled which requires further modification of (\ref{qpcpowergeneral}) to include exotic paths making the classical calculation much harder.   

It is an open issue to determine the complexity class of calculating (\ref{qpcpower}) and (\ref{qpcpowergeneral}) with classical and universal quantum computers. Another open issue is to determine the best method to utilize (\ref{qpcpower}) and (\ref{qpcpowergeneral})  for computational power and  solving appropriate numerical problems.    In Sections \ref{Section7} and \ref{Section8},  the solutions for partial sum of Riemann theta function (Chapter 8 in \cite{osborne2002nonlinear}) and period finding type solution for HSPs  \cite{nc} are provided as examples.   Next, a performance metric is defined for the trade off between energy and complexity.
 
\section{Energy Flow and Complexity}
\label{Section6}
Although Hilbert space of the paths enlarges exponentially,  the total number of large amplitude paths is smaller than the total number of possible paths as numerically analyzed in Section \ref{Section10}. The probability of detection decreases as the particle forwards.   There is a trade off between the particle energy and  the computational complexity (approximated as the number of paths required to be calculated). Magnitude of $n$th path on $j$th plane is defined as follows:
\begin{equation}
\label{pathamp}
I_{path}(n, j)   \equiv \sum_{k=-\infty}^{\infty} T_s \, \big \vert \Psi_{n,j}(k \, Ts) \big \vert^2
\end{equation}
The trade off provides a better representation of interference based Hilbert space as the number of planes increases. Many paths contribute little to a specific sampling position $k \, T_s$ while a large number of them should be taken into account to calculate the final intensity.  A performance metric for path magnitude based dimension ($D_{j}(\epsilon)$) and the probability of detection ($P_{j}$) denoted by $[D_j(\epsilon), \, P_{j}] $ is defined as follows:  
 \begin{align}
\label{hilbertdefinition}
\begin{split}
 D_j(\epsilon): \mbox{count of }  n \mbox{ where  }  I_{path}(n, j) > \epsilon \, I_{path}^{max}(n,j), & \\
P_{j} \equiv \sum_{k = -\infty}^{\infty}\, T_s \,I_{j}(k \, T_s) \hspace{0.7in}  &
\end{split}
\end{align}     
where $I_{path}^{max}(n,j) \equiv \max_{n  \, \in  \, [0, N_{p,j}-1]}\lbrace I_{path}(n, j) \rbrace$,  and $P_{j}$  and $I_{j}(x_j)$ are probability to be detected  and intensity on $j$th plane, respectively.  The number of paths effective on $j$th plane increases as  $\epsilon < 1$ decreases while $\epsilon \,= \, 0$ gives the total number of paths $N_{p,j}$.  The definition in (\ref{hilbertdefinition}) proposes an  interference based  and  energy constrained  Hilbert space where  $D_j(\epsilon) \times P_{j}$ is a novel performance metric. The intensity is normalized  with FPI modeling, i.e., $P_0 = P_1 = 1$.

Different sampling points on the sensor plane require different sets of paths in the summation as shown in numerical analysis in Section \ref{Section10}. Therefore,  reliable value of $\epsilon $ changes with respect to $k \, T_s$ while  $ \epsilon \, =  \, 0$ could be taken the most reliable value forcing the calculation of all the paths for the complete intensity waveform on the sensor plane.  The effects of the paths are analyzed by defining three different cumulative summation methods of the path wave functions with indices sorted with different mechanisms.  $I_{j,s}^a(k \,  T_s)$ and $I_{j,s}^b(k \,  T_s)$ denote the cumulative sum intensities by summing the contributions from the paths indexed by sorting with respect to the descending magnitude of the total probability of the path on the layer and the descending magnitude of the path wave function at the sampling point $k \, T_s$, respectively. Therefore, $I_{j,s}^b(k \,  T_s)$ has a local characteristics tuned to the sampling point. $I_{j,s}^c(k \,  T_s)$ denotes the cumulative sum with the paths indexed by sorting with respect to the paired trajectories almost canceling each other. In other words, the paths are firstly sorted with respect to descending magnitude of the wave functions at the sampling point $k \, T_s$ providing the first sorting outcome. Then, the first path is taken (index of $ \Psi_{n,j}(k \, Ts)$ starting from $n = 1$ instead of $n = 0$) and a search is performed among the remaining paths for minimizing $\vert \Psi_{1,j}(k \, Ts) +   \Psi_{{\color{black}m},j}(k \, Ts) \big \vert^2$ and a new index of $2$ is given to the path with the index ${\color{black}m}$ which almost cancels the wave function of the first path. Then, the next path in the first sorting outcome is taken and consecutive searches are made for each path with the same manner. In this method, we check whether there are specific groups of paths directly canceling each other. However, even if the paths cancel each other, it could be extremely hard to couple the paths among the exponentially large number of trajectories. As a result, the paths are indexed with three different sorting methods denoted with the indices $n_a$, $n_b$ and $n_c$ where the respective cumulative intensities until the path with the index $n_f \leq N_{p,j}$ are defined as follows: 
\begin{equation}
\label{pathampsorted}
I_{j,s}^{type}(n_f, k \,T_s)   \equiv  \bigg \vert \sum_{n_{type}=1}^{n_f} \Psi_{n_{type},j}(k \, Ts) \bigg \vert^2
\end{equation}
where $type$ denotes $a$, $b$ or $c$. Observe that all the sorting types sort the paths with respect to the descending order of the magnitudes. Next, two different methods for exploiting QPC in number theoretical problems are described. 

\section{QPC Solution-1: Partial Sum of Riemann Theta Function}
\label{Section7} 
  
The Riemann theta function designed by Riemann \cite{riemann1857theorie} generalizes Jacobi's theta functions of one variable \cite{jacobi1829fundamenta} to solve the Jacobi inversion problem \cite{deconinck2004computing, mumford1983tata}. It has important applications  in geometry, arithmetic and number theory including the theory of partition functions, representation of integers, evaluation of infinite formal products and modular forms \cite{mumford1983tata}, nonlinear spectral theory for water wave dynamics and oceanography \cite{osborne2002nonlinear},  nonlinear Fourier analysis \cite{wahls2015fast}, conformal field theories, partial differential equations and cryptography \cite{frauendiener2017efficient}. Theta function in $N \, - \, 1$ dimensions is defined as follows \cite{deconinck2004computing}:
\begin{equation}
\label{thetafunction}
\Theta(\mathbf{Y}, \overrightarrow{y}) \equiv \sum_{a_1 \in \mathbb{Z}}\hdots \sum_{a_{N-1} \in \mathbb{Z}} e^{-  \pi \, \overrightarrow{a}^T ( \mathbf{Y}_R  \, + \, \imath \,\mathbf{Y}_I) \, \overrightarrow{a}  \, + \,  2 \, \pi \, (\overrightarrow{y}_R^T  \, + \, \imath \,  \overrightarrow{y}_I^T)\, \overrightarrow{a}  }
\end{equation}
where $\overrightarrow{a}^T  \, = \, [a_1 \, a_2 \, \hdots \, a_{N-1}]$, $\mathbf{Y} \equiv \mathbf{Y}_R \, + \, \imath \, \mathbf{Y}_I$,  $\mathbf{Y}_R$ is a positive definite and symmetric real matrix, $\mathbf{Y}_I$ is a real symmetric matrix, $ \overrightarrow{y} \equiv \overrightarrow{y}_R^T  \, + \, \imath \,  \overrightarrow{y}_I^T$, $\overrightarrow{y}_R$ and $\overrightarrow{y}_I$ are real vectors. The positive definiteness of $\mathbf{Y}_R$ satisfies the convergence of the infinite summation. There are various methods utilized to approximate the series by utilizing partial summation of theta function  defined by limiting the bounds as follows:
\begin{equation}
\label{partialthetafunction}
\Theta_M(\mathbf{Y}, \overrightarrow{y}) \equiv \sum_{a_1 = -M}^{M}\hdots \sum_{a_{N-1}= -M}^{M} e^{-  \pi \, \overrightarrow{a}^T ( \mathbf{Y}_R  \, + \, \imath \,\mathbf{Y}_I) \, \overrightarrow{a}  \, + \,  2 \, \pi \, (\overrightarrow{y}_R^T  \, + \, \imath \,  \overrightarrow{y}_I^T)\, \overrightarrow{a}  }
\end{equation} 
It is exponentially hard to find the summation  in (\ref{partialthetafunction})  with the brute force method on $N-1$ dimensional cubic lattice space of $\overrightarrow{a}$ as thoroughly discussed in \cite{osborne2002nonlinear}.  Approximation methods satisfying special conditions are presented in \cite{osborne2002nonlinear, deconinck2004computing} with Fourier analysis and the summation over lattice spaces of spherical or  ellipsoidal volumes. Therefore, calculation of Riemann theta functions is a significant challenge as an important number theoretical problem.
 
The superposition wave function in (\ref{qpcpower}) is easily converted to Riemann theta function by choosing slit positions on a plane in  a periodic manner and with the constraint of uniform slit width specific to each plane. If $S_j \, = \, M$, $ X_{j,i} = a_j \, \Delta x_j$ for $j \in [1, N-1]$ and $a_j \in [-M, M]$ and $x = k \, T_s$, then  (\ref{qpcpower}) is transformed into $ e^{2 \, A_{N-1} \, k^2 T_s^2} \,\vert \Upsilon_{N} \vert^2 \, \widetilde{I}_{N}^{R}(x) $ where $\widetilde{I}_{N}^{R}(x)$ becomes the following:  
 \begin{align}  
 \label{qpcpowermultitheta}
\begin{split}
\widetilde{I}_{N}^{R}(x)  =  
 \bigg \vert \sum_{a_1 = -M}^{M}\hdots \sum_{a_{N-1}= -M}^{M} e^{  - \,\pi \, \overrightarrow{a}^T \,  \mathbf{\widetilde{Y}}   \, \overrightarrow{a} }   \, e^{2 \, \pi \, \overrightarrow{\widetilde{y}}^T   \overrightarrow{a}}    \bigg \vert^2  = \vert \Theta_M(\mathbf{\widetilde{Y}}, \overrightarrow{\widetilde{y}}) \vert^2    & 
\end{split}
\end{align}  
where $  \mathbf{\widetilde{Y}}  \equiv  - \, \mathbf{D} \, \mathbf{H} \, \mathbf{D} \, / \, \pi $, $\overrightarrow{\widetilde{y}} \equiv  x  \, {\color{black}\mathbf{D}} \,(\overrightarrow{c} \, + \, \imath \, \overrightarrow{d}) \,  / \, (2 \, \pi)$ and $\mathbf{D}$ is the diagonal matrix with the diagonal elements formed of $\lbrace \Delta x_1, \, \Delta x_2, \hdots, \Delta x_{N-1}\rbrace$. The quadratic form of $\overrightarrow{a}^T \,  \mathbf{\widetilde{Y}}   \, \overrightarrow{a}$ allows the calculation with a symmetric matrix  by converting $\mathbf{\widetilde{Y}}$ with $\mathbf{\widetilde{Y}_s} \, \equiv \,(\mathbf{\widetilde{Y}} \, + \, \mathbf{\widetilde{Y}}^T) \, / \, 2$. As a result, QPC setup is utilized to calculate the amplitudes of the specific Riemann theta functions with the matrix and vector input parameters defined by $\mathbf{\widetilde{Y}_s}$ and $\overrightarrow{\widetilde{y}}$, respectively.

On the other hand, QPC performs more complicated functions compared to Riemann theta function including the summation over irrational and non-uniform sampling grid $a_j$ for $j \in [1, N-1]$ compared to the integer and periodic grid of the sampling points. Furthermore, transforming the general wave function $I_{N}^G(x)$  with non-uniform slit widths results in a special form of Riemann theta function having different parameters  $\mathbf{\widetilde{Y}_s}$ and $\overrightarrow{\widetilde{y}}$ for each point on the summation grid. This problem has not any practical and visible method to practically calculate in polynomial time complexity with classical computers. It is an open issue to analyze whether there are classically efficient methods to compute the Riemann theta functions obtained with $\widetilde{I}_{N}^{R}(x)$ and $I_{N}^G(x)$.  Next, the phases of the defined wave forms are utilized for period finding type solutions for specific SDA problems.

\section{QPC Solution-2: Period Finding}
\label{Section8} 

An analogy is presented with QPC based solution and the period finding algorithms in traditional QC algorithms exploiting superposition and entanglement together to realize quantum Fourier transform (QFT).  A special function $f_n(\overrightarrow{x})$ is defined with  periodicity property. The analogy between  tensor product spaces of trajectories  and multiple particle entanglement resources is described in Table \ref{Table1}. Intensity is  sampled   as follows: 
  \begin{align} 
  \label{qphase1} 
\begin{split}
  I^G_N[k]\,  \equiv \, & \bigg \vert \sum_{n=0}^{N_p-1}  e^{  (A_{N-1,n} \, + \, \imath \, B_{N-1,n}) \, (k\,T_s)^2} \Upsilon_{N,n}   \\
  & \, e^{ \overrightarrow{x}_{n}^T \,  \mathbf{H}_{{\color{black}R,N-1,n}}      \, \overrightarrow{x}_{n} }  \, e^{ \overrightarrow{c}_{N-1,n}^T  \,  \overrightarrow{x}_{n} \, k\, T_s} e^{ \imath \,\Theta[n,k]} \bigg \vert^2   \\
=  \, & \bigg \vert \sum_{n=0}^{N_p-1} \gamma_{f,n}\bigg(\frac{T_s }{2 \, \pi} \, \overrightarrow{x}_{n}, k \bigg)  \,  f_n  \bigg(\frac{k\, T_s }{2 \, \pi} \, \overrightarrow{x}_n \bigg) \bigg \vert^2       
\end{split}
\end{align}
 where $x \,= \,k \, T_s$ for integer indices $k \in [-\infty, \infty]$ and sampling period $T_s$,  and $ \Theta[n,k]$, $\gamma_{f,n}(\overrightarrow{x}, k) $ and $f_n (\overrightarrow{x})$ depending on the physical properties of the specific QPC setup are defined as follows:
 \begin{eqnarray}  
  \label{maineqs1}
     \Theta[n,k] \, \equiv \, &&  \,  \overrightarrow{x}_{n}^T \,   \mathbf{H}_{{\color{black}I,N-1,n}} \, \overrightarrow{x}_{n} \, + \, \overrightarrow{d}_{N-1,n}^T  \, \overrightarrow{x}_{n} \, k\, T_s   \\
     \label{maineqs2}
     \gamma_{f,n}(\overrightarrow{x}, k) \, \equiv  \, &&  \, e^{   (A_{N-1,n} \, + \, \imath \, B_{N-1,n}) \, (k\,T_s)^2}\, \Upsilon_{N,n}\,  \nonumber  \\
     && e^{ ( 4\, \pi^2  \, / \,  T_s^2) \, \overrightarrow{x}^T \,  \mathbf{H}_{N-1,n}     \, \overrightarrow{x} }  \, e^{ 2\,\pi \,\overrightarrow{c}^T_{N-1,n}  \,  \overrightarrow{x}  \, k }  \, \, \, \,  \\
     \label{maineqs3}
     f_n (\overrightarrow{x}) \, \equiv \, && e^{\imath \, 2 \, \pi\,\overrightarrow{d}_{N-1,n}^T  \, \overrightarrow{x}}
\end{eqnarray}
The analogy between QC (Section 5.4.1 in   \cite{nc}) and QPC period finding is shown in Table \ref{Table1} and described in detail after defining the following problems:
\begin{myprob}
\label{prob1}
\textbf{Periodicity detection:  Find the  minimum integer  $\widetilde{k} \in \mathbb{Z}^+$}   scaling the given set of $N-1$ dimensional real vectors $\overrightarrow{d}_{N-1,n}$ for a given \textbf{non-uniform lattice}  denoted by $\mathbf{X}_{N-1}^s$  resulting in a \textbf{reciprocal integer lattice} denoted by $\Lambda$ by minimizing the error term $ \epsilon_n$ for $n \in [0, N_p-1]$ in a defined average sense such that $\Lambda \equiv \lbrace  \widetilde{k}\, \, \overrightarrow{d}^T_{N-1,n}  \, \overrightarrow{x}_{n}^s  \, + \, \epsilon_n \in \mathbb{Z}; \, \, \forall \, \, \overrightarrow{x}_{n}^s, \, n \in [0, N_p-1] \rbrace$ where   $\mathbf{X}_{N-1}^s$   formed of a set of real vectors $\overrightarrow{x}_{n}^s$ is defined as follows:   
 \begin{align}
 \label{qlattice}
\begin{split}
  \,\,\,\,\, \overrightarrow{x}_{n}^s  {\color{black}\equiv}  & \, \,   (2 \, \pi)^{-1} \, T_s \, [\overrightarrow{x}_{n}(1)  \hdots \overrightarrow{x}_{n}(N-1) \,] 
\mbox{ \textbf{with}   } \, \,  \overrightarrow{x}_{n}(j)  \in  \lbrace X_{j, -S_j}, \, \hdots, \, X_{j, S_j} \rbrace   \\
&\mbox{ \textbf{s.t.}} \,\, X_{j, i} - X_{j, i+1} >  2 \, \alpha \, \max \lbrace \beta_{j,i},\beta_{j,i+1} \rbrace; \, \, N \geq 2; \, \, \alpha \geq 1\\
&\mbox{\textbf{where}   } \,\,  n \in [0, N_p-1]; \, \, j \in [1, N-1]; \, \, i \in [-S_j, S_j] \\
& \,\,\,\,\,\,\,\,\,\,\,\,\,\,\,\,\,S_j, \, N \in \mathbb{Z}^+; \, \, \beta_{j,i}, \,T_s, \, \alpha \, \in  \mathbb{R}^+;  \, \, X_{j,  i} \in \mathbb{R}
\end{split}
\end{align} 
where  $N_p \equiv  \prod_{j=1}^{N-1} (2 \, S_j\, + \, 1)$, and $\mathbb{Z}$, $\mathbb{Z}^+$, $\mathbb{R}$ and $\mathbb{R}^+$ are the sets of integers, positive integers, real  and  positive real values, respectively.
\end{myprob}  
The condition $X_{j, i} - X_{j, i+1} > 2 \, \alpha \, \max \lbrace \beta_{j,i},\beta_{j,i+1} \rbrace $ for large $\alpha$  satisfies Gaussian slit property. The others  define the physical setup described in Sections \ref{Section2} and \ref{Section4}.  SDA problem presented in  \cite{spa1} is analogical and defined  as follows:  
\begin{myprob}
\label{prob2}
\textbf{SDA: Decide the existence and find the minimum integer  $\widetilde{k} \in \mathbb{Z}^+$}  where $\widetilde{k} \leq K_{pre}$ for some pre-defined $K_{pre} \in \mathbb{Z}^+$ such that it is  SDA solution for the set of real numbers in the set $S_{b} = \lbrace b_0, \, b_1, \hdots, b_{N_p-1} \rbrace$  satisfying the relation $\vert \widetilde{k}\, b[n] - k_n \vert <  \epsilon$  for $n \in [0, N_p-1]$ and for some $k_n \in \mathbb{Z}$ specific to each $n$   where $b[n] \equiv \overrightarrow{d}^T_{N-1,n}   \, \overrightarrow{x}_{n}^s$ and $\epsilon$ is the bounding error term.
\end{myprob}

Polynomial solutions of SDA problem and  performance of  Lenstra, Lenstra Jr., and Lovasz (LLL) algorithm  for large number of inputs  become highly prohibitive for $N_p \gg 1$ \cite{spa1}. Assume that $\vert \vert x \vert \vert$ denotes the distance of the real number $x$ to the closest integer,  the maximum of  $\vert \vert \widetilde{k} \, b[n] \vert\vert$ for $n \in [0, N_p -1]$ is smaller than some pre-defined $\epsilon_p$ and there is some pre-defined bound $M$ with $M > \widetilde{k}$.  LLL algorithm  estimates $\widetilde{k}$ as $\widehat{k}$ satisfying  $1 < \widehat{k} < 2^{N_p \, / \, 2} \, M$ and the maximum  of   $\vert \vert \widehat{k} \, b[n] \vert\vert $ being smaller than $\sqrt{5 \, N_p} \, 2^{(N_p-1)\, / \, 2}\, \epsilon_p$   with the number of operations depending on input size \cite{spa1}.  Error term for  SDA  is defined as $\epsilon[n, \widehat{k}] \equiv   \vert \vert \widehat{k} \, b[n] \vert \vert$  for $n \in [0, N_p-1]$.  Then, $\overline{\epsilon}[\widehat{k}]  \equiv (1 \, / \, N_p) \sum_{n = 0}^{N_p - 1} \epsilon[n, \widehat{k}]$, $\epsilon_{max}[\widehat{k}]  \equiv  \underset{n}{max} \lbrace \epsilon[n, \widehat{k}] \rbrace$  and $\epsilon_{min}[\widehat{k}]  \equiv  \underset{n}{min} \lbrace \epsilon[n, \widehat{k}] \rbrace$ are indicators for observing how $\widehat{k}$ is close to the solution, i.e., $\widetilde{k}$.  
 
 \setlength\tabcolsep{3 pt}    
\newcolumntype{M}[1]{>{\centering\arraybackslash}m{#1}}
\renewcommand{\arraystretch}{1.3}
\begin{table*}[t!]
\caption{The analogy between QC and QPC period finding algorithms}
\begin{center}
\scriptsize
\begin{tabular}{|c|m{3.2cm}|m{0.8cm}|m{4.8cm}|m{1.4cm}|}
\cline{2-5} 
\multicolumn{1}{ c }{ } & \multicolumn{2}{|c|}{QC Period Finding Algorithm \cite{nc}}    & \multicolumn{2}{c|}{QPC Period Finding Algorithm}  \\ 
\cline{2-5} 
\hline
Steps    & Procedure &  \multicolumn{1}{c|}{\makecell[{{m{0.8cm}}}]{$\sharp$ Ops.}}  &   Procedure  &    $\,\,\,\,\,\,\,\sharp$ Ops. \\  
\hline
0 & \multicolumn{1}{c|}{ \makecell[{{m{3.2cm}}}]{  $a.$ The function $f(x)$  \\
$b.$  $x$ is integer, producing single bit output  \\ $c.$ Periodic for $0 < r < 2^L$ integer: $f(x) = f(x+r)$ \hspace{0.16in} \\
$d.$  BB performing $\mathit{U}\ket{x}\ket{y} = \ket{x}\ket{y \oplus f(x)}$  }}     &  \makecell[{{c}}]{0} &  \multicolumn{1}{c|}{\makecell[{{m{4.8cm}}}]{ $a.$  $f_n(\overrightarrow{x}) = e^{\imath \, 2\,\pi\,\overrightarrow{d}_{N-1,n}^T  \, \overrightarrow{x}}$ where $\overrightarrow{x}$  and $\overrightarrow{d}_{N-1,n}$ are tuned  by the setup. \\$b.$  The basis periodicity sets defined as \\
$S_{a}${\color{black}:} $\lbrace \overrightarrow{r}_a = \sum_{n=0}^{N_p-1} a_{n} \overrightarrow{x}_n^s, a_n \in \mathbb{Z},$ \\ $ n \in [0, N_p -1] \rbrace$ for  $\overrightarrow{x}_n^s \in \mathbf{X_{N-1}^s}$  \\
$c.$  $f_n(\overrightarrow{x}) = f_n(\overrightarrow{x} \, + \, \widetilde{k} \, \overrightarrow{r}_a)$  \\ for $ \overrightarrow{r_a} \in S_{a} $  \\ $d.$ QPC setup or BB performing $f_n(k\, \overrightarrow{x}_n^s)$  given $\overrightarrow{x}_n^s$ and integer $k$  }}   &  \makecell[{{c}}]{0}      \\    

\hline
1   & Initial state: $\ket{\mathbf{0}}\ket{\mathbf{0}}$    &  \makecell[{{c}}]{0}  &    $\ket{\Psi_0}${\color{black}:}  coherent Gaussian wave packet & \makecell[{{c}}]{0} \\    
\hline
2   & \multicolumn{1}{c|}{ \makecell[{{c}}]{Superposition: \\$\frac{1}{\sqrt{2^t}}$ $\mathlarger{\sum}\limits_{0}^{2^t-1} \ket{\mathbf{x}}\ket{\mathbf{0}}$ }}  &  \makecell[{{c}}]{0}  &    $N_p$ paths to reach the detector with $\overrightarrow{x}_n^s$  for  $n \in [0, N_p-1]$ and $\mathlarger{\sum}\limits_{n=0}^{N_p-1} \ket{\overrightarrow{x}_n^s}\ket{\Psi_0}$&  \makecell[{{c}}]{0} \\     
\hline  
3   & \multicolumn{1}{c|}{ \makecell[{{c}}]{Black box (BB) $\mathit{U}${\color{black}:}\\ $\frac{1}{\sqrt{2^t}} \mathlarger{\sum}\limits_{0}^{2^t - 1} \ket{\mathbf{x}}\ket{f(\mathbf{x})} $ }}  &  \makecell[{{c}}]{1}  &      \makecell[{{l}}]{ BB params.  $\mathbf{X}_{N-1}^s$,  $S_{\beta}$, $\overrightarrow{L}$, m and $\sigma_0${\color{black}:}  \\ $   \Psi_{N}(k \, T_s)  =   \mathlarger{\sum}\limits_{n=0}^{N_p-1} \Psi_{n,N}(k \, T_s) $  \\ \hspace{0.45in}$   = \, 
    \mathlarger{\sum}\limits_{n=0}^{N_p-1} \gamma_{f,n}(\overrightarrow{x}_n^s, k)  \,  f_n(k\, \overrightarrow{x}_n^s)$ }  &  \makecell[{{c}}]{1}  \\     
\hline
\makecell[{{c}}]{ $4$ $\&$ $5$ }  & \makecell[{{c}}]{ $a.$ $IQFT$:\\ $ (1 \, / \, \sqrt{r})  \mathlarger{\sum}\limits_{0}^{r - 1} \ket{\widetilde{l \, / \, r}}\ket{ \widehat{f}(l)} $  \\  $b$. Measure first register: \\$\widetilde{l \, / \, r}$ }    & \makecell[{{c}}]{    $\mathit{O}(L^2)$}   &   \makecell[{{c}}]{ $a.$ Measure $\vert \Psi_{N}(k \, T_s)\vert^2$ at various $k$   \\ $b.$  $IFFT_M$  at $p$ with  $M \geq \widetilde{k}$: \\ $\mathlarger{\sum}\limits_{h=0}^{\widetilde{k}-1} \Gamma_M^{G}[p \, / \, M,  h \, / \, \widetilde{k}]$  }&   \makecell[{{m{1.0cm}}}]{$\mathit{O}(M \, log M)$}    \\     
\hline
6   & Continued fractions: $r$   &  \makecell[{{c}}]{$\mathit{O}(L^3)$ }   &    Check  IFFT at  $p \in [0, M -1]$ values for   $M \geq \widetilde{k}$  providing an estimation for $h\, / \, \widetilde{k}$ for $h \in [0, \widetilde{k} -1]$ and  resulting in a converging estimation of $\widetilde{k}$ &  \makecell[{{c}}]{Polynomial \\ target}\\
\hline
\end{tabular}
\end{center} 
\label{Table1}  
\end{table*}
\renewcommand{\arraystretch}{1}
\setlength\tabcolsep{6 pt}

Several candidate solution methods requiring more efforts to formally define  and verify the solution algorithms are presented for Problems \ref{prob1} and \ref{prob2}. Besides that, the set of the solutions which can be provided is constrained to the problems implementable with QPC setup without covering all the problems described in Problems \ref{prob1} and \ref{prob2}.  QPC period finding solution  utilizes (\ref{qphase1}-\ref{maineqs3}) in combination with a set of measurements at $x = k \, T_s$.   QC algorithms exploit  superposition generated with Hadamard transforms on two registers initially at $\ket{\mathbf{0}}\ket{\mathbf{0}}$  and  evolution with controlled unitary transforms $\mathit{U}$ in BBs  for a  periodic function $f(x) = f(x \, + \, r)$ \cite{nc}. QPC equation in (\ref{qphase1})  is utilized to find  periodicity in  $f_n(\overrightarrow{x}) \equiv e^{\imath \, 2\,\pi\, \overrightarrow{d}_{N-1,n}^T \, \overrightarrow{x}}$  for specific sets of $\mathbf{X}_{N-1}$ and $\overrightarrow{d}_{N-1,n}$.   The steps of QPC  period finding algorithm are described as follows while the analogy to  QC period finding is shown in Table \ref{Table1}:
\begin{enumerate}
 \setcounter{enumi}{-1}
\item[ $0$  ]   $\overrightarrow{d}_{N-1,n}$ for $n \in [0, N_p-1]$, $\mathbf{X_{N-1}}$ and $\mathbf{X_{N-1}^s}$  are given initially where $\mathbf{X_{N-1}^s}$  is defined in Problem \ref{prob1}. The function  $f_n(\overrightarrow{x}) =e^{\imath \, 2 \, \pi\,\overrightarrow{d}^T_{N-1,n}  \, \overrightarrow{x}}$ has  periodicity  for $\overrightarrow{x}$ with respect to  the unknown period $\widetilde{k}$  and the given basis sets $S_{a}: \lbrace \overrightarrow{r}_a = \sum_{n=0}^{N_p-1} a_{n} \overrightarrow{x}_n^s, a_n \in \mathbb{Z}, n \in [0, N_p -1] \rbrace$ as follows: $f_n(\overrightarrow{x}) = f_n(\overrightarrow{x} \, + \,  \widetilde{k} \, \overrightarrow{r}_a)$  while  the target is to find $\widetilde{k}$.  
\item[ $1$  ]  The wave function $\ket{\Psi_0}$   of coherent source as a Gaussian packet is set up.
\item[ $2$  ] The superposition is due to QPC setup combining $N_p$ paths on the screen and $\ket{\Psi_0}$ where the initial state  is denoted by $\sum_{n=0}^{N_p-1} \ket{\overrightarrow{x}_n^s}\ket{\Psi_0}$.
\item[ $3$  ] BB is the QPC setup with specially designed parameters providing $\overrightarrow{x}_n$ in the grid $\mathbf{X_{N-1}}$ and the vectors $\overrightarrow{d}_{N-1,n}$ while  related parameters  $A_{N-1,n}$, $B_{N-1,n}$,  $\Upsilon_{N,n}$, $\mathbf{H}_{N-1,n}$,  $\overrightarrow{c}_{N-1,n}$, and the setup parameters $S_{\beta}$, $\overrightarrow{L}$,  m, $\sigma_0$ and $T_s$ to be optimally designed for generating $\overrightarrow{x}_n^s$ and the best estimate of $\widetilde{k}$  by using    $I^G_N[k]  =  \vert \sum_{n=0}^{N_p-1} \gamma_{f,n}(\overrightarrow{x}_n^s, k)  \,  f_n(k\, \overrightarrow{x}_n^s)\vert^2$.  
\item [ $4$-$5$ ] A set of $M \geq \widetilde{k}$ samples are taken on  detector plane and  IFFT operation with complexity $\mathit{O}(M \, log M)$ with the output time index $p$ gives information about $p \, / \,\widetilde{k}$ and $h \, / \,\widetilde{k}$ for $h \in [0, \widetilde{k} - 1]$ where $\Gamma_M^{G}[p \, / \, M, h \, / \, \widetilde{k}]$ in (\ref{gammeq}).    
\item [$6$] The number of samples at varying $p$  values is increased for a converging and unbiased estimation of $\widetilde{k}$. The problem is set as a parameter estimation problem for the set of damped sinusoids  \cite{fft1}.  Traditional period finding algorithms are utilized to best estimate $\widetilde{k}$, e.g., $\mathit{O}(M \, log M)$ complexity or polynomial complexity for FFT based solutions in  frequency estimation of damped sinusoidal signals. 
\end{enumerate}

Next, three approaches are  introduced for the final three steps of the algorithm, i.e., Steps 4, 5 and 6. The first approach converts IFFT output to extract information about $h \, / \, \widetilde{k}$ by using the IFFT samples at $p \in [0, M-1]$ in analogy to period finding method for conventional QC \cite{nc} as described in Table \ref{Table1}. The second approach checks the periodicity in the local maximum of $I_{N}^G[k]$ and the third approach models  the problem as a fundamental frequency estimation for a sum of sinusoidal signals.   

\subsection{Conversion of IFFT Output}
IFFT operation with the number of samples $M$ described  in Steps 4$\&$5 is simplified by using (\ref{qphase1}). Define discrete functions of $n$ as $g_{1}[n] \equiv e^{\overrightarrow{c}_{N-1,n}^T\, \overrightarrow{x}_n \, T_s}$, $g_{2}[n] \equiv  \overrightarrow{d}_{ N-1,n }^T  \, \overrightarrow{x}_n \, T_s $, $g_3[n] \equiv  \Upsilon_{N,n} \, e^{\overrightarrow{x}_n^T \, \mathbf{H}_{N-1,n} \, \overrightarrow{x}_n}$ and $g_4[n]$   defined as  $   e^{( A_{N-1,n} \, + \, \imath \,  B_{N-1,n}) \, T_s^2}$.  Since $\overrightarrow{d}_{N-1,n}$ and $(2 \, \pi)^{-1} \, T_s \, \overrightarrow{x}_n$ form an integer lattice for $n \in [0, N_p-1]$ with  integer period $\widetilde{k}$, the expression  $e^{\imath \, \overrightarrow{d}_{N-1,n}^T \, \overrightarrow{x}_n \, k \, T_s } = e^{\imath \, g_2[n] \, k }$ is converted to $e^{\imath \, \widetilde{G}_2[n] \, 2 \, \pi \, k  \, / \, \widetilde{k}}$ due to  periodicity with $\widetilde{k}$ where $\widetilde{G}_2[n]$ is a function mapping the interval $[0, N_p -1]$ into an integer between $[0, \, \widetilde{k}-1]$ while depending on the relation between $\overrightarrow{d}_{N-1,n}$ and $\mathbf{X}_{N-1}^s$. Then,  IFFT output with size $M$  denoted by $IFFT_M\lbrace I^G_N   \rbrace[p]$  becomes as follows:
\begin{align} 
\label{iffteq}
\begin{split}
 & = \, \frac{1}{\sqrt{M}} \,\sum_{k = 0}^{M - 1}   \sum_{n, l = 0}^{N_p - 1}    g_{3,*}[n,l] \,  g_{1,*}^k[n,l]  \, g_{4,*}^{k^2}[n,l] \,   e^{ - \frac{ \imath \, 2 \, \pi \, k \,\Delta G_2[n, l]  }{\widetilde{k}}  }  e^{   \frac{ \imath \, 2 \, \pi \, k  \, p}{M}  }  \\   
 & = \, \sum_{n, l = 0}^{N_p - 1} g_{3,*}[n,l] \, \frac{1}{\sqrt{M}} \,\sum_{k = 0}^{M - 1}  \,g_{1,*}^k[n,l]  \, g_{4,*}^{k^2}[n,l] \,   e^{ - \frac{ \imath \, 2 \, \pi \, k \,\Delta G_2[n, l]  }{\widetilde{k}}  }  e^{   \frac{ \imath \, 2 \, \pi \, k  \, p}{M}  }     
\end{split}
\end{align}
where   $g_{1,*}[n,l] \equiv g_1[n]  \, g_1[l]$, $g_{3,*}[n,l] \, \equiv \, g_3[n] \, g_3^{*}[l]$, $g_{4,*}[n,l] \equiv g_4[n]  \, g_4^{{\color{black}*}}[l]$,   the set of coefficients of $\omega_0 \equiv 2 \, \pi \, / \, \widetilde{k}$ is  $\Delta G_2[n, l] \equiv \widetilde{G}_2[l] - \widetilde{G}_2[n]$. Dividing the set of $[n,l]$ pairs in $[0, N_p-1] \times [0, N_p-1]$  into $\widetilde{k}$ regions with  index $h \in [0, \widetilde{k} - 1]$ denoted by $R_h$  results in   the following equality since $mod(\Delta G_2[n, l], \widetilde{k} )= h$: 
\begin{align}
\label{maingammaeq}
\begin{split}
  IFFT_M\lbrace I^G_N   \rbrace[p]  =  \, & 
 \sum_{h = 0}^{\widetilde{k} - 1}  \Gamma_M^G[\frac{p}{M}, \frac{h}{\widetilde{k}}]  \\
  = \, &  \sum_{h = 0}^{\widetilde{k} - 1} \, \sum_{n,\,l \, \in \,  R_h}  g_{3,*}[n,l] \, \Omega^F_{n,l} \bigg(  \frac{p}{M}  \, - \, \frac{h}{\widetilde{k}} \bigg)  
\end{split}
\end{align}
where $\Gamma_M^G[p \, / \, M,  h \, / \, \widetilde{k}]$ is defined as follows:
 \begin{align} 
\label{gammeq}
\begin{split}
  \Gamma_M^G[\frac{p}{M}, \frac{h}{\widetilde{k}}] =  \, &  \sum_{n,\,l \, \in \,  R_h} \frac{ g_{3,*}[n,l]}{\sqrt{M}}  \,\sum_{k = 0}^{M - 1}  \,g_{1,*}^k[n,l]  \, g_{4,*}^{k^2}[n,l] \,   e^{  \imath \, 2 \, \pi \, k \,  (     \frac{p}{M}  \, - \, \frac{h}{\widetilde{k}}) }       
\end{split}
\end{align}
and $\Omega^F_{n,l}(x)$ is closely related to the discrete approximation  of  the continuous inverse Fourier transform of $g_{1,*}^k[n,l]  \, g_{4,*}^{k^2}[n,l]$ with respect to $k$ by allowing the result at fractional values of the positions defined as follows:
 \begin{align} 
\label{alphafrac}
\begin{split}
\Omega^F_{n,l}(x) \equiv \frac{1}{\sqrt{M}}  \,\sum_{k = 0}^{M - 1}  \,g_{1,*}^k[n,l]  \, g_{4,*}^{k^2}[n,l] \,   e^{  \imath \, 2 \, \pi \, k \, x }  
\end{split}
\end{align}
The structure of the IFFT output is best understood for finding the period of $f(\overrightarrow{x}) = e^{\imath \, 2\,\pi\,\overrightarrow{d}_{N-1}^T  \, \overrightarrow{x}}$ with path independent periodic function obtained with uniform slit widths for each plane. This problem is classically  tractable and numerically analyzed in Section \ref{Section10} as a proof of concept. Assume that the intensity is normalized with $\widetilde{I}_N[k] \, \equiv \, \widetilde{I}[k] \, \equiv \, e^{-2 \, A_{N-1} \, k^2 T_s^2} \, I_{N}[k] $ {\color{black}(obtained from $I_{N}[k] \equiv I_N(k \, T_s)$)} which results in the omission  of the term $g_{4,*}^{k^2}[n,l]$ in (\ref{iffteq}).  Then, the equality in (\ref{iffteq}) is modified as follows  by using the power series summation:  
\begin{align} 
\label{iffteq_constrained}
\begin{split}
  IFFT_M\lbrace \widetilde{I}   \rbrace[p]  \,& = \,\sum_{k = 0}^{M - 1}   \sum_{n, l = 0}^{N_p - 1} \frac{g_{3,*}[n,l]}{\sqrt{M}} \,g_{1,*}^k[n,l]  e^{ - \frac{ \imath \, 2 \, \pi \, k \,\Delta G_2[n, l]  }{\widetilde{k}}  }  e^{   \frac{ \imath \, 2 \, \pi \, k  \, p}{M}  }  \\   
 \,& = \, \sum_{n, l = 0}^{N_p - 1} \frac{g_{3,*}[n,l]}{\sqrt{M}} \,\sum_{k = 0}^{M - 1}    g_{1,*}^k[n,l]   e^{ - \frac{ \imath \, 2 \, \pi \, k \,\Delta G_2[n, l]  }{\widetilde{k}}  }  e^{   \frac{ \imath \, 2 \, \pi \, k  \, p}{M}  }  \\  
 &   = \, \sum_{n = 0}^{N_p - 1}     \sum_{l = 0}^{N_p - 1} \frac{ g_{3,*}[n,l]}{\sqrt{M}}  \dfrac{1 - \gamma_{n,l,p}^M}{1 - \gamma_{n,l,p}}  
 \\  
 &   = \,     \sum_{h = 0}^{\widetilde{k} - 1}  \Gamma_M [\frac{p}{M}, \frac{h}{\widetilde{k}}]
\end{split}
\end{align}
where $\gamma_{n,l,p} \equiv g_{1,*}[n,l] \, e^{ - \frac{ \imath \, 2 \, \pi}{M \, \widetilde{k}} ( \Delta G_2[n, l] \, M\,- \,p \, \widetilde{k})  }$.  After dividing the $(n,l)$ region with $R_h$,  $\Gamma_M[p \, / \, M,  h \, / \, \widetilde{k}]$ is calculated as follows: 
\begin{equation}
\label{gammeq2_constrained}
\Gamma_M[\frac{p}{M}, \frac{h}{\widetilde{k}}] =   \sum_{n,\,l \, \in \,  R_h} \frac{g_{3,*}[n,l] }{\sqrt{M}}     \,  \dfrac{1 -  g_{1,*}^M[n,l]  e^{ - \imath \, 2 \, \pi \frac{ h \, M  }{  \widetilde{k}}  } }{1 -  g_{1,*}[n,l] e^{  \imath \, 2 \, \pi  ( \frac{p}{M}  \, - \,  \frac{h}{\widetilde{k}}    )  } }  
\end{equation}
If $M = \widetilde{k}$, the rational term is $  \big( 1 -    g_{1,*}^{\widetilde{k}}    [n,l] \big)$ $ \big (1 - g_{1,*}[n,l] \, e^{  \imath \, 2 \, \pi \, (p \,- \,h) \,    / \, \widetilde{k}}  \big)^{-1}$. Similar to the Bertocco algorithm for the single sinusoid case \cite{fft1}, it is observed that  exponentially increasing term   ($1 -  g_{1,*}^M[n,l]  e^{ - \imath \, 2 \, \pi   h \, M    / \, \widetilde{k}   } $) in the numerator  results in fast oscillations of the phase for each $h \, \in [0, \widetilde{k} -1]$ if $M < \widetilde{k}$. A  function denoted by $R[M]$  is introduced to utilize in the estimations as follows: 
\begin{equation}
\label{rmeq}
 R[M]\, \equiv \, \dfrac{\big \vert IFFT_M\lbrace \widetilde{I}   \rbrace[0]  \big \vert}{  \frac{1}{M-1} \sum_{k = 1}^{M-1}   \big \vert IFFT_M \lbrace \widetilde{I}    \rbrace[k] \big \vert   }   
\end{equation}
while it is expected to be maximized around $M \approx\widetilde{k}$. High frequency components are averaged  and their mean is compared with  zero frequency component.  Then, checking the samples of $R[M]$ with respect to $M$, i.e., minimizing   high frequency components,  allows roughly determining $\widetilde{k}$. The same periodicity is expected in $R[M]$ since  fluctuations are decreased at multiples of $\widetilde{k}$.  
  
\subsection{Periodicity Detection in Local Maximum of Intensity}

Periodicity $\widetilde{k}$ is heuristically found by checking local maximums in the measurement intensity   $\widetilde{I}[k]$ satisfying the following theorem:
\begin{theorem} 
\label{theorem1}  
Assume that the set of  real vectors $\overrightarrow{c}_{N-1}$ and $\overrightarrow{d}_{N-1}$, and a non-uniform grid $\mathbf{X_{N-1}^s}$ satisfy the following  with the tuned physical setup giving the measurement in (\ref{qphase1}) and the normalized intensity $\widetilde{I}[k]$:  
\begin{enumerate}
\item $\overrightarrow{d}_{N-1}$ and $\mathbf{X_{N-1}^s}$ form an integer lattice   with  $e^{\imath \, g_2[n] \, k } = e^{\imath \, \overrightarrow{d}_{N-1}^T   \overrightarrow{x}_n  T_s \, k }$ to be represented by $e^{\imath \, \widetilde{G}_2[n] \, 2 \, \pi \, k  \, / \, \widetilde{k}}$. 
\item  $\vert H[k,\widetilde{G}_2] \vert <  \vert  H[k,0] \vert$ and $\vert H[k_1,0] \vert > \vert  H[k_2,0] \vert$ where  $ 0  \, <  \, k  \, <  \, \widetilde{k}$, $k_2 < k_1 \leq \widetilde{k}$, and $k, \, k_1, k_2 \in \mathbb{Z}$, and $ H[k, func]$ is defined as follows: 
\begin{equation} 
\sum_{n =0}^{N_p -1} \, g_3[n]  (g_1[n])^{k}  \, e^{ \frac{ \imath \, 2 \, \pi \, func[n] \, k}{\widetilde{k}}  }
\end{equation}
\end{enumerate}
where $func[n] \in [0, \widetilde{k} -1]$ refers to a specific mapping of $n \in [0,N_p-1]$ with a discrete function $func[.]$ and $H[k,0]$ refers to the case where $func[n] = 0$. Then, $\widetilde{I}[\widetilde{k}] >   \widetilde{I}[k]$ is satisfied for $k \in [0, \widetilde{k}-1]$.
\end{theorem} 
The proof is provided in Appendix \ref{proof_theorem1}.  Checking local maximum $\widehat{k}$ with random samples of $\overrightarrow{d}_{N-1}^T \, \overrightarrow{x}_n \, \widehat{k} \, T_s \, / \, (2 \, \pi)$ to verify for integer values determines the periodicity $\widetilde{k}$. The extension of Theorem \ref{theorem1} for $I_{{\color{black}N}}^{{\color{black}G}}[k]$ is required for   the important and  computationally hard problem of $b[n] = \overrightarrow{d}_{N-1,n}^T \, \overrightarrow{x}_n^s$  compared with the classically efficient solutions for the case of $b[n] = \overrightarrow{d}_{N-1}^T \, \overrightarrow{x}_n^s$. The methods for frequency estimation of damped sinusoids as described in \cite{fft1}   is presented next.
    
\subsection{Frequency Estimation for Sinusoidal Signals}
\label{divergingsinusoids}
The problem is considered as finding the fundamental frequency $\omega_0 = 2 \, \pi \, / \, \widetilde{k}$ for the sum of complex sinusoidal signals  \cite{fft1} if (\ref{iffteq}) is transformed as follows:
\begin{equation} 
I_{N}^G[k] =   \sum_{n = 0}^{N_p - 1}  \sum_{l = 0}^{N_p - 1}  \,    g_{3,*}[n,l] \,  g_{1,*}^k[n,l]  \, g_{4,*}^{k^2}[n,l] \,      e^{- \imath \, \Delta G_2[n, l] \,\omega_0\, k }       
\end{equation}
We drop the subscript $N$ in the following and denote the samples of the intensity obtained with a setup composed of non-uniform slit widths as $I^G[k]$ and as $I[k]$ with uniform slit widths for each plane in the constrained case. Then, the effect of the additive white Gaussian noise (AWGN) is modeled as  $I_{n}^G[k] = I^G[k] \, + \, n[k] $ where $n[k]$ is the receiver noise modeled as a Gaussian random process with independent samples. If  Poisson distribution is assumed, then the noise has variance $\sigma_k^2$ proportional to $I^G[k]$. Let us assume that $I[k]$ in the second setup with constrained slit widths is normalized as $\widetilde{I}[k]  = e^{- 2\, A_{N-1} \, (k\,T_s)^2} I[k]$ to exclude the effect of the constant multiplier $A_{N-1}$ which is the same for each path.  If the AWGN output intensity is normalized, then the noise is also amplified with $\widetilde{I}_{n}[k] = \widetilde{I}[k] \, + \, \widetilde{n}[k]$
where $ \widetilde{n}[k] = e^{- 2\, A_{N-1} \, (k\,T_s)^2}\, n[k]$ with the variance $\widetilde{\sigma}^2[k] \equiv  e^{- 4\, A_{N-1} \, (k\,T_s)^2}\, \sigma_k^2$. In the following discussion and Appendix \ref{proof_theorem2}, it is assumed that $I^{*}[k]$ and $n^{*}[k]$ refer to $I^{G}[k]$ and $n[k]$, respectively, for the general setup while referring to the normalized intensity $\widetilde{I}[k]$ and the noise $\widetilde{n}[k]$, respectively, for the constrained setup. Similarly, $\sigma^{*}[k]$ denotes $\sigma_k$ for the general setup while denoting $\widetilde{\sigma}[k]$ for the constrained setup.  Cramer-Rao lower bound for the estimate of $\widetilde{k}$  is provided in the following theorem while the proof is provided in Appendix \ref{proof_theorem2}:
\begin{theorem} 
\label{theorem2}
Cramer-Rao lower bound for  period finding in reciprocal integer lattice of  QPC setup by using a set of intensity measurements in $M$ different positions with sample points $k_p \, T_s$ for $p \in [0, M-1]$ is given as follows: 
\begin{align}
\label{eqtheorem2}
\begin{split}
 CRB(\widetilde{k}) =   \bigg(1 \, + \,  \frac{\delta b(\widehat{k})}{\delta   \widetilde{k} } \bigg)^2   \, / \,    \sum_{p=0}^{M-1} \bigg(\dfrac{1}{\sigma^{*}[p]}   \dfrac{\delta I^{*}[k_p]}{\delta \, \widetilde{k}}\bigg)^2  
\end{split}
\end{align}
where  $b(\widehat{k}) \equiv E \lbrace \widehat{k} \rbrace - \widetilde{k} $ is the bias  while  noise has zero mean. 
\end{theorem}
Open issues in QPC based period finding and SDA solution are described next.

\subsection{Open Issues in Period Finding and SDA Solution}
\label{openissuesofSDA}
The proposed period finding methods for the Steps 4, 5 and 6 in Table \ref{Table1} are heuristic.  An open issue is to best utilize (\ref{maingammaeq}) with samples $p \in [0, M-1]$ by performing polynomial time complexity operations to estimate $\widetilde{k}$ in analogy to IFFT and continued fractions operations in conventional QC period finding algorithm with quantum gates \cite{nc}. 
Formal mathematical proof for determining the group of SDA problems with QPC solution and an exact algorithm finding the solution in Step-6 in Table \ref{Table1} are open issues. It requires to analyze the relation among $A_{N-1,n}$, $B_{N-1,n}$, $\Upsilon_{N,n}$, $\mathbf{X_{N-1}}$, $\mathbf{H}_{N-1,n}$, $\overrightarrow{c}_{N-1,n}$ and $\overrightarrow{d}_{N-1,n}$ in (\ref{qphase1}). SDA  problem solution for a general set of $b[n]$ is NP-hard  \cite{spa1}; however, the proposed $b[n]$ is represented with  $\overrightarrow{d}_{N-1,n}^T  \, \overrightarrow{x}_{n}^s$ as a specific instance limiting the space of the candidate SDA problems with potential solutions. Furthermore, it is an open issue whether the specific group of SDA problems which can be solved with QPC in an efficient manner can also be efficiently solved with classical computers with polynomial complexity of resources.  

The extension of Theorem \ref{theorem1} for $I_{N}^G[k]$ is an open issue which provides detection of periodicity by directly checking the periodicity in intensity. In addition, the existence of $\widetilde{k} \leq K_{pre} \equiv M$ for the proposed simple SDA problem (with $b[n] \equiv \overrightarrow{d}_{N-1}^T  \, \overrightarrow{x}_{n}^s$ and $\widetilde{I}[k]$) is heuristically checked by the existence of fluctuations. If there is no fluctuation, it is assumed as the absence of the bounded error $\epsilon$ such that  the solution  does not exist for $k  \leq M$. If there is a fluctuation, the set of fluctuating points are the candidates for a solution  to be checked. Furthermore,   practical algorithms of fundamental frequency estimation for the sums of sinusoidal signals should be developed for QPC \cite{fft1}. Next,  effects of  non-classical paths discussed in  \cite{exotic, exotic2} are analyzed. 

\section{Effects of Exotic Paths}
\label{Section9}

\begin{figure}[!t]
\centering
\includegraphics[width=4.5in]{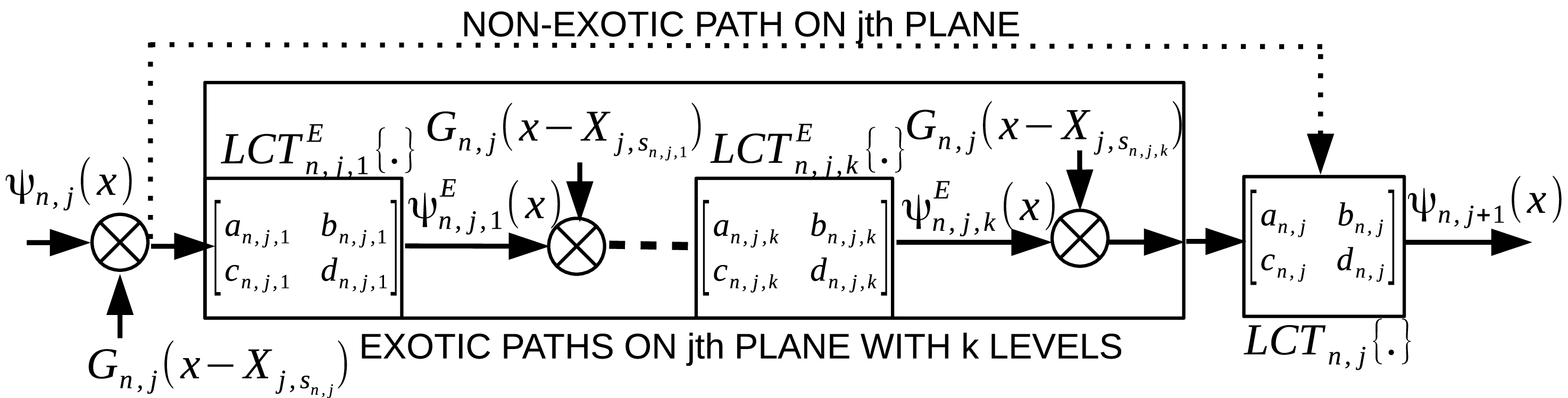}\\
\caption{ The representation of evolution of wave function $\Psi_{n, j} (x)$  on $j$th plane in  $n$th path as consecutive operations of non-classical movements $LCT_{n,j,i}^E\lbrace . \rbrace$  followed by  multiplication of $G_{n, j}(x - X_{j, s_{n,j,i}})$  for $i \in [1, k]$  and finally after $LCT_{n,j}\lbrace . \rbrace$ resulting in  $\Psi_{n, j+1} (x)$.} 
\label{Figure3}
\end{figure}
  
Evolved wave function is calculated by summing contributions from both \textit{non-exotic} (or classical denoting the paths not including non-classical trajectories defined in \cite{exotic, exotic2}) and non-classical paths (trajectories including movements on a single plane) by providing a complete formulation of QPC setup. A sample non-classical path is shown in Fig. \ref{Figure1}(c)  by forming a loop between the slits. Assume that the particle of  $n$th path on  $j$th plane makes $k$ consecutive visits to slits in addition to the first slit with the index $s_{n,j}$ and  position $X_{j, s_{n,j}}$ while the case with $k = 0$ corresponds to the non-exotic path as shown in Fig. \ref{Figure3}.  The wave function in the non-classical path after $k$th slit denoted by $\Psi_{n, j,k}^E (x)$  is explicitly provided in Appendix \ref{proof5} for $k$ bounded by $N_E$. $LCT_{n,j,k}^E\lbrace . \rbrace$  depends on the distance between the slits on $j$th plane defined as  $\Delta_x^E(j, k) \equiv \vert X_{j, s_{n,j,k}} - X_{j, s_{n,j,k-1}} \vert$  where $X_{j, s_{n,j,k}}$ denotes the central position of  $k$th visited slit and $k = 0$ case corresponds to  the position of the first slit on $j$th plane, i.e., $ X_{j, s_{n,j,0}} \equiv X_{j, s_{n,j}}$. Then, setting $N_E$ and finding all paths for $k \in [0,N_E]$ allow to include the effects of all possible non-classical paths. 

Operator formalism for calculating Gouy phase in  \cite{exotic} is utilized to calculate  time durations for the path distance $\Delta_x^E(j, k)$ with $t_{k-1,k}^E(j)$ defined as $  \Delta_x^E(j, k) \, / \, \Delta_v^E(j) = m \, \Delta_x^E(j, k) \, / \, \Delta_p^E(j)$ where $\Delta_p^E(j) =\sqrt{\langle p^2\rangle - \langle p\rangle^2}$ and $\langle p^a\rangle$ for $a \in [1, 2]$ is defined as  $\langle p^a\rangle \equiv$ $ \int_{-\infty}^{\infty}$ $\Psi_j^{*} (x) \big ( (\hbar \, / \, \imath) \,  \delta  \, / \, \delta x \big)^a \Psi_j (x) \,dx$. Total number of different paths  between $j$th and  $(j+1)$th planes including  non-classical movements is denoted by $N_{e,j} = S_{j,T} \, \sum_{k={\color{black}0}}^{N_E}  (S_{j,T}-1)^{k}$ while total number of all  paths on $i$th plane for $i \in [{\color{black}2}, N]$ is given by $N_{p,i}^E \equiv \prod_{j =1}^{i-1} N_{e,j}$. Total number of paths on  sensor plane is denoted by $N_{p,N}^E$  much larger compared with the case including only non-exotic paths, i.e., $N_p$.  Total number of contributions and  effects of the non-classical paths are simulated in Section \ref{Section10}. The first term  $S_{j,T} $ shows different selections of the first slit while the  remaining $k$ different slit movements occur in $(S_{j,T}-1)^{k}$ permutations. Finally, summing the contributions for different $k$ values until $N_E$ results in  $N_{e,j}$.   
 
\section{Numerical Simulations}
\label{Section10}
 
Two different experiments are denoted by $Sim_1$  and $Sim_2$ performed for a simple SDA problem and energy-complexity trade off, respectively, as shown in Table \ref{Table2}. Main system parameters  are shown in Table \ref{Table3} with electron based setup verified for Gouy phase calculations in \cite{exotic}.  In $Sim_1$ and $Sim_2$, it is assumed that the slit widths are constrained to be the same on each plane. Therefore, $\widetilde{I}[k]$, i.e., $e^{-2\, A_{N-1} \, (k\,T_s)^2} I[k] $, denotes the normalized intensity in the simulations as discussed in  Section \ref{divergingsinusoids}. Similarly, $R[M]$   in (\ref{rmeq}) is defined with $\widetilde{I}[k]$. In $Sim_2$,  a highly complex interference setup is realized. The difference between two neighbor slit positions on $j$th plane is chosen as $(9.5 + u) \times \beta_j$ where $u$ is a uniform random variable such that Gaussian slit approximation is satisfied with high accuracy.  $\beta_j$ is increased incrementally in the set $\lbrace 125, 175, 225 \rbrace$ (nm) to reduce computational complexity for finding the desired intensity distributions with such a large number of paths.  
  
\setlength\tabcolsep{3 pt}    
\renewcommand{\arraystretch}{1.5}
\begin{table*}[!t]
\caption{QPC problems and simulation setup parameters}
\begin{center}
\scriptsize
\begin{tabular}{|m{1cm}|m{4cm}|m{6cm}|}
\hline 
  ID & Property & Value \\  
\hline 
\multirow{5}{*}{$Sim_1$} & $N, S_{1}, S_{2} $  & $3,  2, 2$  \\\cline{2-3} 
& $\sigma_0$  (nm) & $500$  \\
\cline{2-3} 
& $\overrightarrow{X}_{1}^T$ (nm), $\overrightarrow{X}_{2}^T$ (nm) & $\left[ -6031.9 \, \, \, \, -2960.6\, \, \, \,  110.7\, \, \, \, 3181.9 \, \, \, \, 6253.2  \right]$, \hspace{0.2in} $\left[ -643.9 \, \, \, \, -327.6\, \, \, \, -11.4\, \, \, \, 304.8 \, \, \, \, 621.1  \right]$   \\
\cline{2-3} 
& $\overrightarrow{d}^T$ $(\mbox{m}^{-2})$ & $\left[ -11825366721.5 \, \, \, \, -114848915118.2 \right]$  \\
\cline{2-3} 
&  $\overrightarrow{L}^T $ (m), $\overrightarrow{\beta}^T$ (nm), $T_s$  ($\mu$m) & $\left[1 \, \, \, \, 400 \times 10^{-6} \, \, \, \, 1\right] $, $\left[ 196.5 \, \, \, \,63.2\right]$, $1$   \\
\hline
\multirow{3}{*}{$Sim_2$} & $N, S_{1,T}, S_{2, T}, S_{3,T}$  & $4,  33, 198, 238$       \\
\cline{2-3} 
&  $\sigma_0$  (nm) & $65$    \\
\cline{2-3} 
&  $\overrightarrow{L}^T $ (m),  $\overrightarrow{\beta}^T$ (nm), $T_s$ (nm) & $\left[0.125 \, \, \, \,1 \, \, \, \, 1 \, \, \, \, 1\right] $, $\left[125  \, \, \, \, 175   \, \, \, \,225   \right]$, $37.69$   \\
\hline
\end{tabular}
\end{center}
\label{Table2} 
\end{table*}
 
\setlength\tabcolsep{2 pt}   
\renewcommand{\arraystretch}{1.2}
\begin{table*}[!t]
\begin{minipage}{0.40\textwidth}
\centering
\caption{Physical parameters} 
\small
\begin{tabular}{|c|c|}
\hline 
Symbol &  Value \\
\hline 
$m$ (kg) & $9.11 \,10^{-31}$     \\
\hline 
$v_z$ (m/s) & $1.46 \,10^{7}$    \\
\hline
$\hbar$  (J $\times$ s) &  $1.05  \, 10^{-34}$ \\
\hline
\end{tabular}
\label{Table3} 
\end{minipage}
\hspace{0.1in}
\setlength\tabcolsep{2 pt}   
\renewcommand{\arraystretch}{1.0}
\begin{minipage}{0.50\textwidth}
\centering
\caption{Path counts on planes for $Sim_1$ }  
\vspace{0.12in}
\begin{tabular}{|c|c|c|}
\hline
\multicolumn{1}{|c|}{Type}           & \multicolumn{1}{c|}{Plane-2} & Sensor   \\ \hline
\multicolumn{1}{|c|}{Non-exotic} & $5$                          & $25$                \\ \hline
\multicolumn{1}{|c|}{$N_E = 1$} & $25$                         & $625$              \\ \hline
\multicolumn{1}{|c|}{$N_E = 2$} & $105$                        & $11025$       \\ \hline
\multicolumn{1}{|c|}{$N_E = 3$} & $425$                        & $180625$     \\ \cline{1-3}
\end{tabular}
\label{Table4} 
\end{minipage}
\end{table*}
\renewcommand{\arraystretch}{1}
\setlength\tabcolsep{6 pt}

\subsection{Simulation-1: Period Finding and SDA Solution}

A simple numerical SDA problem is realized by choosing $b[n] \equiv \overrightarrow{d}^T \, \overrightarrow{x}_n^s$ where the vector $\overrightarrow{d}$ being the same for each path makes the solution classically tractable. In other words, a classically solvable and simple problem is proposed to observe the period finding capability of QPC. In fact, the period of $e^{\imath \, 2 \, \pi \, \overrightarrow{d}^T \, \overrightarrow{x}_n^s \, k}$ can be found classically in an efficient manner by computing the summation $\sum_{x_1^s} \hdots \sum_{x_{N-1}^s } e^{\imath \, 2 \, \pi \, \overrightarrow{d}^T \, \overrightarrow{x}_n^s \, k}$ classically. The summation {\color{black}is} calculated by separating the terms for each $x_j^s$ where $j \in [1, N-1]$   and then multiplying the results at the sampling point $k$. However, for the general case of $n$th path dependent $\overrightarrow{d}_{N-1,n}$, it becomes not possible to separate the summations while requiring to exploit the advantages of QPC summation. The simulation and analysis for more difficult SDA problems are open issues. 

Total number of non-exotic paths is $N_p = 25$ while the number of all paths including  non-classical ones, i.e., $N_{p,N}^E$, for varying $N_{E}$ is shown in Table \ref{Table4}. As $N_E$ increases, $N_{p,N}^E$ becomes significantly large making it difficult to calculate the intensity. The intensity roughly converges as $N_E$ increases to three.
 
\begin{figure*}[!ht]
\centering
\includegraphics[height=3.2cm]{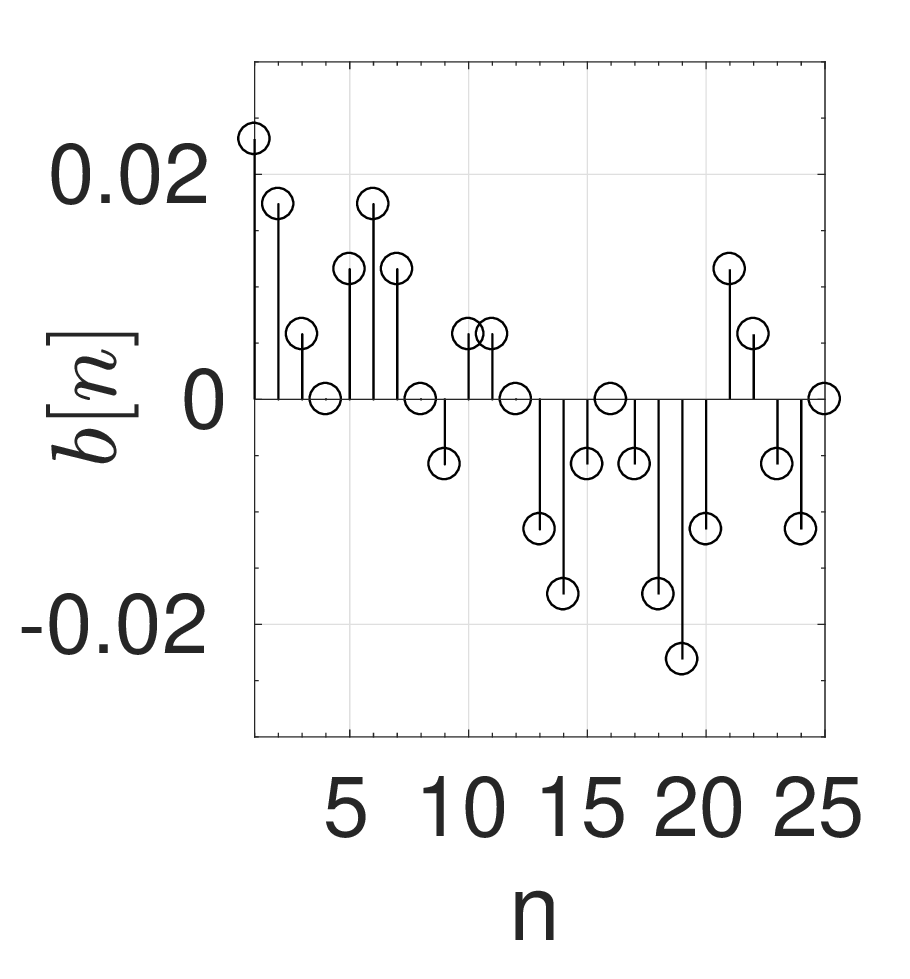} \includegraphics[height=3.2cm]{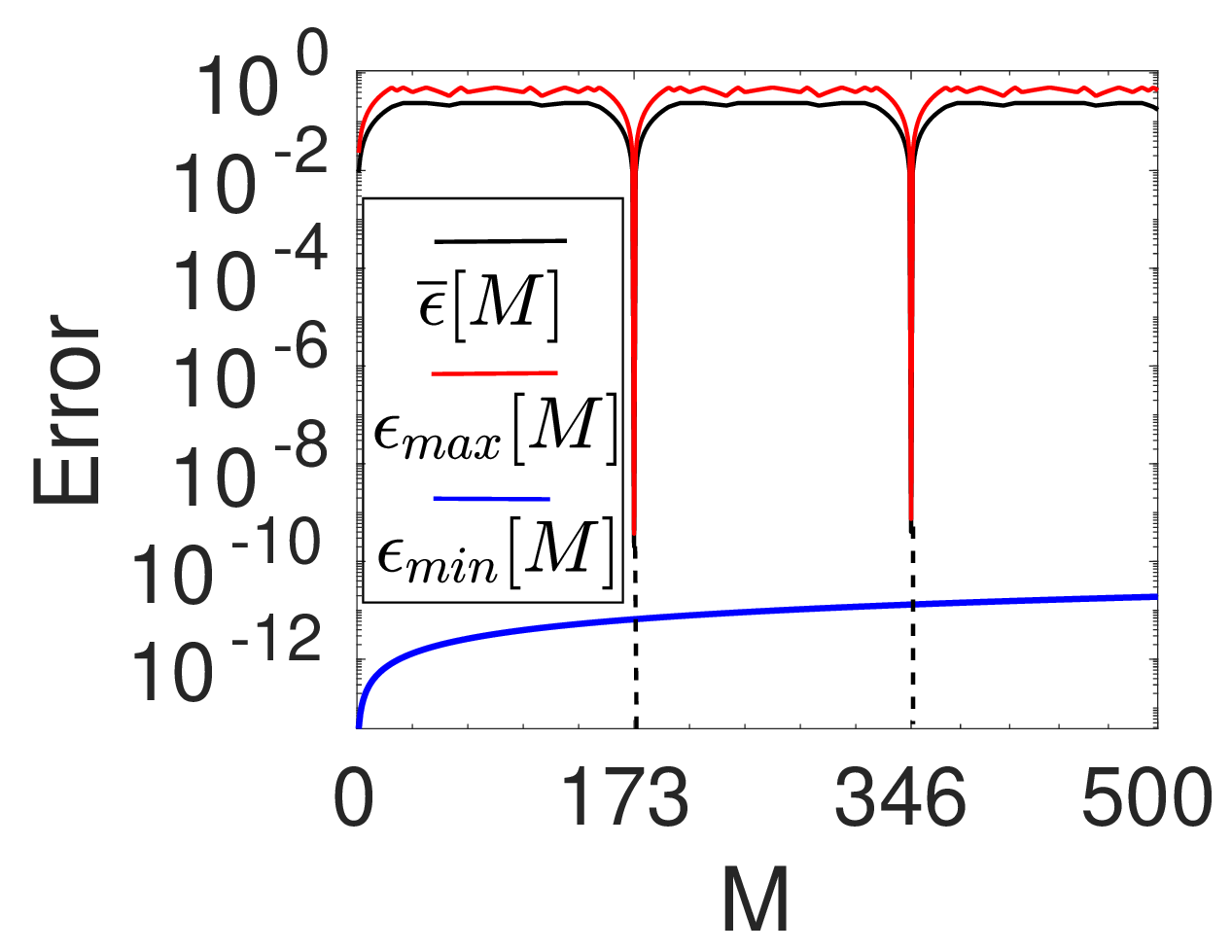} \includegraphics[height=3.2cm]{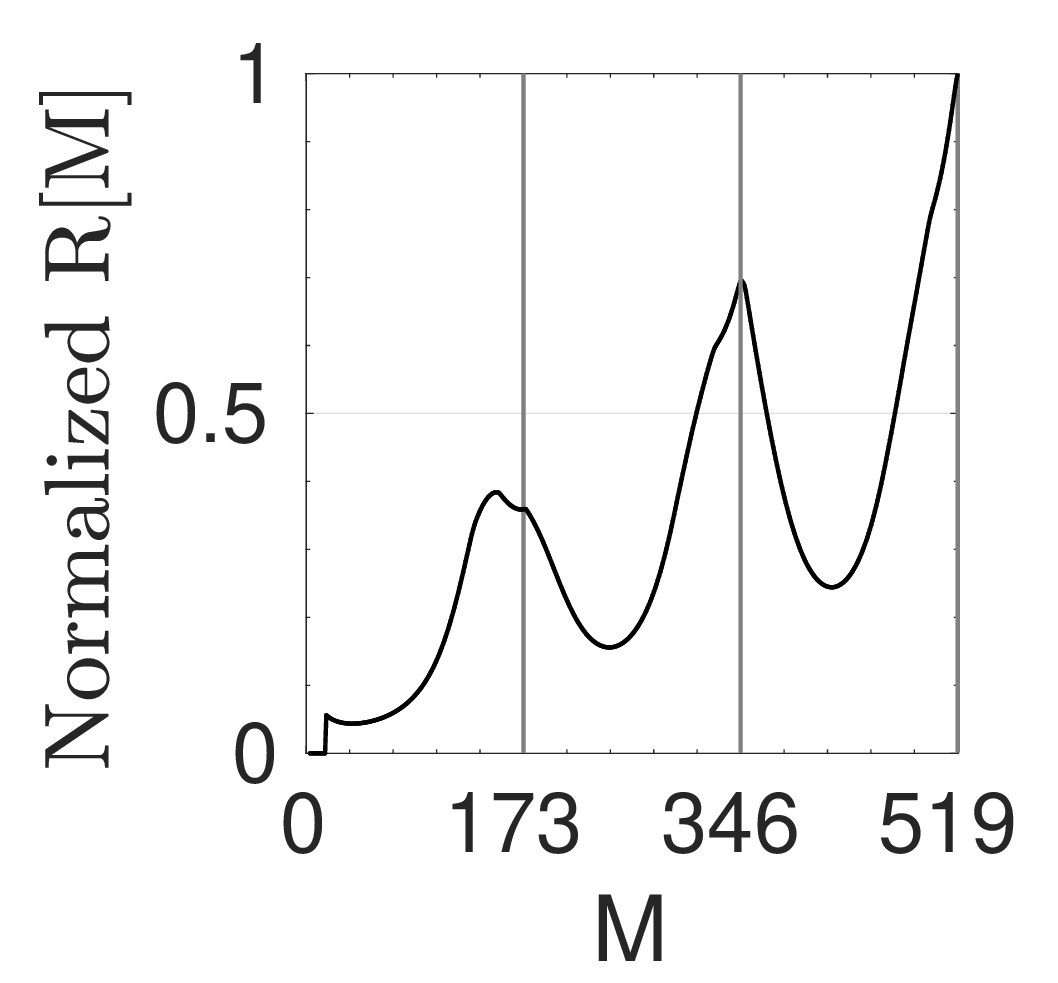}  \\
\hspace{0.2in}  (a) \hspace{1.3in} (b) \hspace{1.25in} (c)\\
\vspace{0.1in}
 \includegraphics[height=3.2cm]{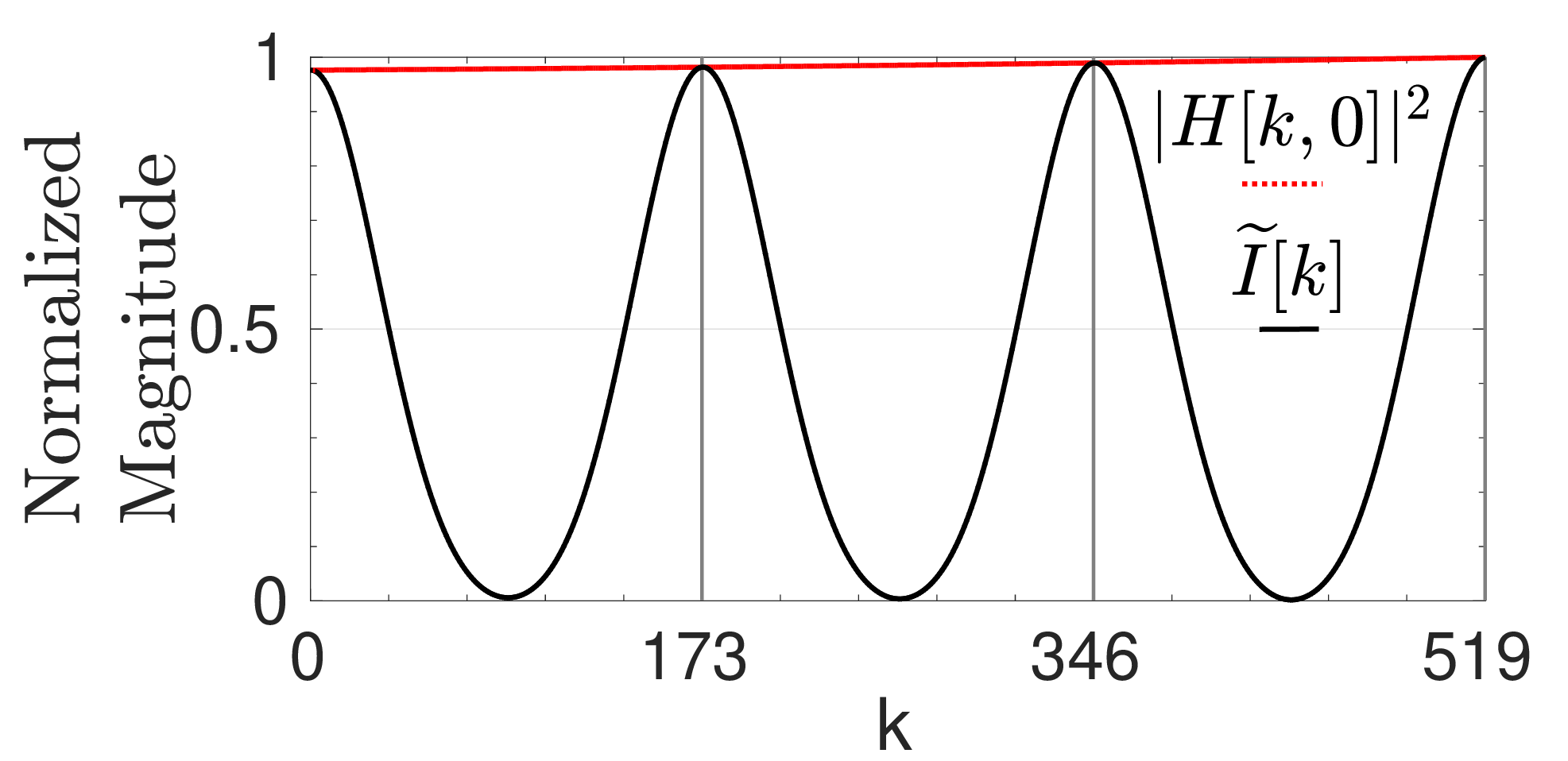} \includegraphics[height=3.2cm]{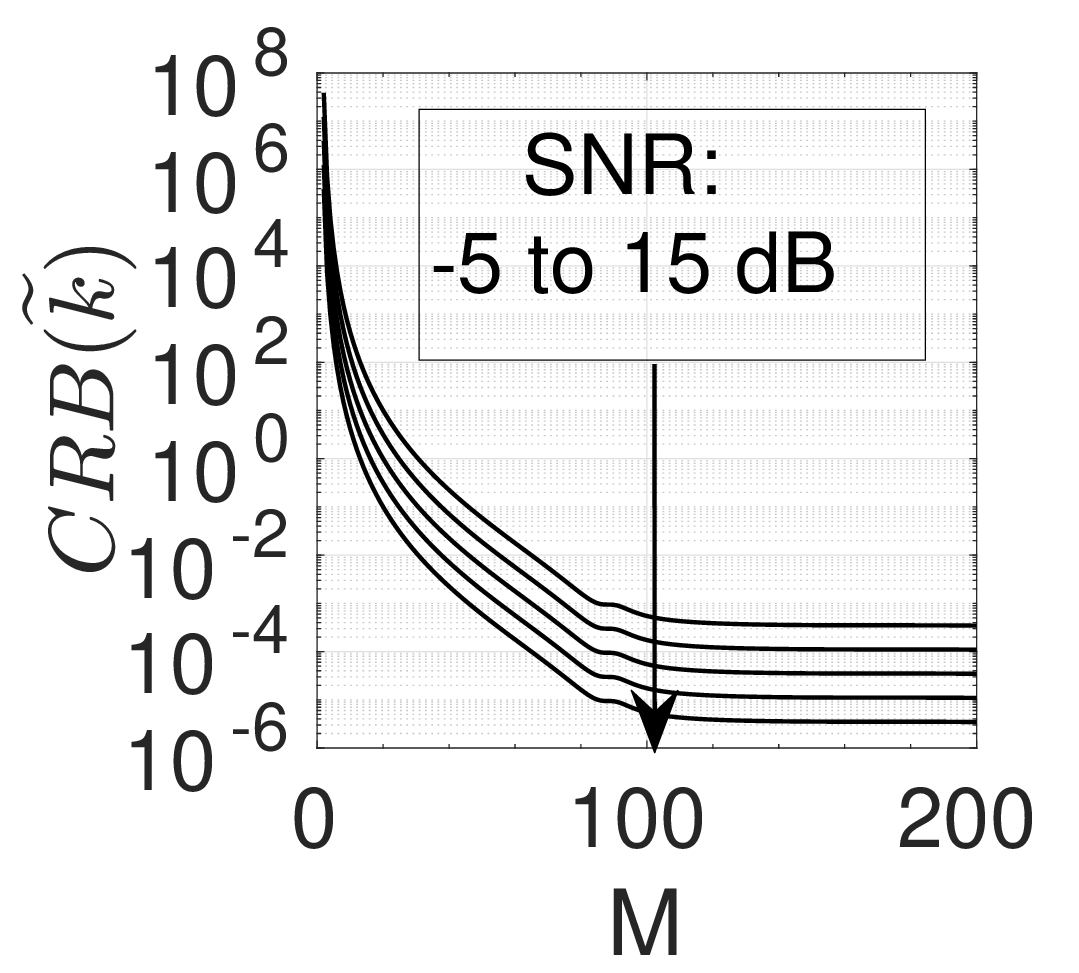}  \\
\hspace{0.9in}  (d) \hspace{1.8in} (e) 
\caption{(a) $b[n]$ for $n \in [0, N_p-1]$ defining  SDA problem of  $Sim_1$,  (b) error terms in SDA problem  including the minimum ones for varying $M$,  (c) $R[M]$ for varying  $M$  with similar periodicity, (d) normalized $\widetilde{I}[k]$  and $\vert H[k,0] \vert^2$ for varying $k$, and (e) Cramer-Rao bound for varying number of samples in $[0, M \, - \,1]$ and SNR in $[-5, \, 0, \, 5, \, 10, \, 15]$ dB.} 
\label{Figure4}
\end{figure*}
 
The fractional numbers forming the SDA problem are shown in Fig. \ref{Figure4}(a). They are chosen to satisfy $\widetilde{k} = 173$. In Fig. \ref{Figure4}(b),  error terms $\overline{\epsilon}[M]$, $\epsilon_{max}[M] $ and $\epsilon_{min}[M]$ are shown for $Sim_1$. The mean error is smaller than $10^{-8}$ for $M  = \widetilde{k} = 173$, assumed to be the SDA solution with  accuracy of eight digits. 
 
In Fig. \ref{Figure4}(d),  normalized $\widetilde{I}[k]$ and $\vert H[k,0]\vert^2$ are shown satisfying the  Theorem \ref{theorem1} such that satisfying $\widetilde{I}[\widetilde{k}] > \widetilde{I}[k]$ and $\widetilde{I}[k] < \vert H[k,0]\vert^2$ for ${\color{black}0} \, {\color{black}<} \, k \, <  \, \widetilde{k}$ with increasing $\vert H[k,0]\vert$.  $IFFT$ based method provides an accurate estimation of $\widetilde{k}$ as shown in Fig. \ref{Figure4}(d). Fluctuations are more visible as $M$ increases at multiples of $173$ while the maximum points of $R[M]$ show  periodicity of $173$ as shown in Fig. \ref{Figure4}(c).   CRB is shown for varying SNR defined as $  I^2[k_p]   \, / \, \sigma_p^2$ in Fig. \ref{Figure4}(e) with a low bound for the number of samples larger than a few tens. Therefore,   estimation methods for damped sinusoids can be applied such as the ones in  \cite{fft1}. 

\begin{figure*}[!ht]
\centering
 \includegraphics[height=3.2cm]{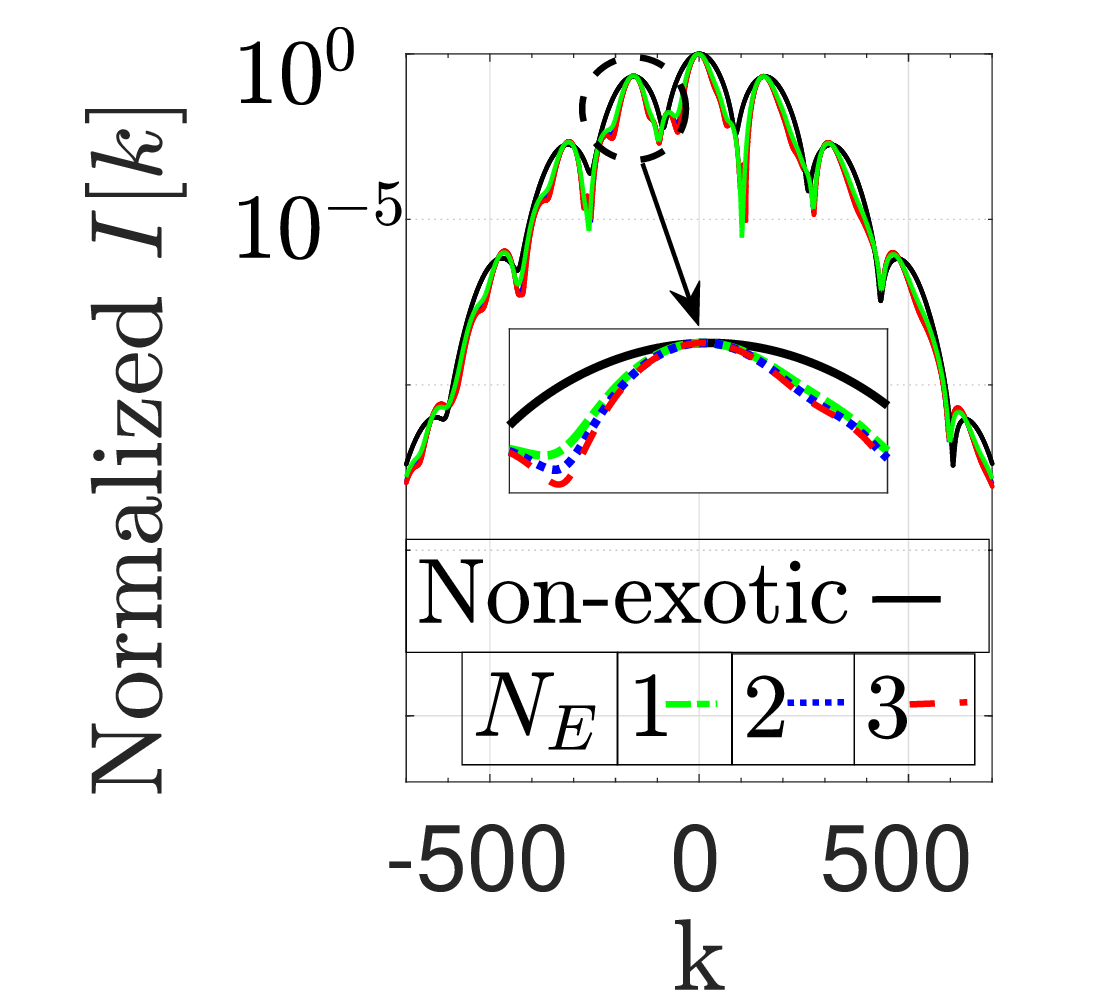} \includegraphics[height=3.2cm]{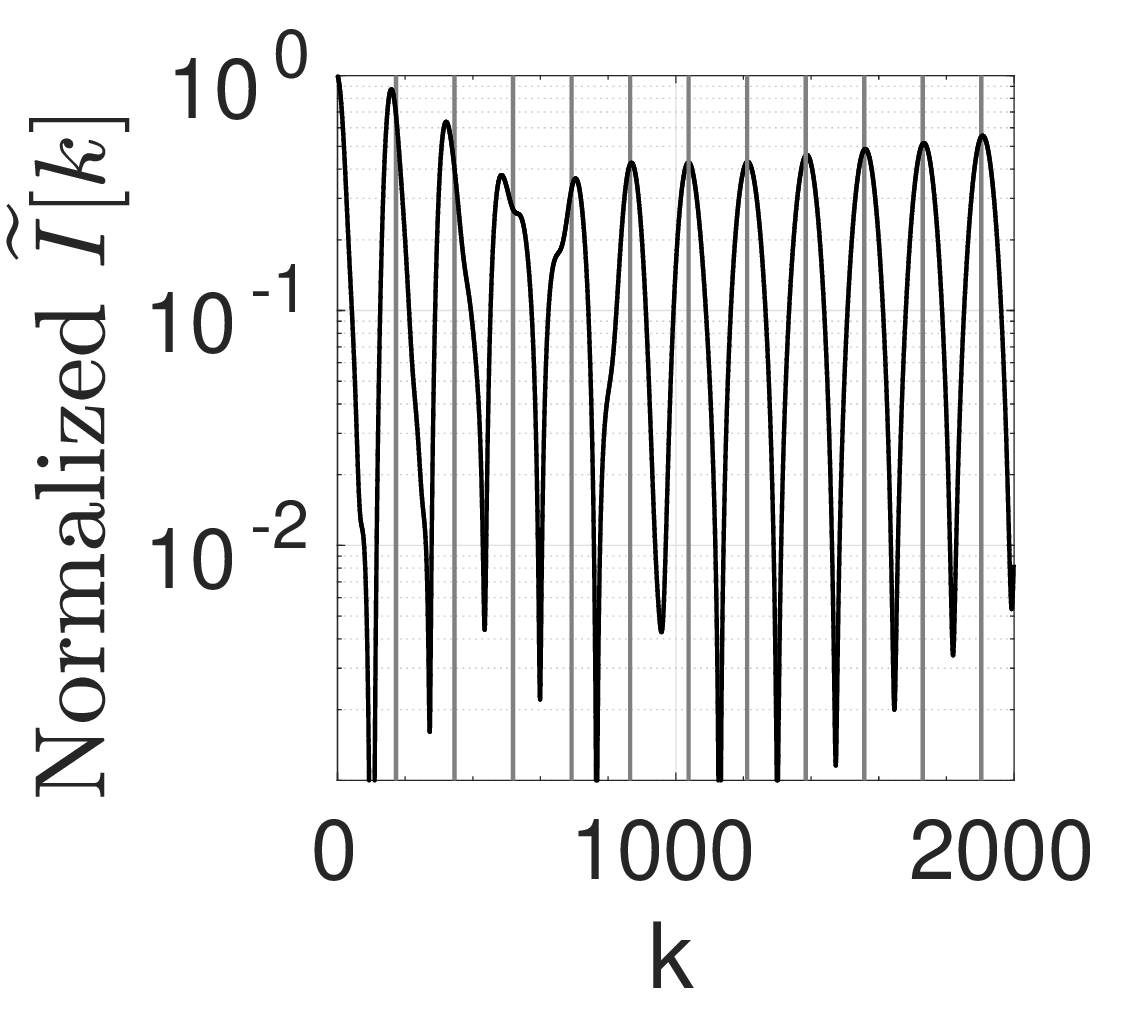} \includegraphics[height=3.2cm]{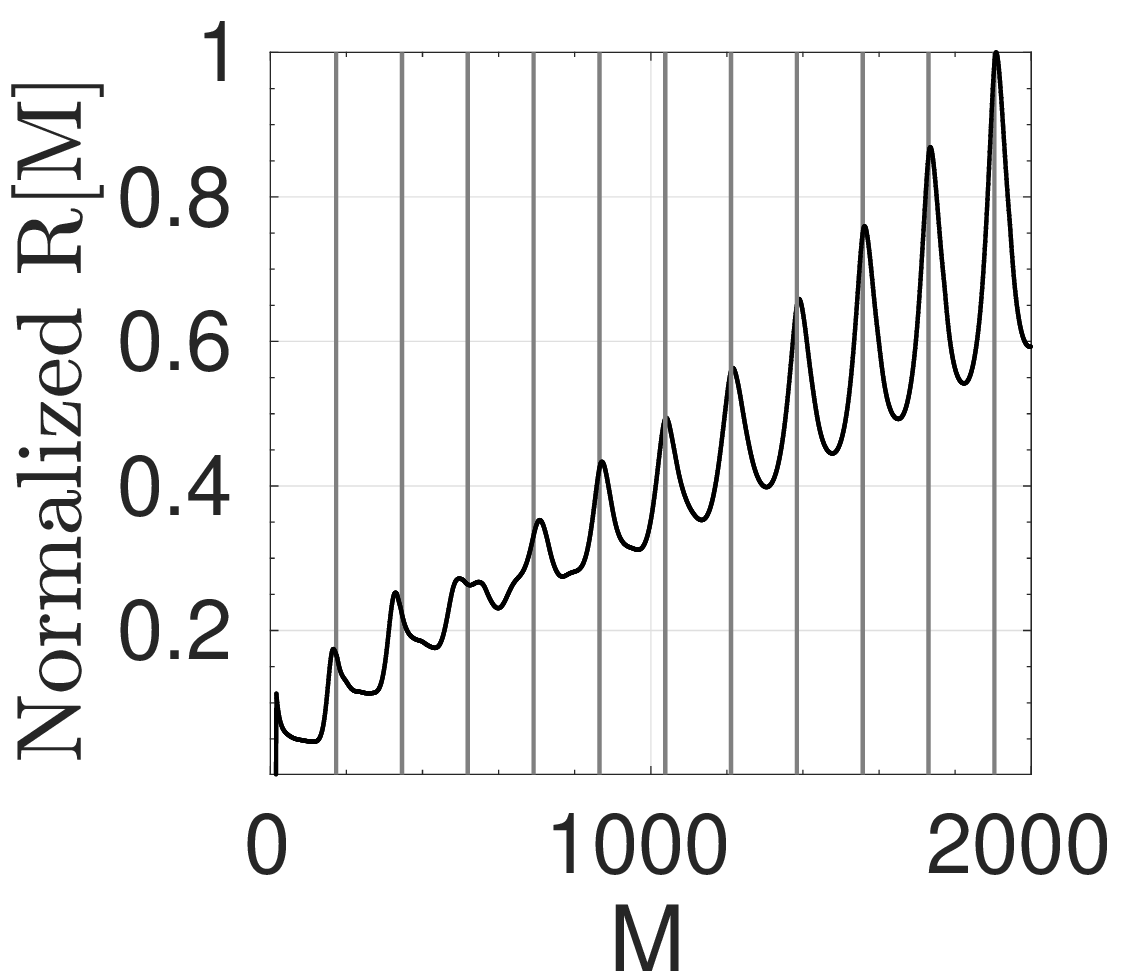} \\
\hspace{0.25in}  (a) \hspace{1.15in} (b)  \hspace{1.23in} (c) \\
\caption{(a)   $I[k]$ (with the maximum normalized to unity) by including non-classical paths and for varying $N_{E}$ where the middle part shows zoomed intensity distribution at the center, (b) normalized $\widetilde{I}[k]$ and (c) $R[M]$ for varying $k$  and $M$, respectively, with the same $x$ axis and periodicity, for the case of $N_E = 3$ where the lines show  multiples of $\widetilde{k} = 173$.} 
\label{Figure5}
\end{figure*}

$I[k]$, i.e., $e^{2\, A_{N-1} \, (k\,T_s)^2} \widetilde{I}[k] $,  is normalized by setting the maximum value to unity as shown in  Fig. \ref{Figure5}(a) while including non-classical paths for varying $N_{E}$. The main structure of the distribution is preserved while  effects for increasing $N_E$ are attenuated as shown in Fig. \ref{Figure5}(b) for the case of $N_E =3$  where normalized   $\widetilde{I}[k]$ periodicity for varying $k$ and the value of $\widetilde{k}$ are still reliably extracted. The same observation is preserved in normalized $R[M]$ for varying $M$ in Fig. \ref{Figure5}(c).   Utilizing   values of  $\widetilde{I}[k]$ for large $k$ requires higher precision measurement instruments due to  significant attenuation  in $I[k]$ at distant sample locations  as shown in Fig. \ref{Figure5}(a) and longer time to collect particles. Special tuning and design of QPC setup  are required for efficient solutions exploiting QPC. 

\subsection{Simulation-2: Energy Flow and Complexity }
 
 \begin{figure*}[!ht]
\centering
\includegraphics[ height = 4cm]{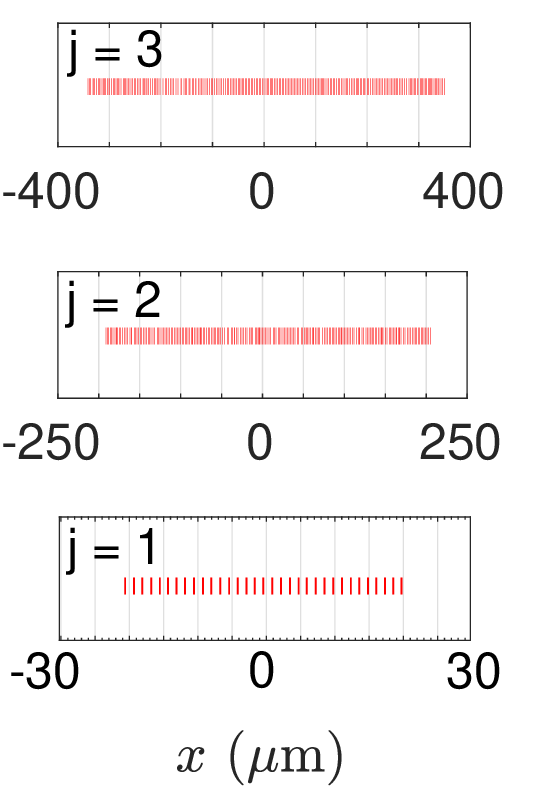}  \includegraphics[  height = 4cm]{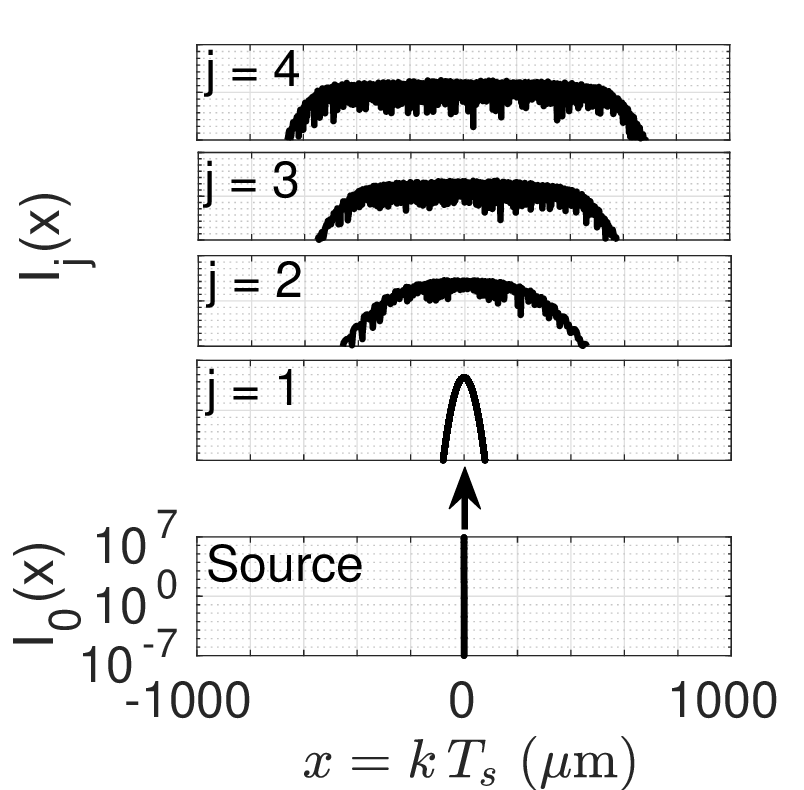}   \includegraphics[  height = 4cm]{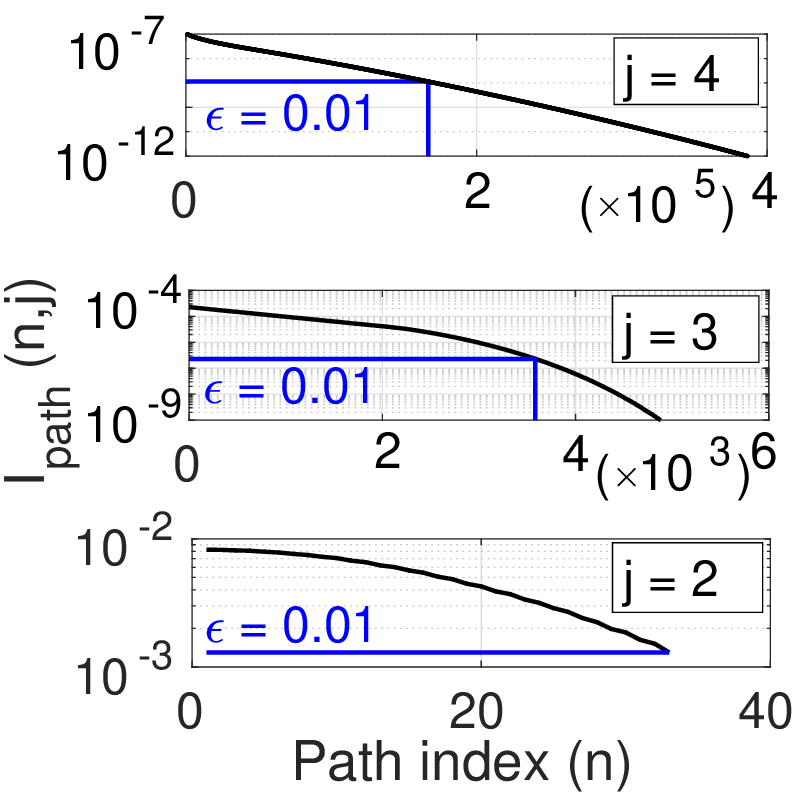} \\ 
   (a)  \hspace{1.1in} (b)  \hspace{1.5in} (c)\\
   \vspace{0.1in}
 \includegraphics[  height = 4cm]{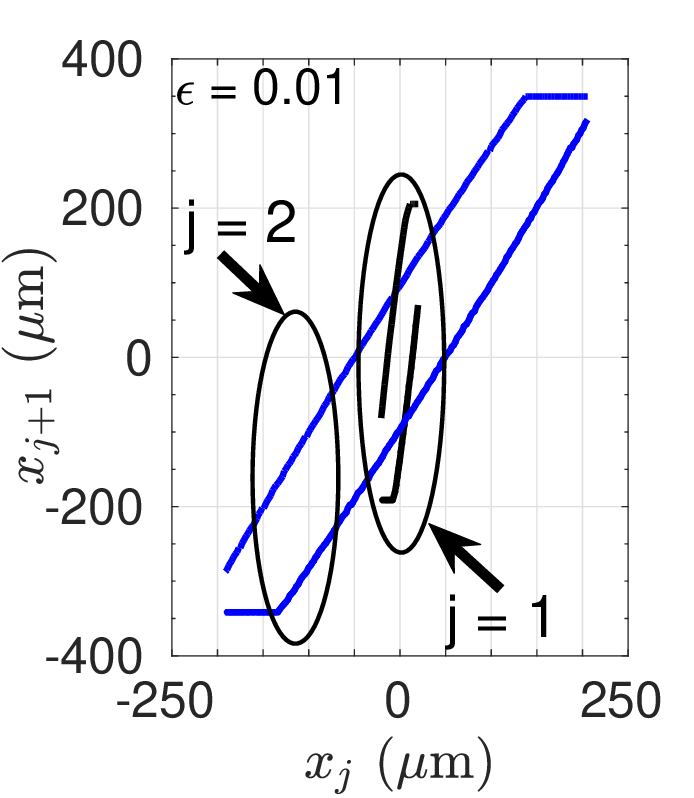} \includegraphics[  height = 4cm]{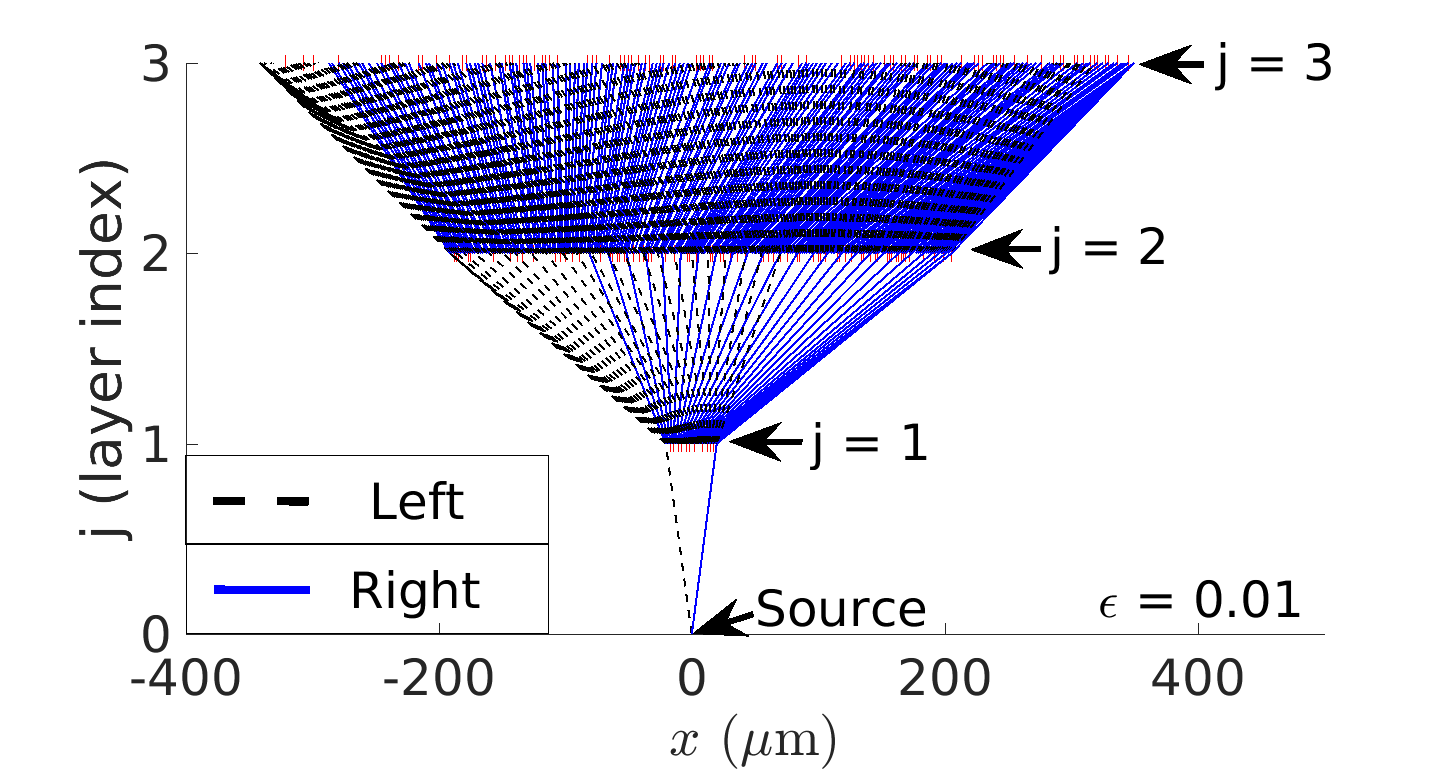}\\
  \hspace{-0.5in}  (d)  \hspace{1.9in} (e) 
\caption{Energy-complexity trade off  where (a) the slit positions for $j \in [1,3]$, (b) $I_j(x)$ for $j \in [1, 4]$, (c) $I_{path}(n,j)$ on $j$th plane for $j \in [2, 4]$, (d) the slit positions on $(j+1)$th plane ($x_{j+1}$) having a large amplitude path  evolving from the slits on $j$th plane with the position $x_j$, (e) the same relation in (d) visually  with the region of slits on $(j+1)$th plane  marked with the left and right boundaries for the paths evolving from the slits on $j$th plane.} 
\label{Figure6} 
\end{figure*}

\begin{figure}[!t]
\centering
\includegraphics[ height = 3.9cm]{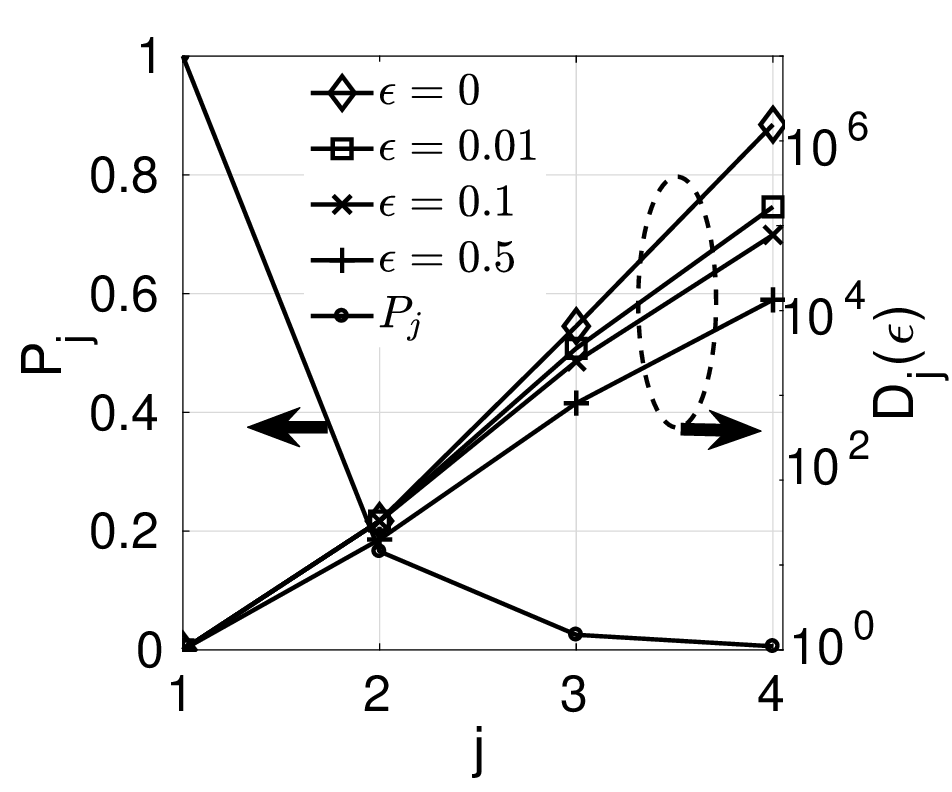}  \includegraphics[ height = 3.9cm]{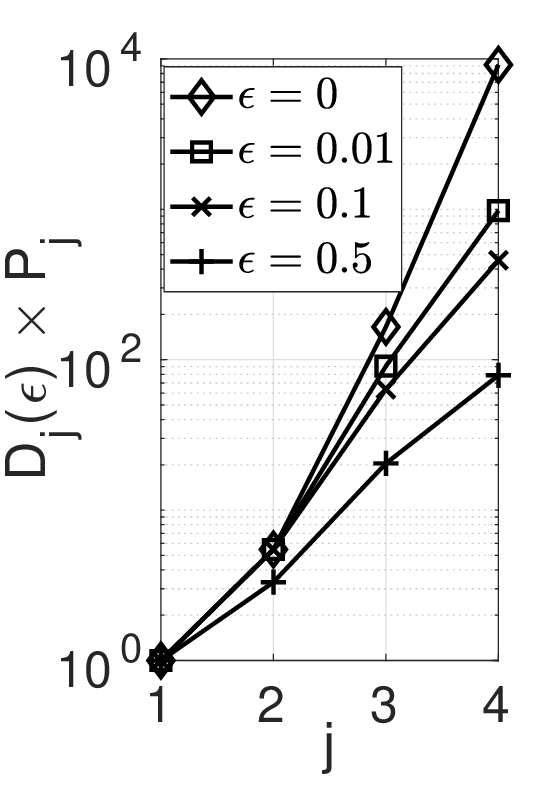} \includegraphics[ height = 3.9cm]{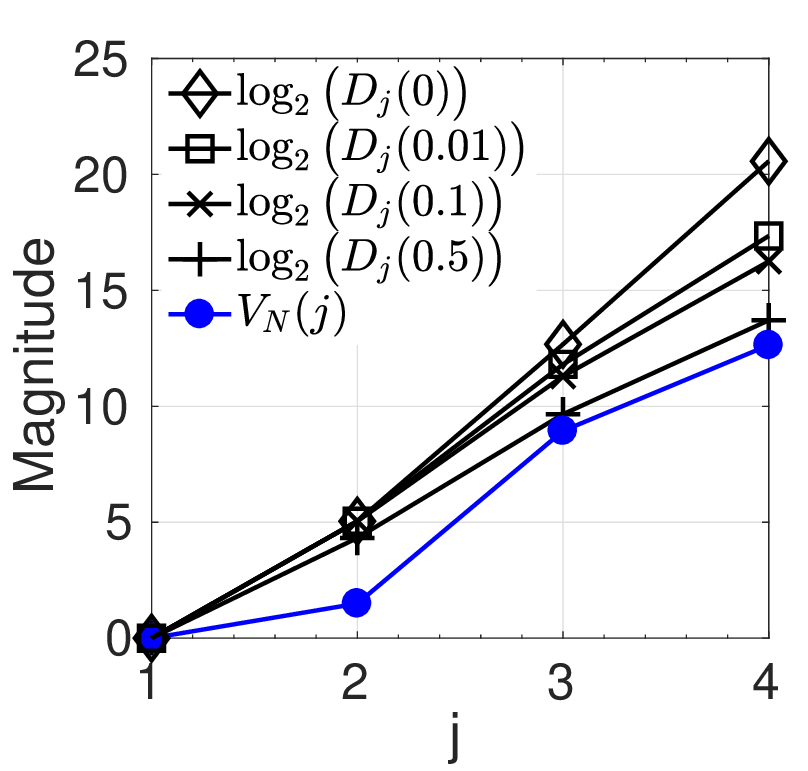}\\
\hspace{0.25in} (a)  \hspace{1.35in} (b) \hspace{1.1in} (c)
\caption{ (a) The probability $P_j$ versus the number of paths $D_{j}(\epsilon)$, (b) $D_{j}(\epsilon) \times P_{j}$ and (c) the comparison of logarithmic number of paths with the negative volume of Wigner  function ($V_N(j)$) for the normalized wave function on each layer for $j  \in [1,4]$ and for varying $\epsilon \in \lbrace 0, 0.01, 0.1, 0.5 \rbrace$ showing a correlated increase in both of them.}
\label{Figure7}
\end{figure}

Energy flow  versus interference complexity trade off for  $Sim_2$ is shown in Figs. \ref{Figure6} and \ref{Figure7}. In Fig. \ref{Figure6}(a), the positions of the slits are shown where a larger number of slits are utilized in consecutive planes to cover the spread intensity distribution. In Fig. \ref{Figure6}(b), complicated interference patterns are shown for the planes  with the indices $j \in [2, 4]$ while the Gaussian source ($j =0$) and the free space propagated version ($j = 1$) are also shown. In Fig. \ref{Figure6}(c), path amplitudes defined in (\ref{pathamp})  are shown for  $j \in [2, 4]$ while the number of paths is marked for $\epsilon = 0.01$ for the definition of energy  constrained Hilbert space in (\ref{hilbertdefinition}) modeled with $D_{j}(\epsilon)$. The regions of the slits on $(j+1)$th plane where a slit on $j$th plane creates a path with high interference amplitude are shown in Figs. \ref{Figure6}(d) and (e). Fig. \ref{Figure6}(d) shows the left and right  boundaries of the slit positions at $x_{j+1}$ corresponding to each slit at $x_j$ for $j = 1$ and $2$. The paths  are shown in Fig. \ref{Figure6}(e) by connecting the slits with a visual representation starting from the source until the slits of the third plane. The left and right boundaries form a region where the slits on   consecutive planes form high amplitude paths. The leftmost slit on  first plane at $x = -20.63 \, \mu$m forms high amplitude paths with the slits having the positions between $(-191.2, -81.19) \, \mu$m (marked as left and right boundaries) including   $\approx 57$ neighbor slits on the second plane.

\begin{figure}[!ht]
\centering
\includegraphics[ width = 5cm]{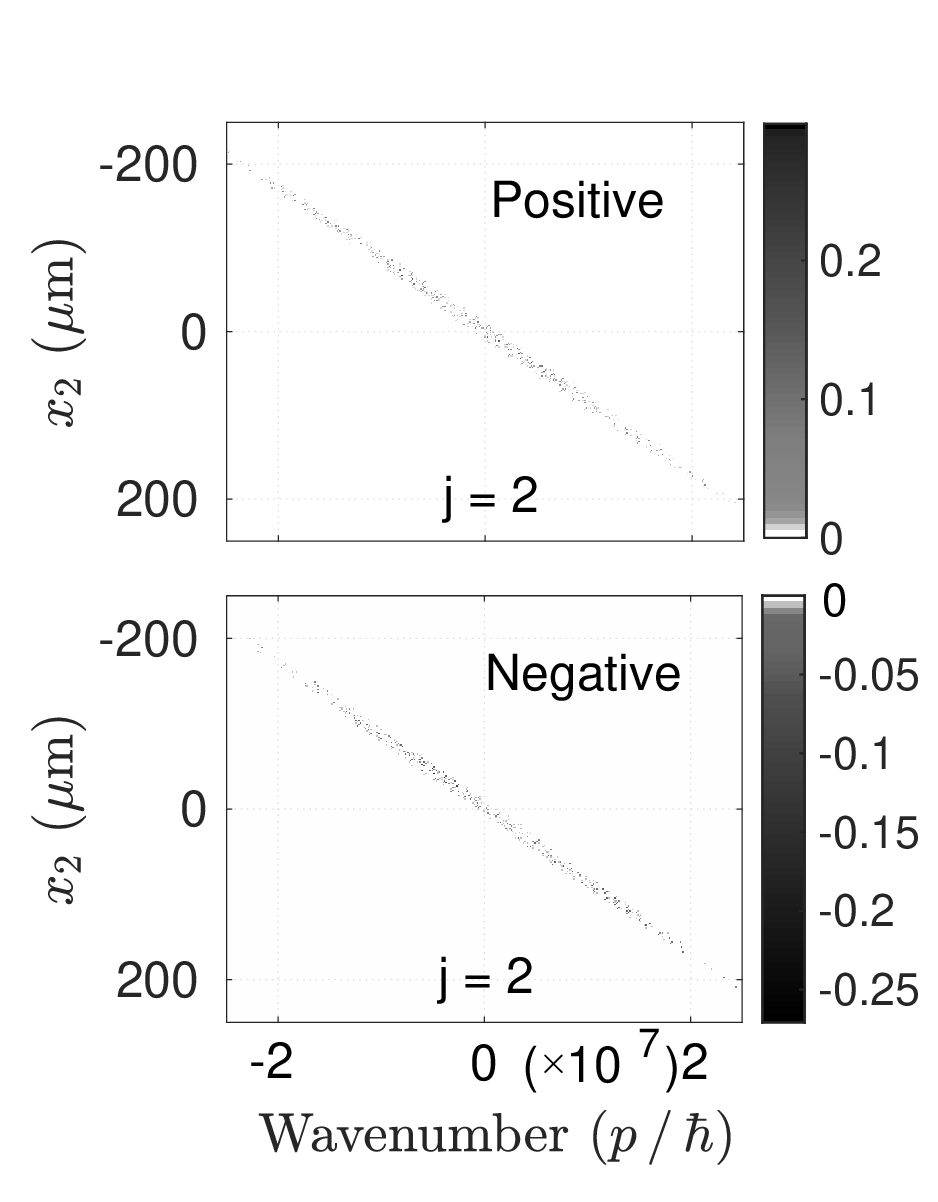} \hspace{0.5cm} \includegraphics[width = 5cm]{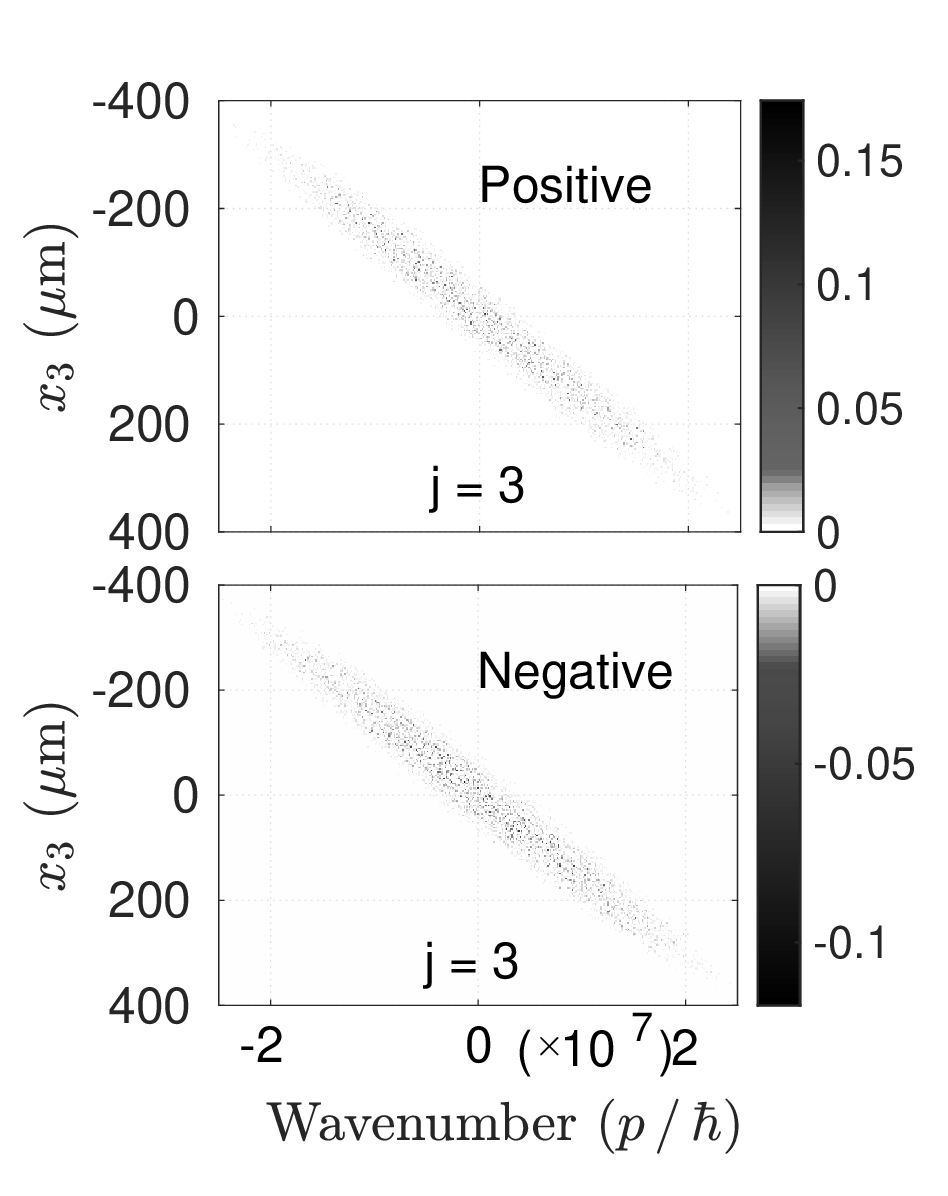} \\
\hspace{0.1in} (a) \hspace{1.95in}  (b) \\
\vspace{0.1in}
\hspace{0.7cm}\includegraphics[ height = 3.6cm]{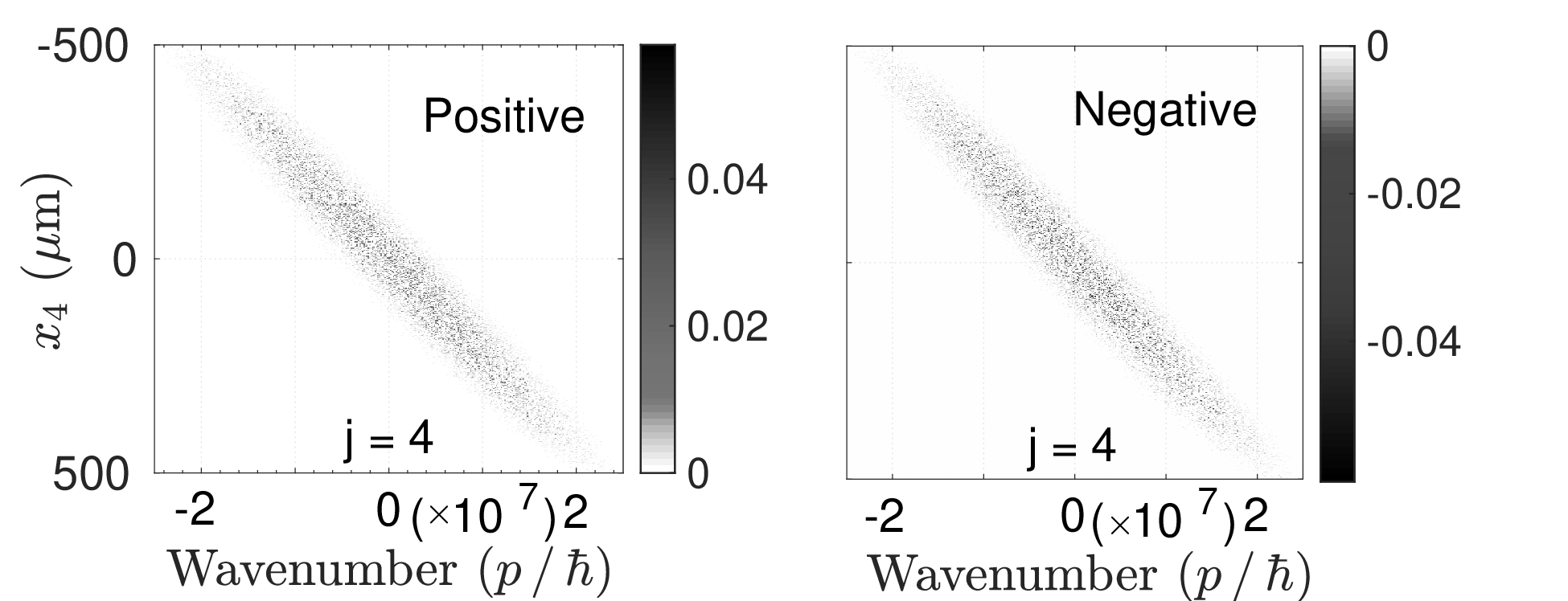}\\
 (c) 
\caption{The positive and negative parts of the Wigner function for the wave functions on (a) the second, (b) the third, and (c) the fourth layers.}
\label{Figure8}
\end{figure}

Fig. \ref{Figure7} shows the results of energy-complexity trade off for varying $\epsilon \in \lbrace 0, \,0.01, \,0.1, \,0.5 \rbrace$. Fig. \ref{Figure7}(a) shows $P_{j}$ versus $D_j(\epsilon)$ for varying $\epsilon$ and the  plane index $j$.  $P_j$ drops to approximately $5.88 \times 10^{-3}$ on the fourth plane while $D_j(\epsilon)$ significantly increases to the values between $1.34 \times 10^{4}$ and $1.55 \times 10^{6}$ for $\epsilon$ between $0$ and $0.5$. The performance parameter  $D_j(\epsilon) \times P_{j}$ is simulated in Fig. \ref{Figure7}(b) showing a significantly increasing size of energy constrained Hilbert space with the number of planes while the exact calculation of intensity requires the calculation of all $33 \times 198 \times 238 = 1555092$ paths.  In Fig. \ref{Figure7}(c), the negative volume of the Wigner distribution ($V_N(j)$) is compared with the $\log_2(.)$ of the number of paths as a complexity performance metric assuming to be implemented with qubits. It is observed that  increasing logarithmic complexity shows a parallel relation with the increasing negative volume of the Wigner function as another supporting observation of the non-classical resource structure of QPC. It requires further analysis for the parametric definition of the amount of resources. The positive and negative parts of the Wigner function for the wave functions on the second, third and fourth planes are shown in Figs. \ref{Figure8}(a), (b) and (c), respectively. Wave functions are normalized on each plane to satisfy  $\int \int W(x,p) \, dx \, dp = 1$. It is observed that as the layer index $j$ increases, the number of highly interfering time-momentum locations increases with a spread in the area of the Wigner function.

\begin{figure}[!ht]
\centering
\includegraphics[ height = 4.2cm]{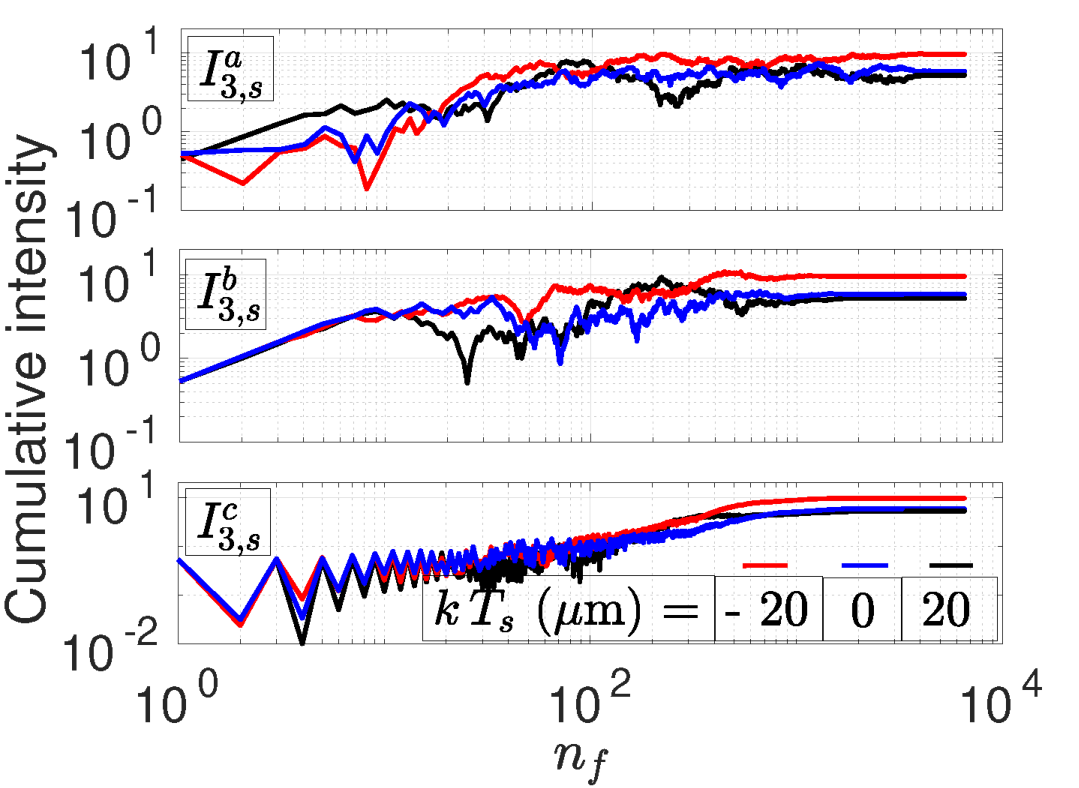}  \includegraphics[ height = 4.2cm]{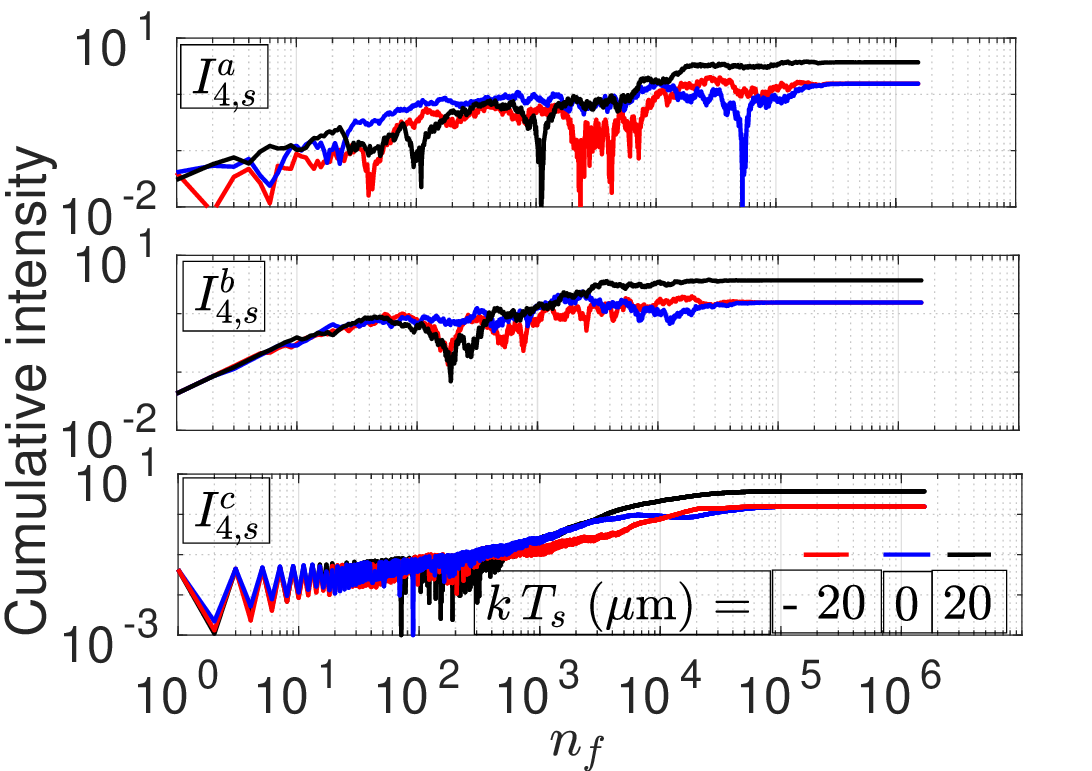}\\
\hspace{0.2in} (a)  \hspace{2.2in} (b)
\caption{ Cumulative summations defined in  (\ref{pathampsorted}) in Section \ref{Section6} for the contributions of the sorted paths with respect to the Type-a, b and c path sorting methods on the (a) third and (b) fourth planes.}
\label{Figure9}
\end{figure}  

In Fig. \ref{Figure9}, the cumulative summations defined in  (\ref{pathampsorted}) in Section \ref{Section6} for the contributions of the sorted paths with respect to the three different types of path sorting methods are shown. The cumulative sums on the third and fourth planes are shown in Figs. \ref{Figure9}(a) and (b), respectively, for the sampling positions of $k \, T_s \in \lbrace -20, \, 0,\, 20 \rbrace \, \mu$m. Cumulative summation with Type-a is calculated by sorting  the paths with respect to the descending probability of the particle to evolve through the specific path. Type-b is calculated by sorting the indices with respect to the descending magnitude of the intensity at the specific sampling location but not with respect to the total probability at all sampling locations. Finally, in Type-c, the paths in Type-b are coupled one by one in the order of descending magnitude to cancel each other. It is clearly observed that Type-b and Type-c having local characteristics tuned to the specific sampling point reach the stable cumulative intensity earlier compared with Type-a. If the paths are chosen to cancel each other in Type-c, then there is still oscillation and a nonlinear increase in the cumulative summation. It still requires a large number of paths to be summed to find the stable level of intensity on the sampling positions. Furthermore, there is not any apparent way to find the correct paths which approximately cancel each other among the exponentially increasing number of paths. Therefore, sorting and limiting the paths with respect to the probability with the complexity term $\epsilon$ do not provide the required sets of the paths applicable at all sampling positions. In other words, $D_j(\epsilon)$ at $\epsilon =0$ can be more reliable to cover all the sampling locations. Therefore, QPC energy-complexity  trade off provides a significantly large energy constrained Hilbert space.

\section{Discussion and Open Issues}
\label{Section11}

QPC system design requires further efforts listed as follows to be experimentally realized and improved for solutions of different numerical problems:
\begin{enumerate}
\item Modifying  (\ref{qpcpowergeneral})  with exponentially increasing number exotic paths.
\item Determining the complexity class of calculating QPC output intensity and comparing with universal classical or quantum computers.
\item  Determining the set of period finding related problems which can be efficiently solved while in particular all the open issues regarding the utilization of QPC for SDA solutions described in Section \ref{openissuesofSDA}.
\item The best utilization of (\ref{qpcpowergeneral}) for computational purposes in addition to the practical problems presented.
\item Experimental verification and realization of Gaussian  slits and mathematical modeling of QPC systems with non-Gaussian arbitrary slit properties.
\item  Minimization of decoherence  due to unintentional interactions with the particles during propagation \cite{divincenzo1998decoherence}.  
\item  Designing novel diffraction geometries in addition to the planar ones to solve different number theoretical problems as a universal system design based on history based resources.
\end{enumerate} 

\section{Conclusion}
\label{Section12}

A low hardware optical QC architecture is presented combining all-in-one targets: practical problem solving capability,  energy efficient processing of sources, using only coherent or classical particle sources including both bosons and fermions, transforming the particle source through the simple classical optics and intensity measurement with simple particle detectors. The method denoted by QPC exploits MPD of the particles by creating exponentially increasing number of propagation paths through the slits in consecutive diffraction planes. Non-Gaussian nature of the propagating wave function is exploited for QC purposes with Feynman's path integral approach and numerical analysis is provided showing increasing negative volume of Wigner function. QPC promises solutions for two practical and hard number theoretical problems: partial sum of Riemann theta function and period finding for solving specific instances of Diophantine approximation problem. Open issues including the best utilization of QPC for computing purposes and modification of diffraction system for covering different problems are discussed. 
     
\setlength\tabcolsep{2 pt}    
\renewcommand{\arraystretch}{2}
\begin{table*}[!t]
\caption{Iteration parameters for FPI modeling}
\begin{center}
\scriptsize
\begin{tabular}{|m{0.7cm}|m{3.9cm}||m{1.37cm}|m{5.2cm}|}
\hline 
  & Formula &  & Formula  \\
\hline 
 $\Psi_0(x)$ & $  \mbox{exp}\big(- \, x^2 \, / \, (2 \, \sigma_0^2) \big) $ $/ $ $ \sqrt{\sigma_0 \,\sqrt{\pi }} $  & $\overrightarrow{p}_1^T$ & $\left[p_{1,1} \, \hdots \, p_{1,N-1} \right] $    \\ 
\cline{1-4} 
 $A_0$ & $ - m^2 \sigma_0^2 \, / \, (2\, \hbar^2 t_{0,1}^2\,+\,2\,m^2 \sigma_0^4)$  & $\overrightarrow{p}_2^T$, $\overrightarrow{p}_3^T$ & $\left[p_{2,2} \, \hdots \, p_{2,N-1} \, 0 \right]$, $\left[p_{3,2}   \hdots   p_{3,N-1} \, 0 \right]$   \\
\cline{1-4} 
 $B_0$ & $ \hbar \, m \,t_{0,1} \,/ \,(2 \,\hbar^2 t_{0,1}^2 \, + \, 2 m^2 \sigma_0^4) $  & $\chi_0$  &   $   \pi^{-1/4}   \, \sqrt{ m \, \sigma_0 \, / \, (m \, \sigma_0^2  \, +  \, \imath \, \hbar \, t_{0,1}) }$   \\
\cline{1-4} 
 $A_1$ &  $  \beta_1^2  \, m^2 \left(2  \, A_{0} \,  \beta_1^2 \, - \, 1 \right) \, / \, (2  \, \zeta_1)$  & $ \chi_{1,n} $  & $\sqrt{\xi_{1}} \,\mbox{exp} \big( p_{1,1}\, X_{1,s_{n,1}}^2 \big)  $     \\
\cline{1-4} 
$B_1 $ & $ (2   B_{0}   \beta_1^4  \, m^2   +  \hbar \,  m \,  t_{1,2}  \varrho_1)\, / \, (2  \, \zeta_1)   $  &   $C_{1,n} $, $D_{1,n} $  &  $   \zeta_{1,c} \, X_{1,s_{n,1}} $, $ \,\,\,  \zeta_{1,d} \, X_{1,s_{n,1}} $ \\
\hline
\end{tabular}
\begin{tabular}{|c|c|m{9.15cm}|}
\hline 
 $j$ & Symbol & Formula \\
\hline 
\multirow{7}{*}{$[1, N-1]$} & $p_{1, j}$  & $- \big(2 \, \hbar \, t_{j, j+1} (A_{j-1} \, + \,\imath \, B_{j-1}) \, +  \, \imath \, m \big) \, / \, (2 \, \imath \, \varsigma_j)  $   \\ 
\cline{2-3} 
 & $\varsigma_j$, $\xi_{j}$ & $ \beta_j^2 \, m \, + \, \hbar \, t_{j, j+1} \,  \big(2  \, \beta_j^2 \, (B_{j-1} \, - \, \imath  \, A_{j-1}) \, + \, \imath \big)$,  $  \beta_j^2 \, m \, / \, \varsigma_j$ \\
\cline{2-3} 
 & $\varrho_j$ & $   4  \, \beta_j^4  \, \left(A_{j-1}^2 \, + \, B_{j-1}^2 \right) \, - \, 4 \, A_{j-1}   \beta_j^2   + \, 1$  \\
\cline{2-3} 
 & $\zeta_j$ & $   4 \, B_{j-1} \, \beta_j^4 \, \hbar \, m \, t_{j, j+1} \, + \, \beta_j^4 \, m^2 \, + \, \hbar^2  \, t_{j, j+1}^2  \, \varrho_{j}$\\
\cline{2-3} 
 & $\zeta_{j,c}$, $\zeta_{j,d}$ & $  (2 \, B_{j-1}  \, \hbar  \, m  \, t_{j, j+1} \,\beta_j^2      + \beta_j^2 \,  m^2) \, / \, \zeta_j$, $   \hbar \,  m  \, t_{j, j+1} \,  \left(2  \, A_{j-1} \,  \beta_j^2 \, - \, 1\right) \, / \, \zeta_j$ \\
\hline
\end{tabular}
\begin{tabular}{|c|m{1.75cm}|m{8.5cm}|}
\hline 
 $j$  & Symbol  &  Formula    \\
\hline 
\multirow{8}{*}{$[2, N-1]$}   &   $p_{2,j}$, $p_{3,j}$    & $  -  \beta_j^2 \, \hbar \, t_{j, j+1} \, / \, ( 2 \, \imath \, \varsigma_j  ) $, $  -  \hbar \, t_{j, j+1}  \, / \, (\imath \, \varsigma_j) $      \\ 
\cline{2-3} 
 &      $p_{4,j}$,  $p_{5,j}$  &   $ \beta_j^2 \, \zeta_{j,c}$,   $  - 2 \,\hbar \, t_{j, j+1}  \,  A_j \, / \,  m  $    \\
\cline{2-3} 
& $A_j$, $B_j$  & $     \beta_j^2 \, m^2 \,  \left(2 \, A_{j-1} \, \beta_j^2 \, -\, 1\right)\, / \, (2 \, \zeta_j) $,  $    (2 \, B_{j-1} \, \beta_j^4 \, m^2  +   \hbar \, m \, t_{j, j+1} \, \varrho_j)    /  (2 \, \zeta_j) $   \\ 
\cline{2-3} 
  & 
  $C_{j,n} $ & $ \zeta_{j,c}  \, X_{j,s_{n,j}}  \, +  \, p_{4,j}  C_{j-1,n}    + \, p_{5,j} \, D_{j-1,n} $ \\
\cline{2-3} 
  & 
$D_{j,n} $ & $ \zeta_{j,d} \,  X_{j,s_{n,j}} \, -  \, p_{5,j}  C_{j-1,n}    + \, p_{4,j} \, D_{j-1,n}$  \\
\cline{2-3} 
  & 
 $\chi_{j,n}$ & \multicolumn{1}{c|}{ \makecell[{{c}}]{ \\ $\sqrt{\xi_{j}}   \, \mbox{exp}\big(p_{1,j}   X_{j,s_{n,j}}^2 \big)  \, \times \, \mbox{exp}\big(p_{2,j}   (C_{j-1,n}  + \, \imath \, D_{j-1,n})^2 \big)$ \\ $\times  \,  \mbox{exp}\big(p_{3,j} \,(C_{j-1,n} \, + \, \imath \, D_{j-1,n}) \,  X_{j,s_{n,j}}\big) $ \\ \,}} \\
\cline{2-3} 
\hline
\end{tabular}
\end{center}
\label{Table5} 
\end{table*} 

\appendix
\section{Iterative Formulation of Path Integrals}
\label{AppendixA} 

Throughout the appendices, various formulations are listed in Table \ref{Table5}. In the following, all the calculations are performed for the specific $n$th path assuming  the corresponding slit width parameters are given by the values of $\beta_j$ on $j$th plane for $j \in [1, N-1]$. Therefore, all the following parameters depend on the path index $n$ without explicitly showing, e.g., $A_{j}$ and $B_{j}$ should be replaced with $A_{j, n}$ and $B_{j, n}$, respectively, for the specific $n$th path. The same is valid for the remaining parameters except the local parameters $C_{j,n}$, $D_{j,n}$ and  $\chi_{j,n}$. They depend directly on the slit positions and accordingly on $n$ such that they are denoted by including the path index $n$ for $j$th plane as $C_{j,n}$, $D_{j,n}$ and  $\chi_{j,n}$. Similarly,  $\overrightarrow{c}_{N-1}$ and $\overrightarrow{d}_{N-1}$ should be replaced by $\overrightarrow{c}_{N-1, n}$ and $\overrightarrow{d}_{N-1, n}$, respectively. The final resulting matrix $\mathbf{H}$ utilized in (\ref{qpcpower}) should be replaced with $\mathbf{H}_{N-1, n}$ as in (\ref{qpcpowergeneral}). Only the parameters $A_0$, $B_0$ and $\chi_0$ are path independent.  Therefore, the corresponding numerical results of the expressions for varying slit widths on each path are obtained by changing $\beta_j$ on $j$th plane for the corresponding path. After integration in (\ref{eq1}), the following is obtained:
  \begin{align} 
\begin{split}
\label{jthplane}
\Psi_{n,j}(x)  \, = \, &  \chi_{0} \, \bigg( \prod_{k=1}^{j-1} \chi_{k,n} \bigg) \, e^{(A_{j-1} \,+ \, \imath \, B_{j-1})\, x^2 \,+ \, (C_{j-1,n}  \,+ \, \imath \, D_{j-1,n})\, x} 
\end{split}
\end{align}
where $j \in [1,  N]$, $x$ corresponds to  the position in $x$-axis on  $j$th plane  and   iterative variables $\chi_{j,n}$, $A_{j}$, $B_{j}$, $C_{j,n}$ and  $D_{j,n}$ are defined in Table \ref{Table5}.  The first integration is obtained with  $\Psi_0(x)$  by free propagation until the first slit plane resulting in  $A_0$, $ B_0  $, $\chi_0$ while $C_0= D_0 = 0$. The second LCT results in  $ A_1 $, $B_1    $, $C_{1,n} $, $D_{1,n}$, $\chi_{1,n}   $ where $p_{1, j} $, $\varsigma_j   $, $\xi_{j} $, $\varrho_j $, $\zeta_j $, $\zeta_{j,c} $ and $\zeta_{j,d}$ are defined for $j  \in[1, N-1]$.  Then, the iterations for $A_j$,  $B_j$, $C_{j,n}$, $D_{j,n}$, $\chi_{j,n}$, $p_{2,j}$, $p_{3,j}$, $p_{4,j}$, $p_{5,j}$ and $\chi_{T,j}  \, \equiv \,  \chi_{0}  \prod_{k=1}^{j} \chi_{k,n}$ are  obtained for $j  \in[2, N-1]$. An iterative relation is obtained as follows: 
\begin{equation}
\bigl[\begin{smallmatrix}
C_{j,n}  \\   D_{j,n} 
\end{smallmatrix} \bigr] = \bigl[\begin{smallmatrix}
\zeta_{j,c}  \\    \zeta_{j,d} 
\end{smallmatrix} \bigr] X_{j,s_{n,j}} + \bigl[\begin{smallmatrix}
 p_{4,j}&p_{5,j}  \\  -p_{5,j}&p_{4,j} 
\end{smallmatrix} \bigr] \bigl[\begin{smallmatrix}
C_{j-1,n}  \\  D_{j-1,n} 
\end{smallmatrix} \bigr]
\end{equation}
Performing iterations results in $C_{N-1,n} = \overrightarrow{c}_{N-1}^T \overrightarrow{x}_{N-1,n}$ and $D_{N-1,n} = \overrightarrow{d}_{N-1}^T \overrightarrow{x}_{N-1,n}$ where $j$th element of the vector $\overrightarrow{x}_{N-1,n}$ of the length $N-1$ is defined as $X_{j,s_{n,j}}$, and the row vectors $\overrightarrow{c}_{j}^T$ and $\overrightarrow{d}_{j}^T$ are defined as  follows:
\begin{equation}
\bigl[\begin{smallmatrix}
\overrightarrow{c}_{j}^T  \\  \overrightarrow{d}_{j}^T 
\end{smallmatrix} \bigr] = \bigl[\begin{smallmatrix}
\overrightarrow{v}_{0,j} & \overrightarrow{v}_{1,j}  & \hdots  &   \overrightarrow{v}_{j-1,j}
\end{smallmatrix} \bigr]
 \end{equation}
where   $\overrightarrow{v}_{k, j}$ for $k \in [0, j-1]$, utilized to obtain $C_{j,n} = \overrightarrow{c}_{j}^T \overrightarrow{x}_{j, n}$ and $D_{j, n} = \overrightarrow{d}_{j}^T \overrightarrow{x}_{j, n}$, is given as follows:
\begin{equation}
\overrightarrow{v}_{k, j} \, \equiv \, \bigg(  \prod_{i=1}^{j-1-k} \bigl[\begin{smallmatrix}
p_{4,j+1-i}       & p_{5,j+1-i}  \\ -p_{5,j+1-i}       & p_{4,j+1-i}   
\end{smallmatrix} \bigr] \bigg) \bigl[\begin{smallmatrix}
\zeta_{k+1,c}   \\  \zeta_{k+1,d} 
\end{smallmatrix} \bigr]
 \end{equation}
and the  matrix multiplication symbol $\prod_{i=1}^{k} \mathbf{U}_i$ denotes $\mathbf{U}_1 \, \mathbf{U}_2 \hdots \mathbf{U}_k$ for any matrix $\mathbf{U}_i$ for $i \in [1,k]$.  The following is obtained  after inserting the resulting expressions of $C_{N-1,n}$ and $D_{N-1,n}$ into $\chi_{T, N-1}$:
\begin{align} 
\begin{split}
\chi_{T,N-1}  \equiv &  \,  \chi_0  \bigg( \prod_{j=1}^{N-1}  \sqrt{\xi_{j}} \bigg) \,  e^{ \overrightarrow{p}_1^T \big(  \overrightarrow{x}_{N-1, n}  \odot  \overrightarrow{x}_{N-1,n}  \big) }  \times \, e^{ \overrightarrow{p }_2^T\big( \big(\mathbf{G} \, \overrightarrow{x}_{N-1, n} \big) \odot  \big(\mathbf{G} \, \overrightarrow{x}_{N-1, n} \big) \big) }     \\
& \times \,  e^{\overrightarrow{p}_3^T \big( \big(\mathbf{G} \, \overrightarrow{x}_{N-1, n} \big) \odot \big( \mathbf{E}_{1}\, \overrightarrow{x}_{N-1, n} \big) \big) } 
\end{split}
\end{align}
where $\odot$ denotes the point-wise product,   $\mathbf{G}$ and $\mathbf{E_1}$ are defined as follows:
\begin{equation}
\mathbf{G} \equiv \bigl[\begin{smallmatrix}
\mathbf{E}_{2} \mathbf{V}_{L}        &\mathbf{0_{N-2}}  \\  \mathbf{0^T_{N-2}}      & 0   
\end{smallmatrix} \bigr]; \, \,    \mathbf{E_1} \equiv \bigl[\begin{smallmatrix}
\mathbf{0_{N-2}} & \mathbf{I_{N-2}}  \\   0      &  \mathbf{0^T_{N-2}}  
\end{smallmatrix} \bigr]
 \end{equation}
while  $\mathbf{E}_{2}$ has $j$th row as $\bigl[\begin{smallmatrix}
\mathbf{0^T_{2(j-1)}}  &  1 & \imath &  \mathbf{0^T_{2(N-2-j)}}
\end{smallmatrix} \bigr]$, $\mathbf{V}_{L}$ is the matrix whose $j$th column is given by $\bigl[\begin{smallmatrix}
\mathbf{0^T_{2(j-1)}}  &  \overrightarrow{v}_{j-1,j}^T & \hdots  &  \overrightarrow{v}_{j-1,N-2}^T 
\end{smallmatrix} \bigr]^T$,  $\mathbf{0_{k}}$ is the column vector of zeros of length $k$, the sizes of $\mathbf{E}_{2}$ and $\mathbf{V}_{L}$ are $ (N - 2) \times  (2 N - 4)$ and $(2 N - 4) \times (N - 2)$, respectively, and  $\mathbf{G}$ and $\mathbf{E}_1$ are $(N - 1) \times (N - 1)$. Then,  the resulting wave function $\Psi_{n,N}(x) $ is given by the following:
  \begin{align} 
\begin{split}
\Psi_{n,N}(x) =    & \,\Upsilon_{N}\,   e^{\sum_{k = 1}^3 \overrightarrow{p}_k^T \bigg( \big(\mathbf{M}_{1,k} \, \overrightarrow{x}_{N-1, n} \big) \odot \big(\mathbf{M}_{2,k} \, \overrightarrow{x}_{N-1, n} \big) \bigg) }  \\
 & \times  \,  e^{(A_{N-1} \, + \, \imath \, B_{N-1}) \, x^2 \,  + \, (\overrightarrow{c}_{N-1}^T  \, + \, \imath \, \overrightarrow{d}_{N-1}^T) \, \overrightarrow{x}_{N-1, n} \, x }  \, \, \, \, \, \, \, 
\end{split}
\end{align} 
where $\Upsilon_{j} = \chi_0 \, \big ( \prod_{i=1}^{j-1} \sqrt{\xi_{i}}  \big )$, $\mathbf{M}_{1,1} = \mathbf{M}_{2,1} = \mathbf{I_{N-1}}$, $\mathbf{M}_{1,2} = \mathbf{M}_{2,2} = \mathbf{G}$, $\mathbf{M}_{1,3} = \mathbf{G}$, $\mathbf{M}_{2,3} = \mathbf{E}_{1}$, $\mathbf{I_{k}}$ is  identity matrix of size $k$,  complex valued column vectors $\overrightarrow{p}_{k}$ for $k \in [1, 3]$,  real valued iterative variables $A_j$ and $B_j$, and  complex valued iterative variable $\xi_j$ are defined in  Table \ref{Table5}, and $\overrightarrow{x}_{N-1, n} \equiv  [X_{1,s_{n,1}} \, X_{2,s_{n,2}} \,  \hdots  \, X_{{\color{black}N-1},s_{n, N-1}}  ]^T$. Intensity distribution on screen is  $I_N(x) = \big \vert  \Psi_{N}(x) \big \vert^2 $ which is equal to   $\big \vert \sum_{n=0}^{N_p-1} \, e^{ (A_{N-1} \,   +\, \imath \, B_{N-1} )\, x^2 } $ $ \Upsilon_N \, e^{r \lbrace \overrightarrow{x}_{n}\rbrace } \,e^{\overrightarrow{c}^T   \overrightarrow{x}_{n} \, x}  \,e^{\imath \, \overrightarrow{d}^T    \overrightarrow{x}_{n} \, x} \big \vert^2$ where $I_j(x) =  \big \vert  \Psi_{j}(x) \big \vert^2$ denotes the intensity or the probability of detection on $j$th plane for $j \in [0, N]$, $I_0(x) \equiv \vert \Psi_{0}(x)\vert^2$, the subscript $N - 1$ is dropped from the vectors to simplify the notation, e.g., $ r\lbrace \overrightarrow{x}_{n} \rbrace \, \equiv \,     \sum_{k = 1}^3 \overrightarrow{p}_k^T \big( (\mathbf{M}_{1,k} \, \overrightarrow{x}_{n} ) \odot  (\mathbf{M}_{2,k} \,\overrightarrow{x}_{n} ) \big)$, $\overrightarrow{x}_{n} \equiv \overrightarrow{x}_{N-1, n}$, $\overrightarrow{c} \equiv \overrightarrow{c}_{N-1}$ and $\overrightarrow{d} \equiv \overrightarrow{d}_{N-1}$. It can be easily shown that $r \lbrace \overrightarrow{x}_{n}\rbrace $ is equal to $\overrightarrow{x}_{n}^T \, \mathbf{H}  \, \overrightarrow{x}_{n} $   where the proof is in Appendix \ref{AppendixB}  and the matrix $\mathbf{H}$ is given as  
$\mathbf{H} = \sum_{k=1}^{3} \mathbf{M}_{2,k}^T \, \mbox{diag}\lbrace \overrightarrow{p}_k  \rbrace \, \mathbf{M}_{1,k}$ where $\mbox{diag}\lbrace \overrightarrow{y} \rbrace$ is the operator creating a diagonal matrix with the elements composed of the vector $\overrightarrow{y}$.

\section{Generation of the H-matrix}
\label{AppendixB}
$\sum_{k = 1}^3 \overrightarrow{p}_k^T \big( (\mathbf{M}_{1,k}  \overrightarrow{x}_{n} ) \odot (\mathbf{M}_{2,k}  \overrightarrow{x}_{n} ) \big)$ is transformed to four different equalities. Firstly, it equals to $ \overset{1}{=} \sum_{k = 1}^3 $ $ \mathbf{Tr} \big  \lbrace \mbox{diag}\lbrace  \overrightarrow{p}_k \rbrace $ $  \mathbf{M}_{1,k} $ $ \overrightarrow{x}_{n}  \overrightarrow{x}_{n}^T $ $  \mathbf{M}_{2,k}^T \big  \rbrace$ where the equality is obtained by transforming the inner and  point-wise product  combination into a trace. Then, $\overset{2}{=} $ $ \sum_{k = 1}^3  \mathbf{Tr} $ $\big \lbrace \mathbf{M}_{2,k}^T \, \mbox{diag}\lbrace  \overrightarrow{p}_k \rbrace \, \mathbf{M}_{1,k} \, \overrightarrow{x}_{n} \, \overrightarrow{x}_{n}^T   \big \rbrace $ and $ \overset{3}{=} $ $ \mathbf{Tr} $ $\big \lbrace $ $ \big ( \sum_{k = 1}^3   \mathbf{M}_{2,k}^T   $ $ \mbox{diag}\lbrace  \overrightarrow{p}_k \rbrace $ $  \mathbf{M}_{1,k} \big ) $ $ \overrightarrow{x}_{n} \, \overrightarrow{x}_{n}^T   \big \rbrace$ are obtained due to the permutation and the addition properties of the trace, respectively. Finally, $  \overset{4}{=} $ $ \mathbf{Tr} \big \lbrace $ $ \overrightarrow{x}_{n}^T $ $\big ( \sum_{k = 1}^3   \mathbf{M}_{2,k}^T \, \mbox{diag}\lbrace  \overrightarrow{p}_k \rbrace \, \mathbf{M}_{1,k} \big ) $ $ \overrightarrow{x}_{n} \,    \big \rbrace$ is obtained with the permutation property. Then, the quadratic form is obtained.

\section{Proof of Theorem 1}
\label{proof_theorem1}
The intensity at $ \widetilde{k} - k$ for $k \in [1, \widetilde{k}]$ is  given as follows due to the definition in (\ref{qphase1}) and the first condition in Theorem \ref{theorem1}: 
\begin{equation}
 \widetilde{I}[\widetilde{k} - k]     \overset{1}{ = }    \bigg  \vert  \sum_{n = 0}^{N_p - 1}      g_3[n]  \,  (g_1[n])^{\widetilde{k} - k}   \, e^{  -  \imath \,   2 \, \pi \,  \widetilde{G}_2[n] \, k  \, / \, \widetilde{k}  }   \bigg  \vert^2  = \vert H[ \widetilde{k} \, - \, k,\widetilde{G}_2] \vert^2
 \end{equation}
  Then, $\widetilde{I}[\widetilde{k} - k]  $ $\overset{2}{ < }   $ $ \big \vert $ $  \sum_{n = 0}^{N_p - 1}   $ $    g_3[n]  \,  (g_1[n])^{\widetilde{k} - k}  \big \vert^2 \,  \overset{3}{ < } $ $  \widetilde{I}[\widetilde{k}] = \big \vert  \sum_{n = 0}^{N_p - 1}       g_3[n]  \,  (g_1[n])^{\widetilde{k}} \big \vert^2 $ are obtained with the second  condition  in Theorem \ref{theorem1}. 

\section{Proof of Theorem 2}
\label{proof_theorem2} 

The conditional probability for the sample at $k_p$ is given by the following:
 \begin{equation}
 p(I^{*}_n[k_p] \big \vert \widetilde{k})  = (2 \, \pi \, \widetilde{\sigma}^2_p)^{-1/2} e^{- (I^{*}_n[k_p]  - I^{*}[k_p])^2  \, / \, (2 \, \widetilde{\sigma}^2_p)} 
 \end{equation}
where $\widetilde{\sigma}_p \equiv  \sigma^{*}[k_p]$. Then, denoting the noisy and noise-free intensity vectors by  $\overrightarrow{I}^{*}_n = [I^{*}_n[k_0] \hdots I^{*}_n[k_{M-1}]]^T$ and $\overrightarrow{I}^{*} = [I^{*}[k_0] \hdots I^{*}[k_{M-1}]]^T$, respectively, the log likelihood function is given as  $ \mbox{log}\big( p(\overrightarrow{I }^{*}_n \big \vert \widetilde{k}) \big) = - \frac{M}{2} \, \mbox{log}(2 \, \pi) \,  - \, \frac{1}{2} \sum_{p = 0}^{M-1}  \mbox{log}(\widetilde{\sigma}_p^2)  \,
- \, (\overrightarrow{I}^{*}_{n, \widetilde{\sigma}} - \overrightarrow{I}^{*}_{\widetilde{\sigma}})^T \cdot (\overrightarrow{I}^{*}_{n, \widetilde{\sigma}} - \overrightarrow{I}^{*}_{\widetilde{\sigma}}) $  where $ I^{*}_{n, \widetilde{\sigma}}[k_p] \equiv I^{*}_n[k_p]\, / \, (\widetilde{\sigma}_p  \, \sqrt{2})$ and $I^{*}_{\widetilde{\sigma}}[k_p] \equiv  I^{*}[k_p]\, / \,( \widetilde{\sigma}_p\, \sqrt{2})$. Fisher information matrix is given as follows:
\begin{equation}
I_F[\widetilde{k}]\equiv E \bigg \lbrace \bigg(\delta \, \mbox{log}\big( p(\overrightarrow{I}^{*}_n  \big \vert \widetilde{k}) \big) \,/\,\delta \widetilde{k} \bigg)^2 \bigg\rbrace \,= \,-  \,E \lbrace    \delta^2   \mbox{log}\big( p(\overrightarrow{I}^{*}_n  \big \vert \widetilde{k}) \big)  \,  / \,\delta \widetilde{k}^2 \rbrace  
\end{equation}
 where $\delta (.) \, / \, \delta \widetilde{k}$ denotes the partial derivative of $(.)$ with respect to $\widetilde{k}$. If the zero mean random variable is assumed at each sample point, then $I_F[\widetilde{k}]$ is obtained after simple calculations as   $I_F[\widetilde{k}] = \sum_{p=0}^{M-1}  \widetilde{\sigma}_p^{-2}$ $ \big( \delta I^{*}[k_p] \, / \, \delta \, \widetilde{k}\big)^2$
 which depends on the square of the derivative of the intensity on the period $\widetilde{k}$. Then, assuming an estimation method denoted by $\widehat{k}$ has a bias $b(\widehat{k}) \equiv E \lbrace \widehat{k} \rbrace - \widetilde{k} $, the Cramer-Rao Bound, i.e., $CR(\widetilde{k})$, satisfies $\mbox{Var}(\widehat{k})  \, \geq \, CR(\widetilde{k})$ for the variance of estimation where $ CR(\widetilde{k})$ is given by the following:
\begin{equation}
  CR(\widetilde{k}) \, \equiv \, (1 \, + \,  \delta b(\widehat{k}) \, / \, \delta  \widetilde{k})^2  \, / \,I_F[\widetilde{k}] \, = \,  (1 \, + \,  \delta b(\widehat{k}) \, / \, \delta  \widetilde{k})^2  \bigg( \sum_{p=0}^{M-1}  \widetilde{\sigma}_p^{-2}  \big(  \delta I^{*}[k_p] \,  / \, \delta \, \widetilde{k} \big)^2 \bigg)^{-1}
    \end{equation}
Furthermore, assuming $\widetilde{\sigma}_p^2 \leq \widetilde{\sigma}_{max}^2$, the maximum of the minimum variance bound is given by   the following:
\begin{equation}
CR(\widetilde{k})  \leq  
    \widetilde{\sigma}_{max}^2  \, (1 \, + \,  \delta b(\widehat{k}) \, / \, \delta  \widetilde{k})^2    \bigg(\sum_{p=0}^{M-1}    \big(  \delta I^{*}[k_p] \, / \, \delta \, \widetilde{k}\big)^2 \bigg)^{-1} 
\end{equation} 
while with $\Delta G_2[n, l] \equiv \widetilde{G}_2[l]\,-\,\widetilde{G}_2[n]$, $\delta I^{*}[k_p] \, / \, \delta \, \widetilde{k}$ becomes as follows for $I^{*}[k_p] \equiv I^G[k_p]$:
\begin{equation}
  \sum_{n = 0}^{N_p - 1}     \sum_{l = 0}^{N_p - 1}     g_{3,*}[n,l] \,  g_{1,*}^{k_p}[n,l]  \, g_{4,*}^{k_p^2}[n,l]    e^{ -\Delta G_2[n, l] \, \imath \, \, 2 \, \pi \, k_p \, / \, \widetilde{k} }   \big (\Delta G_2[n, l] \, \imath \,   2 \, \pi \, k_p  \, / \, \widetilde{k}^2 \big) 
 \end{equation}
and it is represented as follows for the normalized wave function  $I^{*}[k_p] \equiv e^{- 2\, A_{N-1} \, (k_p\,T_s)^2} I[k_p]$:
\begin{equation}
\sum_{n = 0}^{N_p - 1}     \sum_{l = 0}^{N_p - 1}     g_{3,*}[n,l] \,  g_{1,*}^{k_p}[n,l]   e^{ -\Delta G_2[n, l] \, \imath \, \, 2 \, \pi \, k_p \, / \, \widetilde{k} }   \big (\Delta G_2[n, l] \, \imath \,   2 \, \pi \, k_p  \, / \, \widetilde{k}^2 \big) 
 \end{equation}
 
\section{Path Integral with Exotic Paths}
\label{proof5} 
The evolution of the wave function in  $n$th  path after  the non-classical travels of $k$  slits with $k \in [1, N_E]$ as shown in Fig. \ref{Figure3} is  given as   $\Psi_{n,j,k}^E(x^E_{j,k}) =    \int_{x_j} 
 f^E_{n,k}(x^E_{j,k},  x_{j})  \, \Psi_{n,j,0}^E(x^E_{j,0}) \, \diff  x_j$
where  $\Psi_{n,j,0}^E(x^E_{j,0}) \equiv  G_{n,j}(x_{j}  - X_{j, s_{n,j}})   \, \Psi_{n, j}(x_j)$,  $x_{j, 0}^E \equiv x_j$, $f^E_{n,1}(x^E_{j,1},  x_{j}) \equiv K(x^E_{j,1}, t^E_{j,1}; x_{j}, t_{j}) $ and $f^E_{n,k}(x^E_{j,k},  x_{j})$ for $ k \geq 2$ is defined as follows:
\begin{equation}
\int_{\overrightarrow{x}^E_{j, k}}   \diff  \overrightarrow{x}^E_{j, k} \, K(x^E_{j,1}, t^E_{j,1}; x_{j}, t_{j})  \prod_{p = 2}^{k}  K(x^E_{j,p}, t^E_{j,p}; x^E_{j,p-1}, t^E_{j,p-1})  \, G_{n,j}(x^E_{j,p-1}  - X_{j, s_{n,j,p-1}})
 \end{equation}
while $k = 0$ case corresponds to the wave function evolution without any non-classical path, i.e., $\Psi_{n,j,0}^E(x^E_{j,0})$,    $t^E_{j,k} \equiv \sum_{p = 1}^{k}  t_{p-1,p}^E(j) \, + \, t_j$ is the time after visiting $k$th slit on $j$th plane, $t_j$ corresponds to the time at the beginning of the non-classical movements and $\overrightarrow{x}^E_{j, k}  \equiv [ x_{j, 1}^E \, \,  x_{j,2}^E \, \,  \hdots \, \,  x_{j, k-1}^E ]$.  If it is assumed that the $n$th path performs $k \geq 1$ consecutive visits to the slits on $j$th plane while the entrance slit is $X_{j, s_{n,j}}$ and the wave function at the position $x_j$ is $\Psi_{n, j}(x_j)$, then the wave function on the next plane, i.e., $\Psi_{n,j+1}(x_{j+1})$, is calculated as follows:
\begin{equation}
\int K(x_{j+1}, t^E_{j,k}   + \, t_{j,j+1}; x^E_{j,k}, t^E_{j,k})  \, G_{n,j}(x^E_{j,k}  - X_{j, s_{n,j,k}})   \, \Psi_{n,j,k}^E(x^E_{j,k}) \, \diff  x^E_{j,k}
\end{equation}

\begin{acknowledgements}
{\color{black}I would like to thank the referees for very helpful comments and suggestions.} 
\end{acknowledgements}


\end{document}